\documentclass[10pt,oneside,letterpaper]{article}
\usepackage{epsfig}
\usepackage{bm}
\usepackage{amsfonts}
\usepackage{amsmath,amssymb}
\usepackage{dcolumn}
\PassOptionsToPackage{linktocpage}{hyperref}
\usepackage[draft=false]{hyperref}
\usepackage{threeparttable}
\usepackage[draft=false]{hyperref}
\usepackage{threeparttable}
\def\case#1/#2{\textstyle\frac{#1}{#2}}

\def\bea{\begin{eqnarray}}
\def\eea{\end{eqnarray}}
\def\case#1/#2{\textstyle\frac{#1}{#2}}
\newtheorem{thm}{Theorem}

\newtheorem{rem}{Remark}
\newtheorem{prop}{Proposition}
\newtheorem{defn}{Definition}

\newcommand{\vphi}{\varphi}

\newcommand{\be}{\begin{equation}}
\newcommand{\ee}{\end{equation}}
\newcommand{\ben}{\begin{eqnarray}}
\newcommand{\een}{\end{eqnarray}}

\begin{document}

\title{Phase-space of flat Friedmann-Robertson-Walker models with both a scalar field coupled to
matter and radiation}
\author{Genly Leon\thanks{Email address: genly@uclv.edu.cu},
Pavel Silveira\thanks{Email address: pavel.silveira@gmail.com}
and Carlos R. Fadragas\thanks{Email address: fadragas@uclv.edu.cu}
\\
Department of Mathematics, \\
Universidad Central de Las Villas, Santa Clara \enskip CP 54830, Cuba and\\
Department of Physics, \\
Universidad Central de Las Villas, Santa Clara \enskip CP 54830, Cuba\\
}

\date{\today}
\maketitle
\thispagestyle{empty}
\setcounter{page}{1}

\begin{abstract}
We investigate the phase-space of a flat FRW universe including both a scalar field, $\phi,$ coupled to matter, and radiation. The model is inspired in scalar-tensor theories of gravity, and thus, related with $F(R)$ theories through conformal transformation.
The aim of the chapter is to extent several results to the more realistic situation when radiation is included in the cosmic budget particularly for studying the early time dynamics.  Under mild conditions on the potential we prove that the equilibrium points corresponding to the non-negative local minima for $V(\phi)$ are asymptotically stable.  Normal forms are employed to obtain approximated solutions associated to the inflection points and the strict degenerate local minimum of the potential. We prove for arbitrary potentials and arbitrary coupling functions $\chi(\phi),$ of appropriate differentiable class, that the scalar field almost always diverges into the past.
It is designed a dynamical system adequate to studying the stability of the critical points in the limit $|\phi|\rightarrow\infty.$ We obtain there: radiation-dominated cosmological solutions; power-law scalar-field dominated inflationary cosmological solutions; matter-kinetic-potential scaling solutions and radiation-kinetic-potential scaling solutions.
Using the mathematical apparatus developed here, we investigate the important examples of higher order gravity theories $F(R) = R + \alpha R^2$ (quadratic gravity) and $F(R) =R^n.$ We illustrated both analytically and numerically our principal results. In the case of quadratic gravity we prove, by an explicit computation of the center manifold, that the equilibrium point corresponding to de Sitter solution is locally asymptotically unstable (saddle point).

\end{abstract}

\vspace{2in}


\vspace{.08in} \noindent \textbf{Keywords:} Relativity and
gravitation, Dark energy, Asymptotic structure.

%% Other situations:
%\noindent \textbf{Key Words}: inflation, cosmological parameters
\vspace{.08in} \noindent \textbf{AMS Subject Classification:}
83F05, 83C05, 83C30.

\tableofcontents

\section{Introduction}

Modern Cosmology is a broad and promising area of research in applied mathematics and applied physics based on the observed data available in the modern astronomical literature. One of the most celebrated discoveries of observational cosmology is that the observable universe is now known to be accelerating \cite{Perlmutter:1998np,Bennett:2003bz,Tegmark:2003ud,Allen:2004cd}, and this feature led physicists to follow two directions in order to explain it. The first is to introduce the concept of dark energy (see the reviews \cite{Copeland:2006wr,Caldwell:2009ix,Turner:2007qg,Alam:2004jy,Leon:2009ce} and references therein) in the right-hand-side of the field equations, which could either be the simple cosmological constant $\Lambda$ \footnote{This choice is seriously plagued by the well known coincidence and fine tuning problems \cite{Sahni:2002kh,Carroll:1991mt,Sahni:1999gb,Weinberg:1988cp}.} or, one or several scalar fields \cite{Tsujikawa:2006mw,Tsujikawa:2005ju,Amendola:2004be,Beck:2003wj,Maor:2002rd}. The second is looking for alternative models \cite{Sami:2009dk,Alcaniz:2006ay,Mannheim:2005bfa}. Alternative approaches to dark energy are the so-called Extended Theory of Gravitation (ETG) and, in particular, higher-order theories of gravity (HOG) \cite{Carroll:2004de,Carroll:2003wy,Capozziello:2003gx,Capozziello:2003tk,Capozziello:2007ec,Wands:1993uu,Capozziello:2010wt,Capozziello:2009yj,Capozziello:2009nq,Capozziello:2007zz,Capozziello:2008gu,Capozziello:2002rd,Faraoni:2005vk,Ruggiero:2006qv,Ruggiero:2006qv,delaCruzDombriz:2006fj,Poplawski:2006kv,Brookfield:2006mq,Song:2006ej,Li:2006ag,Sotiriou:2006hs,Bertolami:2009ac,Bertolami:2009cd,Bertolami:2009ac,Bertolami:2007gv,Briscese:2006xu,Leon:2010pu}. Such an approach can still be in the spirit of General Relativity Theory (GRT) since the only request the Hilbert-Einstein action should be generalized (by including non-linear terms in the Ricci curvature $R$ and/or involving combinations of derivatives of $R$
\cite{Kerner:1982yg,Duruisseau:1986ga,Teyssandier:1989dw,Magnano:1987zz}) asking for a gravitational interaction acting, in principle, in different ways in both cosmological
\cite{Capozziello:2003gx,Capozziello:2003tk,Capozziello:2006uv,Maeda:2004vm,Maeda:2004hu,Ohta:2004wk,Akune:2006dg} and astrophysical \cite{Capozziello:2006uv,Capozziello:2006ph} scales. In this case the field equations can be recast in a way that the higher order corrections are written as an energy-momentum tensor of geometrical origin describing an ``effective" source term on the right hand side of the standard Einstein field equations \cite{Capozziello:2003gx,Capozziello:2003tk}. These models have been studied from the dynamical systems viewpoint in
\cite{Leon:2010pu,Carloni:2004kp,Carloni:2007eu,Carloni:2009nc,Carloni:2007br,Carloni:2004kp,Leach:2006br,Leach:2007ss,Goheer:2007wx,Goheer:2007wu,Goheer:2008tn,Goswami:2008fs,Miritzis:2009bi,Abdelwahab:2007jp}.
On the other hand, scalar fields have played an essential role as models for the early-time universe. In the inflationary universe scenarios (mainly based on GRT) matter is modeled, usually, as a scalar field, $\phi,$ with potential $V(\phi)$, which must meet the requirements necessary to lead to the early-time accelerating expansion \cite{Billyard:1999a,Copeland:2004b,Copeland:1993,Kolb:1995,Lidsey:1997}.
If the potential is constant, i.e., if $V (\phi) = V_0$, space-time is de Sitter and expansion is exponential. If the potential is exponential, i.e., $V (\phi) = V_0 \exp[-\lambda\phi]$, we get an inflationary powerlaw solution \cite{Lucchin:1984yf,Burd:1988}.
``Extended'' inflation models \cite{Barrow:1990nv,Faraoni:2006ik,Liddle:1991am,La:1989za}, on the other hand, use the Brans-Dicke theory (BDT) \cite{Brans:1961sx} as the correct theory of gravity, and in this case the vacuum energy leads directly to a powerlaw solution \cite{Mathiazhagan:1984vi} while the exponential expansion can be obtained if a cosmological constant is explicitly inserted into the field equations \cite{Barrow:1990nv,Kolitch:1994kr,Romero:1992ci}.

Several gravity theories consider multiple scalar fields with exponential potential, particularly assisted inflation scenarios \cite{Chimento:1998,Guo:2003b,Coley:2000,Copeland:1999,Malik:1999},
quintom dark energy paradigm \cite{Guo:2004fq,Zhang:2005eg,Lazkoz:2006pa,Lazkoz:2007mx} and others. Also, have been considered positive and negative exponential potentials \cite{Heard:2002}, single exponential and double exponential
\cite{Huang:2006ku,Arias:2003,Gonzalez:2007,Gonzalez:2006}, etc.
Other generalizations with multiple scalar fields are available \cite{vandenHoogen:2000cf,Damour:1992d}.

In BDT, a scalar field, $\chi,$ acts as the source for the gravitational coupling with a varying Newtonian 'constant'  $G\sim \chi^{-1}.$ BDT is the first propotype of the so-called Scalar-Tensor theories (STT) of gravity \cite{Wagoner:1970vr,O'Hanlon:1972my,O'Hanlon:1972hq,Bekenstein:1977rb,Bergmann:1968ve,Nordtvedt:1970uv}.
It is worthy to mention that BDT survive several observational tests including Solar System tests \cite{Abramovici:1992ah} and Big-Bang nucleosynthesis constraints \cite{Serna:2002fj,Serna:1995tr}. More general STT with a non-constant BD parameter $\omega(\chi),$ and non-zero self-interaction potential $V(\chi),$ have been formulated, and also survive astrophysical tests
\cite{Will:2005va,Will:1993ns,Barrow:1996kc}.

To our knowledge, the dynamical behavior of space-times based on GRT is so far known for a large variety of models with  scalar fields with non-negative potential
\cite{Foster:1998sk,Miritzis:2003ym,Miritzis:2005hg,Rendall:2004ic,Rendall:2006cq,Hertog:2006rr}.
In reference \cite{Hertog:2006rr}, have been extended many of the results obtained in \cite{Miritzis:2003ym} considering arbitrary potentials. In \cite{Foster:1998sk} it has been shown that for a large class of FRW cosmologies with scalar fields with arbitrary potential, the past attractor is a family of solutions in one-to-one correspondence with exactly integrable cosmologies with a massless scalar field. This result has been extended somewhat in \cite{Leon:2008de} to FRW cosmologies based on STTs. In this reference was investigated a general model of coupled dark energy with arbitrary potential $V$ and coupling function $\chi$.  It was proved there, by using dynamical systems techniques that if the potential and the coupling function are sufficiently smooth functions; the scalar field almost always diverges into the past.
Under some regularity conditions for the potential and for the coupling function in that limit, it was constructed a dynamical system well suited to investigate the dynamics near the initial singularity. The critical points therein were investigated and the cosmological solutions associated to them were characterized. There was presented asymptotic expansions for the cosmological solutions near the initial space-time singularity, which extend previous results of \cite{Foster:1998sk}. On the other hand, in \cite{Giambo':2009cc} it was investigated flat and negatively curved Friedmann-Robertson-Walker (FRW) models with a perfect fluid matter source and a scalar field arising in the conformal frame of $F(R)$ theories nonminimally coupled to matter. It was proved there that, for a general class of potentials $V,$ the equilibrium corresponding to non-negative local minima for $V$ are asymptotically stable, as well as horizontal asymptotes approached from above by $V$. For a nondegenerate minimum of the potential with zero critical value they prove in detail that if $\gamma>1$, then there is a transfer of energy from the fluid to the scalar field and the later eventually dominates in a generic way. As we will see in next sections the results in \cite{Giambo':2009cc} and in \cite{Leon:2008de} can be obtained by investigating a general class of models containing both STTs and $F(R)$ gravity.

The aim of the chapter is to extent several results in \cite{Foster:1998sk,Miritzis:2003ym,Leon:2008de,Giambo':2009cc} to the more realistic situation when radiation is included in the cosmic budget (particularly for investigating the early-time dynamics). We will focus mainly in a particular era of the universe where matter and radiation coexisted. Otherwise the inclusion of radiation complicates the study in an unnecessary manner, since, assuming a perfect barotropic fluid with an arbitrary barotropic index $\gamma$, for $\gamma= 4/3$, this matter source corresponds to radiation.
Thus we will consider both ordinary matter described by a perfect fluid with equation of state $p=(\gamma-1)\rho$ (coupled to a scalar field) and radiation with energy density $\rho_r$. We are interested in investigate all possible scaling solutions in this regime. Although we are mainly interest in describing the early time dynamics of our model, for completeness we will focus also in the late-time dynamics. As in \cite{Giambo':2009cc} we obtain for flat FRW models sufficient conditions under the potential, to establish the asymptotic stability of the non-negative local minima for $V(\phi).$ Normal forms are employed to obtain approximated solutions associated to the local degenerated minimum and the inflection points of the potential. We prove for arbitrary potentials and arbitrary coupling functions $\chi(\phi),$ of appropriate differentiable class, that the scalar field almost always diverges into the past generalizing the results in \cite{Foster:1998sk,Leon:2008de}). It is designed a dynamical system adequate to studying the stability of the critical points in the limit $|\phi|\rightarrow\infty.$ We obtain there: radiation-dominated cosmological solutions; power-law scalar-field dominated inflationary cosmological solutions; matter-kinetic-potential scaling solutions and radiation-kinetic-potential scaling solutions. It is discussed, by means of several worked examples, the link between our results and the results obtained for specific $F(R)$ frameworks by using appropriated conformal transformations. We illustrated both analytically and numerically our principal results. Particularly, we investigate the important examples of higher order gravity theories $F(R) = R + \alpha R^2$ (quadratic gravity) and $F(R) =R^n.$  In the case of quadratic gravity we prove, by an explicit computation of the center manifold, that the equilibrium point corresponding to \emph{de Sitter} solution is locally asymptotically unstable (saddle point). This result complements the result of the proposition discussed in \cite{Miritzis:2005hg} p. 5, where it was proved the local asymptotic instability of the de Sitter universe for positively curved FRW models with a perfect fluid matter source and a scalar field which arises in the conformal frame of the $R+\alpha R^2$ theory.
Finally, we investigate a general class of potentials containing the cases investigated in \cite{Copeland:1997et,vandenHoogen:1999qq}. In order to provide a numerical elaboration for our analytical results for this class of models, we re-examine the model with power-law coupling and Albrecht-Skordis potential $V(\phi )=e^{-\mu \phi }{\left( A+(\phi -B)^2\right)}$ investigated in \cite{Leon:2008de} in presence of radiation.

\section{Brief review on dynamical systems analysis of cosmological models}

Qualitative methods have been proved to be a powerful scheme for investigating the physical behavior of cosmological models.
It has been used three different approaches: approximation by parts, Hamiltonian methods, and dynamical systems methods \cite{WE}. In the third case the Einstein's field equations of Bianchi's cosmologies and its isotropic subclass (FLRW models), can be written as an autonomous system of first-order differential equations whose solution curves partitioned to $\mathbb{R}^ n$ in orbits, defining a dynamical system in
$\mathbb{R}^ n$. In the general case, the elements of the phase space partition (i.e., critical points, invariant sets, etc.) can be listed and described. This study consists of several steps: determining critical points, the linearization in a neighborhood of them, the search for the eigenvalues of the associated Jacobian matrix, checking the stability conditions in a neighborhood of the critical points, the finding of the stability and instability sets and the determination of the basin of attraction, etc. In some occasions, in order to do that, it is needed to simplify a dynamical system. Two approaches are applied to this objective: one, reduce the dimensionality of the system and two, eliminate the nonlinearity. Two rigorous mathematical techniques that allow substantial progress along both lines are center manifold theory and the method of normal forms. We will use both approaches in our analysis. We submit the reader to sections \ref{sectionCM} and \ref{sectionNF} for a summary about such techniques.

The most general result to be applied in order to determine the asymptotic stability of a equilibrium point, $a,$ is Liapunov's stability theorem. Liapunov's stability method provides information not only about the asymptotic stability of a given equilibrium point but also about its basin of attraction. This cannot be obtained by the usual methods found in the literature, such as linear stability analysis or first-order perturbation techniques. Moreover, Liapunov's method is also applicable to non-autonomous systems. To our knowledge, there are few works that use Liapunov's method in cosmology  \cite{Setare:2010zd,Cardoso:2008bp,Lavkin:1990gu,Charters:2001hi,Aref'eva:2009xr}. In \cite{Charters:2001hi} it is investigated the general asymptotic behavior of Friedman-Robertson-Walker (FRW) models with an inflaton field, scalar-tensor FRW cosmological models and diagonal Bianchi-IX models by means of Liapunov's method. In \cite{Aref'eva:2009xr} it is investigated the stability of isotropic cosmological solutions for two-field models in the Bianchi I metric. The author proved that the conditions sufficient for the Lyapunov stability in the FRW metric provide the stability with respect to anisotropic perturbations in the Bianchi I metric and with respect to the cold dark matter energy density fluctuations. Sufficient conditions for the Liapunov's stability of the isotropic fixed points of the system of the Einstein equations are also found (these conditions coincided with the previously obtained in \cite{Lazkoz:2007mx} for the quintom paradigm without using Liapunov's technique).
To apply Liapunov's stability method it is required the construction of the so-called strict Liapunov's function, i.e., a $C^1$ function $V:U\subset\mathbb{R}^n\rightarrow \mathbb{R}$ defined in a neighborhood $U$ of $a$ such that $V(a)=0,
V(x)>0, x\neq a$ and $\dot V(x)\leq 0$ ($<0$) in $U\setminus \{a\}.$
The construction of such V is laborious, and sometimes impossible. One alternative way is to follow point (5) in section \ref{Procedure}.

For the investigation of hyperbolic equilibrium points of autonomous vector fields we can use Hartman-Grobman's theorem (theorem 19.12.6 en \cite{wiggins} p. 350) which allows for analyzing the stability of an equilibrium point from the linearized system around it. For isolated nonhyperbolic equilibrium points we can use normal forms theorem (theorem 2.3.1 in \cite{arrowsmith}), which contains Hartman-Grobman's theorem as particular case. The aim of the normal form calculation is to construct a sequence of non-linear transformations which successively remove the non-resonant terms in the vector field of order $r$ in the Taylor's expansion, ${\bf X}_r,$ starting from $r=2$. Normal forms (NF) theory has been used in the context of cosmological models in order to get useful information about isolated nonhyperbolic critical points. In \cite{Leon:2009rc} was investigated Ho\v{r}ava-Lifshitz cosmology from the dynamical system view point. There was proved the stability of de Sitter solution (corresponding to a non-hyperbolic critical point) in the case where the detailed-balance condition is relaxed, using NF expansions. In \cite{Miritzis:2007yn} were investigated closed isotropic models in second order gravity. There the normal form of the dynamical system has periodic solutions for a large set of initial conditions. This implies that an initially expanding closed isotropic universe may exhibit oscillatory behavior.

The more relevant concepts of qualitative theory are the concept of flow, and the concept of invariant manifold. The invariant manifold theorem (theorem 3.2.1 en \cite{wiggins}), that claims for the existence of local stable and unstable manifolds (under suitable conditions for the vector field), only allows to obtain partial information about the stability of equilibrium points and does not gives a method for determining the stable and unstable manifolds. Sometimes it is required to consider higher order terms in the Taylor's expansion of the vector field (e.g. normal forms).
For the investigation of the asymptotic states of the system the appropriated concepts are $\alpha$ and $\omega$-limit sets of
$x\in \mathbb{R}^n$ (definition 8.1.2 en \cite{wiggins} p. 105).
To characterize these invariant sets one can use the LaSalle's
Invariance Principle or the Monotonicity Principle (\cite{WE},
p. 103; \cite{LeBlanc:1994qm} p. 536). To apply the Monotonicity
Principle it is required the construction of a monotonic function.
In some cases a monotonic function is suggested by the form of the differential equation (see equation \eqref{Eq1} below) and in some
cases by the Hamiltonian formulation of the field equations. In
\cite{Heinzle:2009zb} is given a prescription of how to find
monotonic functions for Bianchi cosmologies using Hamiltonian
techniques, merely as a convenient tool for an intermediate step;
the final results are described in terms of scale-automorphism
invariant Hubble-normalized reduced state vector, which is
independent of a Hamiltonian formulation. Also, one can use the
Poincar\'e-Bendixson (\cite{Coley:2003mj} p. 22, theorem 2
\cite{Coley:1999uh} p. 6, \cite{Hirsch}) theorem and its
collorary to distinguishing among all possible $\omega$-limit sets
in the plane. From the Poincar\'e-Bendixon Theorem follows that
any compact asymptotic set is one of the following; 1.) a critical
point, 2.) a periodic orbit, 3.) the union of critical points and
heteroclinic or homoclinic orbits. As a consequence if the
existence of a closed (i.e., periodic, heteroclinic or homoclinic)
orbit can be ruled out it follows that all asymptotic behavior is
located at a critical point. To ruled out a closed orbits for
two-dimensional systems we can use the Dulac's criterion (theorem
3 \cite{Coley:1999uh} p. 6, se also \cite{WE}, p. 94, and
\cite{Coley:2003mj}). It requires the construction of a Dulac's
function. A Dulac's function, $B,$ is a $C^1$  function defined in
a simply connected open subset $D \subseteq R^2$ such that
$\nabla(Bf) = \frac{\partial}{\partial \sigma_2}
    (Bf_1)+\frac{\partial}{\partial \sigma_3}(Bf_2) > 0, $ or $(<0)$
for all $x \in D$ (see \cite{Coley:1999uh}).

Finally, in order to the chapter to be self-contained, let us formulate (without proof) some of the results outlined above that we shall use in our proofs.

\subsection{Some terminology and results from the Dynamical Systems Theory}\label{appendixA}

In the following $\phi(\tau,\mathbf{x})$ denotes the flow generated by the vector field (or differential equation)

\begin{equation}{\mathbf{x}}'(\tau)=\mathbf{f}(\mathbf{x}(\tau)),\,\mathbf{x}(\tau)\in\mathbb{R}^n,\label{vectorfield}\end{equation} where the prime denote derivative with respect to $\tau.$

\subsubsection{Limit sets}

\begin{defn}[definition 8.1.1, \cite{wiggins} p. 104]\label{omegalimitpoint} A point $\mathbf{x}_0\in\mathbb{R}^n$ is called an $\omega$-limit point of
$\mathbf{x}\in\mathbb{R}^n,$ denoted $\omega(\mathbf{x}),$ if there exists a sequence $\{\tau_i\},\,\tau_i\rightarrow\infty$ such that
$\phi(\tau_i,\mathbf{x})\rightarrow \mathbf{x}_0.$ $\alpha$-limits are defined similarly by taking a sequence $\{\tau_i\},\,\tau_i\rightarrow -\infty.
$

\end{defn}

\begin{defn}[definition 8.1.2, \cite{wiggins} p. 105]\label{omegalimitset} The set of all $\omega$-limit points of a flow or map is called a
$\omega$-limit set. The $\alpha$-limit is similarly defined.

\end{defn}

\begin{prop}[proposition 8.1.3 \cite{wiggins}, p. 105]\label{omegalimitsetproperties} Let $\phi(\tau,\cdot)$ be a flow generated by a vector field and
let $M$ be a positively invariant compact set for this flow (see
definition 3.0.3 p. 28 \cite{wiggins}). Then for $\mathbf{p}\in M,$ we
have

\begin{enumerate}
\item[i)] $\omega(\mathbf{p})\neq\emptyset$
\item[ii)] $\omega(\mathbf{p})$ is closed
\item[iii)] $\omega(\mathbf{p})$ is invariant under the flow, i.e., $\omega(\mathbf{p})$ is a union of orbits.
\item[iv)] $\omega(\mathbf{p})$ is connected.
\end{enumerate}

\end{prop}

\subsubsection{Monotone functions and Monotonicity Principle}

\begin{defn}[definition 4.8 \cite{WE}, p. 93]\label{Definition 4.8} Let $\phi(\tau,\mathbf{x})$ be a flow on $\mathbb{R}^n,$ let $S$ be an invariant set of $\phi(\tau,\mathbf{x})$ and let $Z:S\rightarrow\mathbb{R}$ be a continuous function. $Z$
is monotonic decreasing (increasing) function for the flow
$\phi(\tau,\mathbf{x})$ means that for all $\mathbf{x}\in S,$ $Z(\phi(\tau,\mathbf{x}))$ is a monotonic decreasing (increasing) function of $\tau.$

\end{defn}

\begin{prop}[Proposition 4.1, \cite{WE}, p. 92]\label{Prop 4.1}
Consider a differential equation \eqref{vectorfield} with
flow $\phi(\tau,\mathbf{x}).$ Let $Z:\mathbb{R}^n\rightarrow \mathbb{R}$ be a $C^1\left(\mathbb{R}^n\right)$ function which satisfies
$Z'=\alpha Z,$ where $\alpha: \mathbb{R}^n\rightarrow \mathbb{R}$
is a continuous function. Then, the subsets of
$\mathbb{R}^n$ defined by  $Z>0,$ $Z=0,$ or $Z<0$ are invariant sets for $\phi(\tau,\mathbf{x}).$
\end{prop}

\begin{thm}[Monotonicity Principle]\label{theorem 4.12}

Let $\phi(\tau,\mathbf{x})$ be a flow on $\mathbb{R}^n$ with $S$ an invariant
set. Let $Z: S\rightarrow\mathbb{R}$ be a
$C^1\left(\mathbb{R}^n\right)$ function whose range is the
interval $(a,\;b)$ where $a\in \mathbb{R} \cup \{-\infty\},$ $b\in
\mathbb{R} \cup \{+\infty\},$ and $a<b.$ If $Z$ is decreasing on
orbits in $S,$ then for all $\mathbf{x}\in S$, $\omega(\mathbf{x})\subset\{\mathbf{s}\in \bar
S\setminus S|lim_{\mathbf{y}\rightarrow \mathbf{s}} Z(\mathbf{y})\neq b \}$ and
$\alpha(\mathbf{x})\subset\{\mathbf{s}\in \bar S\setminus S|lim_{\mathbf{y}\rightarrow \mathbf{s}} Z(\mathbf{y})\neq a
\}.$
\end{thm}

\subsubsection{Results for planar systems}

\begin{thm}[Poincar\'e-Bendixon Theorem]\label{PBT} Consider the differential equation \eqref{vectorfield} on $\mathbb{R}^2$, with $\mathbf{f} \in C^2$, and suppose that there are at most a finite number of singular points (i.e., no non-isolated singular points).  Then any compact asymptotic set is one of the following;
1. a singular point, \enskip
2. a periodic orbit, \enskip
3. the union of singular points and heteroclinic or homoclinic orbits.
\end{thm}

\begin{thm}[Dulac's Criterion]\label{DC}  If $D \subseteq R^2$ is a simply connected open set and  $B$ is a Dulac's function on $D$, then the differential equation \eqref{vectorfield} on $\mathbb{R}^2$,
with $\mathbf{f} \in C^1$ has no periodic (or closed) orbit which is contained in $D$.
\end{thm}

\subsubsection{Normal Forms}\label{sectionNF}

In this section we offer the main techniques for the construction of normal forms for vector fields in $\mathbb{R}^n$. We follow the approach in \cite{arrowsmith}.

Let ${\bf X}:\mathbb{R}^n\rightarrow \mathbb{R}^n$ be a smooth vector field satisfying ${\bf X}({\bf 0})={\bf 0}.$ We can formally construct the Taylor expansion of ${\bf x}$ around ${\bf
0},$ namely, ${\bf X}={\bf X}_1+{\bf X}_2+\ldots +{\bf
X}_k+{O}(|{\bf x}|^{k+1}),$ where ${\bf X}_r\in H^r,$ the real vector space of vector fields whose components are homogeneous polynomials of degree $r$. For $r=1$ to $k$ we write

\ben
&{\bf X}_r({\bf x})=\sum_{m_1=1}^{r}\ldots\sum_{m_n=1}^{r}\sum_{j=1}^{n}{\bf X}_{{\bf m},j}{{\bf x}}^{{\bf m}}{\bf e}_j,\nonumber\\
& \sum_i m_i=r, \een

Observe that ${\bf X}_1={\bf DX(\mathbf{0})}{\bf x}\equiv {\bf
A}{\bf x},$ i.e., the Jacobian matrix.

The aim of the normal form calculation is to construct a sequence
of transformations which successively remove the non-linear term
${\bf X}_r,$ starting from $r=2.$

The transformation themselves are of the form

\be {\bf x}={\bf y}+{\bf h}_r ({\bf y}),\label{htransform}\ee
where ${\bf h}_r\in H^r,\,r\geq 2.$

The effect of (\ref{htransform}) in ${\bf X}_1$ is as follows
\cite{arrowsmith}: Observe that ${\bf x}={O}(|{\bf y}|).$ Then,
the inverse of (\ref{htransform}) takes the form \be {\bf y}={\bf
x}-{\bf h}_r({\bf x})+{O}(|{\bf
x}|^{r+1}).\label{inversehtransform}\ee

By applying total derivatives in both sides, and assuming ${\bf
x}'={\bf A}{\bf x}+{\bf X}_r({\bf x})$, we find

\be {\bf y}'={\bf A} {\bf y}-{\bf L_A} {\bf h}_r ({\bf y})+{\bf
X}_r({\bf y})+{O}(|{\bf y}|^{r+1})\label{evoly}\ee where ${\bf
L_A}$ is the linear operator that assigns to ${\bf h(y)}\in H^r$
the Lie bracket of the vector fields ${\bf A y}$ and ${\bf h(y)}$:

\ben {\bf L_A}: H^r& &\rightarrow H^r\nonumber\\
     {\bf h}  & & \rightarrow  {\bf L_A} {\bf h (y)}={\bf D h(y)} {\bf A y}- {\bf A h(y)}.
\een

Both ${\bf L_A}$ and ${\bf X}_r\in H^r,$ so that the deviation of
the right-hand side of (\ref{evoly}) from ${\bf A y}$ has no terms
of order less than $r$ in $|{\bf y}|.$ This means that if ${\bf
X}$ is such that ${\bf X}_2=\ldots {\bf X}_{r-1}=0,$ they will remain zero under the transformation (\ref{htransform}). This makes clear how we may be able to remove ${\bf X}_r$ from a suitable choice of ${\bf h}_r.$

The proposition 2.3.2 in \cite{arrowsmith} states that if the inverse of ${\bf L_A}$ exists, the differential equation \be {\bf
x}'={\bf A}{\bf x}+{\bf X}_r({\bf x})+{O}(|{\bf
x}|^{r+1})\label{ODE1}\ee with ${\bf X}_r\in H^r,$ can be transformed to \be {\bf y}'={\bf A y}+{O}(|{\bf
y}|^{r+1})\label{ODE2}\ee by the transformation (\ref{htransform})
where \be {\bf h}_r({\bf y})={\bf L_A}^{-1} {\bf X}_r(\bf
y)\label{Transform2}\ee

The equation \be {\bf L_A}{\bf h}_r({\bf y})={\bf X}_r(\bf y)\ee
is named the homological equation.

If ${\bf A}$ has distinct eigenvalues $\lambda_i,\, i=1,2,3,$ its eigenvectors form a basis of $\mathbb{R}^n.$ Relative to this eigenbasis, ${\bf A}$ is diagonal. It can be proved (see proof in
\cite{arrowsmith}) that ${\bf L_A}$ has eigenvalues $\Lambda_{{\bf
m},i}={\bf m}\cdot {\mathbf{\lambda}}-\lambda_i=\sum_j
m_j\lambda_j-\lambda_i$ with associated eigenvectors ${\bf
x^m}{\bf e}_i.$ The operator, ${\bf L_A}^{-1},$ exists if and only
if the $\Lambda_{{\bf m},i}\neq 0,$ for every allowed ${\bf m}$
and $i=1\ldots r.$

If we were able to remove all the nonlinear terms in this way,
then the vector field can be reduced to its linear part
$${\bf x}'={\bf X}({\bf x})\rightarrow {\bf y}'={\bf A}{\bf y}.$$
Unfortunately, not all the higher order terms vanishes by applying these transformations. It is the case if resonance occurs.

The n-tuple of eigenvalues ${\bf
\lambda}=(\lambda_1,\ldots,\lambda_n)^T$ is resonant of order $r$
(see definition 2.3.1 in \cite{arrowsmith}) if there exist some
${\bf m}=(m_1,m_2,\ldots m_n)^T$ (a n-tuple of non-negative
integers) with $m_1+m_2+\ldots m_n=r$ and some $i=1\ldots n$ such
that $\lambda_i={\bf m}\cdot \lambda,$ i.e., if $\Lambda_{{\bf
m},i}=0$ for some ${\bf m}$ and some $i.$

If there is no resonant eigenvalues, and provided they are different, we can use the eigenvectors of ${\bf A}$ as a basis for
$H^r.$ Then, we can write ${\bf h}_r$ as
$${\bf h}_r({\bf x})=\sum_{{\bf m},i,\sum m_j=r}h_{{\bf m},i}{\bf x}^{\bf m}{\bf e}_i$$ and any vector
field ${\bf X}\in H^r$ as $${\bf X}({\bf x})=\sum_{{\bf m},i,\;
\sum m_j=r} {\bf X}_{{\bf m},i}{\bf x}^m {\bf e}_i$$ where  ${\bf
m}=(m_1,m_2,\ldots,m_n)^T,$ ${\bf
x}^m=x_1^{m_1}\,x_2^{m_2}\,\ldots\,x_n^{m_n}$ and ${\bf
e}_i,\,i=1,\ldots n$ stands for the canonical basis in
$\mathbb{R}^n.$ If the eigenvalues of ${\bf A}$ are not resonant
of order $r,$ then $$h_{{\bf m},i}=X_{{\bf m},i}/\Lambda_{{\bf
m},i}.$$ This gives ${\bf h}_r$ explicitly in terms of ${\bf
X}_r.$

In case that resonance occurs, we proceed as follows. If ${\bf
A}$ can diagonalized, then the eigenvectors of ${\bf L}_{\bf A}$ form a basis of $H^r.$ The subset of eigenvectors of ${\bf L}_{\bf
A}$ with non-zero eigenvalues then form a basis of the image, $B^r,$ of $H^r$ under ${\bf L}_{\bf A}.$ It follows that the component of ${\bf X}_r$ in $B^r$ can be expanded in terms of these eigenvectors and ${\bf h}_r$ chosen such that $$h_{{\bf
m},i}=X_{{\bf m},i}/\Lambda_{{\bf m},i}.$$ to ensure the removal
of these terms. The component, ${\bf w}_r,$ of ${\bf X}_r$ lying in the complementary subspace, $G^r,$ of $B^r$ in $H^r$ will be unchanged by the transformations ${\bf x}={\bf y}+{\bf h}_r({\bf
y})$ obtained from $B^r.$

Since $$ {\bf X}_r({\bf y}+{\bf h}_{r+k}({\bf y}))={\bf X}_r({\bf
y})+{O}(|{\bf y}|^{r+k+1}),r\geq 2, k=1,2,\ldots,$$ these terms are not changed by subsequent transformations to remove non-resonant terms of higher order.

The above facts are expressed in

\begin{thm}[theorem 2.3.1 in \cite{arrowsmith}]\label{NFTheorem}
Given a smooth vector field $\bf X({\bf x})$ on $\mathbb{R}^n$
with ${\bf X(0)=0},$ there is a polynomial transformation to new
coordinates, ${\bf y},$ such that the differential equation ${\bf
x}'={\bf X}({\bf x})$ takes the form ${\bf y}'={\bf J}{\bf
y}+\sum_{r=1}^N {\bf w}_r({\bf y})+{O}(|{\bf y}|^{N+1}),$ where
${\bf J}$ is the real Jordan form of ${\bf A}={\bf D X}({\bf 0})$
and ${\bf w}_r\in G^r,$ a complementary subspace of $H^r$ on
$B^r={\bf L_A}(H^r).$
\end{thm}

\subsubsection{Center Manifold theory}\label{sectionCM}

In this section we offer the main techniques for the construction of center manifolds for vector fields in $\mathbb{R}^n$. We follow the approach in \cite{wiggins} chapter 18.

The setup is as follows. We consider vector fields in the form

\begin{align}&\mathbf{x}'=\mathbf{A x}+\mathbf{f}(\mathbf{x},\mathbf{y}),\nonumber\\
&\mathbf{y}'=\mathbf{B x}+\mathbf{g}(\mathbf{x},\mathbf{y}),  \; (\mathbf{x},\mathbf{y})\in \mathbb{R}^c\times\mathbb{R}^s,\label{basiceqs}\end{align}
where \begin{align}& \mathbf{f(0,0)}=\mathbf{0}, \mathbf{Df(0,0)}=\mathbf{0},\nonumber\\&\mathbf{g(0,0)}=\mathbf{0}, \mathbf{Dg(0,0)}=\mathbf{0}.\end{align}
In the above, $\mathbf{A}$ is a $c\times c$ matrix having eigenvalues with zero real parts,
$\mathbf{B}$ is an $s\times s$ matrix having eigenvalues with negative real parts, and $\mathbf{f}$ and $\mathbf{g}$ are $C^r$ functions ($r\geq 2$).

\begin{defn}[Center Manifold]\label{CMdef}
An invariant manifold will be called a center manifold for \eqref{basiceqs} if it can locally be represented as follows
\[
W^{c}\left(  \mathbf{0}\right)  =\left\{  \left( \mathbf{x},\mathbf{y}\right)
\in\mathbb{R}^c\times\mathbb{R}^s:\mathbf{y}=\mathbf{h}\left( \mathbf{x}\right)
,\;\left\vert \mathbf{x}\right\vert <\delta\right\}  ;\;\;\;\mathbf{h}\left( \mathbf{0}\right)
=\mathbf{0},\;D\mathbf{h}\left( \mathbf{0}\right)  =\mathbf{0},
\]
for $\delta$ sufficiently small (cf. \cite{wiggins} p. 246, \cite{Perko},p. 155).
\end{defn}

The conditions $\mathbf{h}\left( \mathbf{0}\right)
=\mathbf{0},\;D\mathbf{h}\left( \mathbf{0}\right)  =\mathbf{0}$ imply that $W^{c}\left(  \mathbf{0}\right)$ is tangent  to $E^c$ at $\left(\mathbf{x},\mathbf{y}\right)=(\mathbf{0},\mathbf{0}),$ where $E^c$ is the generalized eigenspace whose corresponding eigenvalues have zero real parts. The following three theorems (see theorems 18.1.2, 18.1.3 and 18.1.4  in \cite{wiggins} p. 245-248) are the main results to the treatment of center manifolds.
The first two are existence and stability theorems of the center manifold for \eqref{basiceqs} at the origin.
The third theorem allows to compute the center manifold to any desired degree accuracy by using Taylor series to solve a quasilinear partial differential equation that
$\mathbf{h}\left( \mathbf{x}\right)$ must satisfy. The proof of those results is given in \cite{Carr:1981}.

\begin{thm}[Existence]\label{existenceCM}
There exists a $C^r$ center manifold for \eqref{basiceqs}. The dynamics of \eqref{basiceqs} restricted to the center manifold is, for $\mathbf{u}$ sufficiently small,
given by the following c-dimensional vector field
\be\label{vectorfieldCM}
\mathbf{u}'=\mathbf{A u}+\mathbf{f}\left(\mathbf{u},\mathbf{h}\left(\mathbf{u}\right)\right),\; \mathbf{u}\in\mathbb{R}^c.
\ee
\end{thm}

The next results implies that the dynamics of \eqref{vectorfieldCM} near $\mathbf{u}=0$ determine the dynamics of \eqref{basiceqs} near $\left(\mathbf{x},\mathbf{y}\right)=(\mathbf{0},\mathbf{0})$
(see also Theorem 3.2.2 in \cite{Guckenheimer}).

\begin{thm}[Stability]\label{stabilityCM}
i) Suppose the zero solution of \eqref{vectorfieldCM} is stable (asymptotically stable) (unstable); then the zero solution of \eqref{basiceqs} is also stable (asymptotically stable) (unstable).
Then if $(\mathbf{x}(\tau),\mathbf{y}(\tau))$ is a solution of \eqref{basiceqs} with $(\mathbf{x}(0),\mathbf{y}(0))$ sufficiently small, then there is a solution $\mathbf{u}(\tau)$ of \eqref{vectorfieldCM} such that, as $\tau\rightarrow\infty$
\begin{align*}
& \mathbf{x}(\tau)=\mathbf{u}(\tau)+{\cal O}(e^{-r t}),\\
& \mathbf{x}(\tau)=\mathbf{h}\left(\mathbf{u}(\tau)\right)+{\cal O}(e^{-r t}),
\end{align*} where $r>0$ is a constant.
\end{thm}

{\bf Dynamics Captured by the center manifold}

Stated in words, this theorem says that for initial conditions of the \emph{full system} sufficiently close to the origin, trajectories through them asymptotically approach a trajectory on the center manifold. In particular, equilibrium points sufficiently close to the origin, sufficiently small amplitude periodic orbits, as well as small homoclinic and heteroclinic orbits are contained in the center manifold.

The obvious question now is how to compute the center manifold so that we can use the result of theorem \ref{stabilityCM}? To answer this question we will derive an equation that $\mathbf{h}(\mathbf{x})$ must satisfy in order to its graph to be a center manifold for \eqref{basiceqs}.

Suppose we have a center manifold \[
W^{c}\left(  \mathbf{0}\right)  =\left\{  \left(  \mathbf{x},\mathbf{y}\right)
\in\mathbb{R}^c\times\mathbb{R}^s:\mathbf{y}=\mathbf{h}\left(  \mathbf{x}\right)
,\;\left\vert \mathbf{x}\right\vert <\delta\right\}  ;\;\;\;\mathbf{h}\left( \mathbf{0}\right)
=\mathbf{0},\;D\mathbf{h}\left(  \mathbf{0}\right)  =\mathbf{0},
\] with $\delta$ sufficiently small. Using the invariance of $W^{c}\left(  \mathbf{0}\right)$ under the dynamics of \eqref{basiceqs}, we derive a quasilinear partial differential equation that $\mathbf{h}\left( \bf{x}\right)$ must satisfy. This is done as follows:

\begin{enumerate}
\item The $(\mathbf{x},\mathbf{y})$ coordinates of any point on $W^{c}\left(  \mathbf{0}\right) $ must satisfy \be\mathbf{y}=\mathbf{h}(\mathbf{x})\label{coordCM}\ee
\item Differentiating \eqref{coordCM} with respect to time implies that the $(\mathbf{x}',\mathbf{y}')$ coordinates of any point on  $W^{c}\left(  \mathbf{0}\right) $ must satisfy
\be\mathbf{y}'=D\mathbf{h}\left( \mathbf{x}\right)\mathbf{x}'\label{totalderivative}\ee
\item Any point in $W^{c}\left(  \mathbf{0}\right) $ obey the dynamics generated by \eqref{basiceqs}. Therefore substuting
\begin{align*}
\mathbf{x}'=\mathbf{A x}+\mathbf{f}\left(\mathbf{x},\mathbf{h}(\mathbf{x})\right),\\
\mathbf{y}'=\mathbf{B}\mathbf{h}(\mathbf{x})+\mathbf{g}\left(\mathbf{x},\mathbf{h}(\mathbf{x})\right)
\end{align*} into \eqref{totalderivative} gives
\be{\cal N}\left(\mathbf{h}(\mathbf{x})\right)\equiv D\mathbf{h}(\mathbf{x})\left[\mathbf{A x}+\mathbf{f}\left(\mathbf{x},\mathbf{h}(\mathbf{x})\right)\right]
-\mathbf{B}\mathbf{h}(\mathbf{x})-\mathbf{g}\left(\mathbf{x},\mathbf{h}(\mathbf{x})\right)=0\label{MaineqcM}.\ee
\end{enumerate}

Equation \eqref{MaineqcM} is a quasilinear partial differential that $\mathbf{h}(\mathbf{x})$ must satisfy in order for its graph to be an invariant center manifold.
To find the center manifold, all we need to do is solve \eqref{MaineqcM}.

Unfortunately, it is probably more difficult to solve \eqref{MaineqcM} than our original problem; however the following theorem give us a method for computing an approximated solution of \eqref{MaineqcM} to any desired degree of accuracy.

\begin{thm}[Approximation]\label{approximationCM}
Let $\mathbf{\Phi}:\mathbb{R}^c\rightarrow\mathbb{R}^s$ be a $C^1$ mapping with $\mathbf{\Phi}(\mathbf{0})=\mathbf{0}$ and $D\mathbf{\Phi}(\mathbf{0})=\mathbf{0}$ such that
${\cal N}\left(\mathbf{\Phi}(\mathbf{x})\right)={\cal O}(|\mathbf{x}|^q)$ as $\mathbf{x}\rightarrow \mathbf{0}$ for some $q>1.$ Then,
$|\mathbf{h}(\mathbf{x})-\mathbf{\Phi}(\mathbf{x})|={\cal O}(|\mathbf{x}|^q)$ as $\mathbf{x}\rightarrow \mathbf{0}$
\end{thm}

This theorem allows us to compute the center manifold to any desired degree of accuracy by solving \eqref{MaineqcM} to the same degree of accuracy. For this task power series expansions will work nicely. Let us consider a concrete example further in section \ref{stabilityP4}.

\subsection{Standard procedure to analyze the properties of the flow of a vector field}\label{Procedure}

Given a cosmological dynamical system determined by the differential equation

\begin{eqnarray}
\frac{\mathrm{d}y}{\mathrm{d}\tau}&=&f(y), y\in \mathbb{R}^n\label{Eq1},\\
g(y)&=&0\label{eq2},
\end{eqnarray}
the standard procedure to analyze the properties of the flow
generated by \eqref{Eq1} subject to the constraint(s) \eqref{eq2}
(see, for example, the reference \cite{WE}) is the following:

\begin{enumerate}
\item Determine whether the state space, as defined by \eqref{eq2}, is compact.
\item Identify the lower-dimensional invariant sets, which contains the orbits of more special classes of models with additional symmetries.
\item Find all the equilibrium points and analyze their local stability. Where possible identify the stable and unstable manifolds of the equilibrium points, which may coincide with some of the invariant sets in point (2).
\item Find Dulac's functions or monotone functions in various invariant sets where possible.
\item Investigate any bifurcation that occur as the equation of state parameter $\gamma$ (or any other parameters) varies. The bifurcations are associated with changes in the local stability of the equilibrium points.
\item Having all the information in the points (1)-(5) one can hope to formulate precise conjectures about the asymptotic evolution, by identifying the past and the future attractors. The past attractor will describe the evolution of a typical universe near the initial singularity while the future attractor will play the same role at late times. The monotone functions in point (4) above, in conjunction with theorems of dynamical systems theory, may enable some of the conjectures to be proved.
\item Knowing the stable and unstable manifolds of the equilibrium points it is possible to construct all possible heteroclinic sequences that join the past attractor, thereby gaining insight into the intermediate evolution of cosmological models.
\end{enumerate}

\section{Flat FRW models
with both a scalar field coupled to matter and radiation}

There exist a number of physical theories which predict the presence of a scalar field coupled to matter. For example, in
string theory the dilaton field is generally coupled to matter
\cite{Gasperini:2007zz}. Nonminimally coupling occurs also in STT
of gravity \cite{Fujii:2003pa,Faraoni:2004pi}, in HOG theories
\cite{Capozziello:2007ec} and in models of chameleon gravity
\cite{Waterhouse:2006wv}. Coupled quintessence was investigated
also in
\cite{Amendola:1999er,TocchiniValentini:2001ty,Billyard:2000bh} by using dynamical systems techniques. The cosmological dynamics of
scalar-tensor gravity have been investigated in
\cite{Carloni:2007eu,Tsujikawa:2008uc}. Phenomenological coupling
functions were studied for instance in \cite{Boehmer:2008av} which
can describe either the decay of dark matter into radiation, the
decay of the curvaton field into radiation or the decay of dark
matter into dark energy (see section III of \cite{Boehmer:2008av}
for more information and for useful references).  In the reference
\cite{Tsujikawa:2008uc}, the authors construct a family of viable scalar-tensor models of dark energy (which includes pure $F(R)$ theories and quintessence). By investigating a phase space the authors obtain that the model posses a phase of late-time acceleration preceded by a standard matter era, while at the same time satisfying the local gravity constraints (LGC).

The aim of the section is to extent several results in
\cite{Foster:1998sk,Miritzis:2003ym,Leon:2008de,Giambo':2009cc} to
the more realistic situation when radiation is included in the cosmic budget (particularly for the study of the early time dynamics). We will focus mainly in a particular era of the universe where matter and radiation coexisted. We are interested in investigate all possible scaling solutions in this regime.

\subsection{The action}

The action for a general class of STT, written in the so-called Einstein frame (EF), is given by \cite{Kaloper:1997sh}:
\begin{align}&S_{EF}=\int_{M_4} d{ }^4 x \sqrt{|g|}\left\{\frac{1}{2} R-\frac{1}{2}(\nabla\phi)^2-V(\phi)+\chi(\phi)^{-2}
\mathcal{L}_{matter}(\mu,\nabla\mu,\chi(\phi)^{-1}g_{\alpha\beta})\right\}\label{eq1}
\end{align}
In this equation $R$ is the curvature scalar, $\phi$ is the a
scalar field, related via conformal transformations with the
dilaton field, $\chi.$ \footnote{For a discussion about the regularity of the conformal transformation, or the equivalence issue of the two frames, see for example
\cite{Magnano:1990qu,Cotsakis:1993vm,Teyssandier:1995wr,Schmidt:1995ws,Cotsakis:1995wt,Capozziello:1996xg,Magnano:1993bd,Faraoni:1998qx,Faraoni:2007yn,Faraoni:2006fx} and references therein.} $V(\phi)$ is the quintessence self-interaction potential, $\chi(\phi)^{-2}$ is the coupling
function, $\mathcal{L}_{matter}$ is the matter Lagrangian, $\mu$
is a collective name for the matter degrees of freedom. The
energy-momentum tensor of background matter is defined by
\begin{equation}T_{\alpha
\beta}=-\frac{2}{\sqrt{|g|}}\frac{\delta}{\delta g^{\alpha
\beta}}\left\{\sqrt{|g|}
 \chi^{-2}\mathcal{L}(\mu,\nabla\mu,\chi^{-1}g_{\alpha
 \beta})\right\}.\label{Tab}\end{equation}
In the STT written in the EF as (\ref{eq1}), the conservation equations for matter/energy are expressed in
$$Q_\beta\equiv\nabla^\alpha T_{\alpha
\beta}=-\frac{1}{2}T\frac{\partial_\phi\chi(\phi)}{\chi(\phi)}\nabla_{\beta}\phi,$$
where $T=T^\alpha_\alpha$ denotes the trace of the energy-momentum tensor of background matter.
For HOG theories derived from Lagrangians of the form
\begin{equation}
L=\frac{1}{2}F\left(  R\right)
\sqrt{-g}+\mathcal{L}_{matter}(\mu,\nabla\mu,g_{\alpha \beta}),
\label{lagr}%
\end{equation} it is well known that under the conformal transformation, $\widetilde{g}%
_{\mu\nu}=F^{\prime}\left(  R\right)  g_{\mu\nu}$, the field
equations reduce to the Einstein field equations with a scalar
field $\phi$ as an additional matter source, where
\begin{equation}
\phi=\sqrt{\frac{3}{2}}\ln F^{\prime}\left(  R\right)  . \label{scfi}%
\end{equation}

Assuming that (\ref{scfi}) can be solved for $R$ to obtain a function $R\left(  \phi\right) ,$ the potential of the scalar field is given by
\begin{equation}
V\left(  R\left(  \phi\right)  \right)  =\frac{1}{2\left(
F^{\prime}\right)
^{2}}\left(  R F^{\prime}-f\right)  , \label{pote}%
\end{equation}
and quadratic gravity, $F(R)=R+\alpha R^2$ with the potential
$$V\left( \phi\right) =\frac{1}{8\alpha}\left(
1-e^{-\sqrt{2/3}\phi}\right) ^{2}$$ is a typical example. The restrictions on the potential in the papers
\cite{Rendall:2004ic,Rendall:2005if,Rendall:2005fv,Rendall:2006cq} were used in \cite{Macnay:2008nw} to impose conditions on the function $f\left(  R\right)  $ with corresponding potential (\ref{pote}). The conformal equivalence can be formally obtained by conformally transforming the Lagrangian (\ref{lagr}) and the resulting action becomes \cite{Bean:2006up},
\begin{equation}
\widetilde{S}=\int d^{4}x\sqrt{-\widetilde{g}}\left\{\frac{1}{2}
\widetilde{R}-\left(  \nabla\phi\right)  ^{2}-V\left( \phi\right)
 +e^{-2\sqrt{2/3}\phi}\mathcal{L}_{matter}\left(\mu,\nabla\mu,
e^{-\sqrt{2/3}\phi
}\widetilde{g}\right)  \right\}. \label{action}%
\end{equation}
It is easy to note that the model arising from the action
\eqref{action} can be obtained from \eqref{eq1} with the choice $\chi(\phi)=e^{\sqrt{2/3}\phi}.$

\subsection{The field equations}

In this section we consider a phenomenological model inspired in
the action \eqref{eq1} for FRW space-times with flat spatial
slices, modelled by the metric:
\begin{align}d s^2 = -d t^2 +a(t)^2\left({d r^2}+r^2\left(d \theta^2+\sin^2\theta d \varphi^2\right)\right).
\label{flatFRW}
\end{align} We use a system of units in which $8\pi G=c=\hbar=1.$
We assume that the energy-momentum tensor \eqref{Tab} is in the
form of a perfect fluid
$$T^\alpha_\beta=\text{diag} \left(-\rho,p,p,p\right),$$ where $\rho$ and
$p$ are respectively the isotropic energy density and the isotropic pressure (consistently with FRW metric, pressure is necessarily isotropic \cite{Trodden:2004st}). For simplicity we will assume a barotropic equation of state $p=(\gamma-1)\rho.$ We include radiation in the cosmic budget since it is an important matter source in the early universe.

The cosmological equations for flat FRW models with both a scalar field coupled to matter and radiation are given by

\ben \dot H &=& -\frac{1}{2}\left(\gamma
\rho
+\frac{4}{3}\rho_r+\dot\phi^2\right),\label{Raych}\\
\dot \rho&=& -3\gamma H \rho-\frac{1}{2}(4-3\gamma)\rho\dot\phi
\frac{\mathrm{d}\ln
\chi(\phi)}{\mathrm{d}\phi},\label{consm}\\
\dot \rho_r &=& -4 H \rho_r,\label{consr}\\
\ddot \phi&=& -3 H \dot \phi-\frac{\mathrm{d}
V(\phi)}{\mathrm{d}\phi}+\frac{1}{2}(4-3\gamma)\rho
\frac{\mathrm{d}\ln
\chi(\phi)}{\mathrm{d}\phi}\label{KG},\\
3H^2 &=& \frac{1}{2}\dot\phi^2+V(\phi)+\rho+\rho_r. \label{Fried}
\een where $a$ is the scale factor, $H=\dot a/a$ is the Hubble parameter, $\rho$ denotes the energy density of barotropic matter,
$\rho_r$ denotes de energy density of radiation, $\phi$ denotes the scalar field and $V(\phi)$ and $\chi(\phi)$ are, respectively, the potential and coupling functions. To maintain the analysis as general as possible, we will not specify the functional forms of the potential and the coupling function from the beginning. Instead we consider the general hypothesis $V(\phi)\in C^3, V(\phi)>0, \chi(\phi)\in C^3$ and $\chi(\phi)>0.$ We impose they in order to obtain dynamical systems of class $C^2.$ However, to derive some of our results we will relax some of this hypothesis, or consider further assumptions (they will be clearly stated when applicable).  We consider also $\rho\geq 0$ and $0<\gamma<2, \gamma\neq \frac{4}{3}.$ This hypothesis for background matter are the usual. We assume $\gamma\neq \frac{4}{3}$ to exclude the possibility that the background matter behaves as radiation. The energy momentum tensor for radiation ($\gamma=4/3$) is traceless, so it is automatically decoupled from a scalar field non-minimally coupled to dark matter in the Einstein frame. Radiation source is added by hand in order to model also the cosmological epoch when barotropic matter and radiation coexisted since we want to investigate the possible scaling solutions in the radiation regime. We neglect ordinary (uncoupled) barotropic matter.

\subsection{Late time behavior}

The purpose of this section is to formulated a proposition that extent in some way (we are considering only flat
FRW models) the proposition 1 of \cite{Giambo':2009cc}, which gives a characterization of the future attractor
of the system \eqref{Raych}, \eqref{consm},
\eqref{consr}, \eqref{KG}.

In the following, we study the late time behavior of solutions of
\eqref{Raych}, \eqref{consm}, \eqref{consr}, \eqref{KG}, which are
expanding at some initial time, i. e., $H(0)>0.$ The state vector
of the system is $\left(\phi, \dot\phi, \rho, \rho_r, H\right).$
Defining $y:=\dot\phi,$ we rewrite the autonomous system as \ben
\dot\phi&=&y,\label{dotphi}\\
\dot y&=& -3 H y-\frac{\mathrm{d}
V(\phi)}{\mathrm{d}\phi}+\frac{1}{2}(4-3\gamma)\rho
\frac{\mathrm{d}\ln
\chi(\phi)}{\mathrm{d}\phi}\label{newKG},\\
\dot \rho&=& -3\gamma H \rho-\frac{1}{2}(4-3\gamma)\rho y
\frac{\mathrm{d}\ln
\chi(\phi)}{\mathrm{d}\phi},\label{newconsm}\\
\dot \rho_r &=& -4 H \rho_r,\label{newconsr}\\
\dot H &=& -\frac{1}{2}\left(\gamma \rho
+\frac{4}{3}\rho_r+y^2\right),\label{newRaych} \een subject to the
constraint \be 3H^2 = \frac{1}{2}y^2+V(\phi)+\rho+\rho_r.
\label{newFried}\ee

\begin{rem}\label{decreasing}
Using standard arguments of ordinary differential equations
theory, follows from equations \eqref{newconsm} and \eqref{newconsr}
that the signs of $\rho$ and $\rho_r,$
respectively, are invariant. This means that if $\rho>0$ and $\rho_r>0$
for some initial time $t_0,$ then $\rho(t)>0,$ and $\rho_r(t)>0$ throughout the solution.
From \eqref{newRaych} and \eqref{newFried} and only if additional conditions are
assumed, for example $V(\phi)\geq 0$ and $V(\phi_*)= 0$ for some $\phi_*$, follows that the sign of $H$ is invariant.
From \eqref{newconsr} and \eqref{newRaych} follows that $\rho_r$ and
$H$ decreases. Also, defining $\epsilon=\frac{1}{2} y^2+V(\phi),$
follows from \eqref{newKG}-\eqref{newconsm} that
\be\dot\epsilon+\dot\rho=-3 H(y^2+\gamma \rho)\label{decr}.\ee
Thus, the total energy density contained in the dark sector is
decreasing.
\end{rem}

Let us formalize notion of degenerate local minimum introduced in \cite{Giambo':2009cc}:

\begin{defn}
The function $V(\phi)$ is said to have a degenerate local minimum
at $\phi_\star$ if  $$V'(\phi),V''(\phi), \dots V^{(2n-1)}$$
vanish at $\phi_*,$ and $V^{(2n)}(\phi_*)>0,$ for some integer
$n.$
\end{defn}

\begin{prop}\label{thmIII} Suppose that $V(\phi)\in C^2(\mathbb{R})$ satisfies the following conditions \footnote{See assumptions 1 in
\cite{Giambo':2009cc}.}
\begin{itemize}
\item[(i)] The (possibly empty) set $\{\phi: V(\phi)<0\}$ is bounded;
\item[(ii)] The (possibly empty) set of critical points of $V(\phi)$ is finite.
\end{itemize}
Let $\phi_*$ a strict local minimum for $V(\phi),$ possibly
degenerate, with non-negative critical value. Then ${\bf
p}_*:=\left(\phi_*,y_*=0,\rho_*=0,\rho_r=0,
H=\sqrt{\frac{V(\phi_*)}{3}}\right)$ is an asymptotically stable
equilibrium point for the flow of \eqref{dotphi}-\eqref{newRaych}.
\end{prop}

{\bf Proof.}

We adapt the demonstration in \cite{Giambo':2009cc} (for flat FRW
cosmologies) to the case where radiation is considered.
\footnote{From physical considerations we can neglect radiation
for the analysis of the future attractor, but we prefer to offer
the complete proof.}

First let us consider the case $V(\phi_*)>0.$ Let
$\tilde{V}>V(\phi_*)$ be a regular value for $V$ such that the
connected component of $V^{-1}\left((-\infty,\tilde{V}]\right)$
that contains $\phi_*$ is a compact set in $\mathbb{R}.$ Let us
denote this set by $A$ and define $\Psi$ as
$$\Psi=\left\{(\phi,y,\rho,\rho_r,H): \phi\in A,
\epsilon+\rho\leq \tilde{V}, \rho\geq 0, \rho_r\in
\left[0,\tilde{W}\right]\right\},$$ where $\tilde{W}$ is a
positive constant. We can show that $\Psi$ is a compact set as
follows.

\begin{itemize}
\item[(i)] $\Psi$ is a closed set in $\mathbb{R}^5;$
\item[(ii)] $V(\phi_*)\leq V(\phi)\leq \tilde{V},\, \forall \phi\in A;$
\item[(iii)] $\frac{1}{2}y^2+V(\phi_*)\leq \frac{1}{2}y^2+V(\phi)+\rho\leq \tilde{V}$, and therefore $y$ is bounded;
\item[(iv)] $\rho\leq \tilde{V}-\frac{1}{2}y^2-V(\phi)\leq \tilde{V}-V(\phi_*)$ and then $\rho$ is bounded;
\item[(v)] Finally, from \eqref{newFried}, $\frac{V(\phi_*)}{3}\leq H^2\leq \tilde{V}+\tilde{W}.$
\end{itemize}

Let $\Psi_+\subseteq \Psi$ be the connected component of $\Psi$
containing ${\bf p}_*.$ Following similar arguments as in
\cite{Giambo':2009cc} it can be proved that $\Psi_+$ is positively
invariant with respect to  \eqref{dotphi}-\eqref{newRaych}, i.e,
all the solutions with initial data at $\Psi_+$ remains at
$\Psi_+$ for all $t>0.$ Indeed, let ${\bf x}(t)$ be such a
solution and $$\bar{t}=\sup\left\{t>0:
H(t)>0\right\}\in\mathbb{R}\cup\{+\infty\}.$$ When $t<\bar{t},$
equation \eqref{decr} imply that $\epsilon+\rho$ decreases
monotonically (cf remark \ref{decreasing}). Moreover, it can be
proved by contradiction that \be\phi(t)\in A\, \forall
t<\bar{t}\label{phiinA},\ee otherwise there would exists some
$t<\bar{t}$ such that $V(\phi(t))>\tilde{V},$ but then
$$\tilde{V}<V(\phi(t))\leq \frac{1}{2}y(t)^2+V(\phi(t))+\rho(t)\leq \tilde{V},$$
a contradiction. Thus, \eqref{phiinA} holds. But since $\rho_r\geq
0$ along the flow (cf remark \ref{decreasing}), it follows that
$$H(t)^2\geq \frac{1}{3}\left(\frac{1}{2}y(t)^2+V(\phi(t))+\rho(t)\right)\geq \frac{V(\phi(t))}{3}\geq \frac{V(\phi_*)}{3}>0.$$

We have proved that as long as $H$ remains positive, it is strictly bounded away from zero; thus $\bar{t}=+\infty,$ and from this can be deduced that ${\bf x}(t)$ remains in $\Psi_+$ for all $t>0.$

From all the above $\Psi_+$ satisfies the hypothesis of LaSalle's invariance theorem \cite{wiggins}. If we consider the monotonic decreasing functions $\epsilon+\rho$ and $\rho_r$ defined in
$\Psi_+$ then follows that, every solution with initial data at
$\Psi_+$ must be such that
$H\left(y^2+\gamma\rho\right)\rightarrow 0$ and
$H\rho_r\rightarrow 0$ as $t\rightarrow +\infty.$ Since $H$ is
strictly bounded away from zero in $\Psi_+$ follows that
$(y,\rho,\rho_r)\rightarrow (0,0,0)$ and
$H^2-\frac{V(\phi)}{3}\rightarrow 0$ as $t\rightarrow +\infty.$

Since $H$ is monotone decreasing (cf remark \ref{decreasing}) and
it is bounded away from zero it must have a limit. This means that
$V(\phi)$ also admits a limit. This limit has to be $V(\phi_*);$
otherwise $V'(\phi)$ would tend to a positive value and so would
the righthand side of \eqref{newKG}, a contradiction. Therefore
the solution approaches the equilibrium point ${\bf p}_*$.

If $V(\phi_*)=0$, the above argument can be easily adapted. In
this case the set $\Omega$ is connected and we choose $\Omega_{+}$
to be its subset characterized by the property $H\geq0$. The only
point in $\Omega_{+}$ with $H=0$ is exactly the equilibrium point
${\bf p}_*$, and so if $H(t)\rightarrow 0$ the solution is forced
to approach the equilibrium since $H$ is monotone; if by
contradiction $H(t)$ had a strictly positive limit, we could argue
as before to find $y\rightarrow0,\,\rho\rightarrow0$ and
$\rho_r\rightarrow0$ and so $H$ must necessarily converge to zero. $\blacksquare$

\subsection{Early time behavior}

In order to analyze the initial singularity (and also, the late time behavior) it is convenient to normalize the variables, since in the vicinity of an hypothetical initial singularity, the physical variables would typically diverge, whereas at late times they commonly vanish \cite{Wainwright:2004cd}. In this section we rewrite equations \eqref{Raych}, \eqref{consm}, \eqref{consr}, \eqref{KG} as an autonomous system defined on a state space by introducing Hubble-normalized variables. These variables satisfy an inequality arising from the Friedmann equation \eqref{Fried}. We analyze the cosmological model by investigating the flow of the autonomous system in a phase space by using dynamical systems tools.

\subsubsection{Normalized variables and dynamical system}

Let us introduce the following dimensionless variables

\be \sigma_1=\phi,
\sigma_2=\frac{\dot\phi}{\sqrt{6}H},\,\sigma_3=\frac{\sqrt{\rho}}{\sqrt{3}H},\,\sigma_4=\frac{\sqrt{V}}{\sqrt{3}H},
\,\sigma_5=\frac{\sqrt{\rho_r}}{\sqrt{3}H}\label{vars}\ee and the time coordinate \be \mathrm{d}\tau=3 H \mathrm{d}t.\ee We considered the scalar field itself as a dynamical variable.

Using these new coordinates equations \eqref{Raych}-\eqref{KG} recast as an autonomous system satisfying an inequality arising from the Friedmann
equation \eqref{Fried}. This system is given by

\begin{align}
&\sigma_1'=\sqrt{\frac{2}{3}} \sigma_2 \label{eq0phi}\\
&\sigma_2'=\sigma_2^3+\frac{1}{6}\left(3\gamma \sigma_3^2+4 \sigma_5^2-6\right)\sigma_2
-\frac{\sigma_4^2}{\sqrt{6}}\frac{\mathrm{d}\ln V(\sigma_1)}{\mathrm{d}\sigma_1}
+\frac{\left(4-3\gamma\right)\sigma_3^2}{2\sqrt{6}}\frac{\mathrm{d}\ln \chi(\sigma_1)}{\mathrm{d}\sigma_1},\label{eq0x1}\\
&\sigma_3'=\frac{1}{6}\sigma_3\left(6\sigma_2^2+3\gamma\left(\sigma_3^2-1\right)+4 \sigma_5^2\right)
-\frac{\left(4-3\gamma\right)\sigma_2 \sigma_3}{2\sqrt{6}}\frac{\mathrm{d}\ln \chi(\sigma_1)}{\mathrm{d}\sigma_1},\label{eq0x2}\\
&\sigma_4'=\frac{1}{6}\sigma_4\left(6\sigma_2^2+3\gamma
\sigma_3^2+4 \sigma_5^2\right)
+\frac{\sqrt{6}}{6}\sigma_2 \sigma_4 \frac{\mathrm{d}\ln V(\sigma_1)}{\mathrm{d}\sigma_1},\label{eq0x3}\\
&\sigma_5'=\frac{1}{6}\sigma_5\left(6\sigma_2^2+3\gamma
\sigma_3^2+4 \sigma_5^2-4\right).\label{eq0x4}
\end{align}
The system \eqref{eq0phi}-\eqref{eq0x4} defines a flow in the phase space

\be \Sigma :=\left\{\sigma\in\mathbb{R}^5: \sum_{j=2}^5
\sigma_j^2=1, \sigma_j\geq 0, j=3, 4, 5\right\}\label{Sigma}.\ee

\subsubsection{The topological properties of the phase space}

Let us define the sets $\Sigma_0:=\{\sigma\in\Sigma: \sigma_5=0\}$
and $\Sigma_+=\{\sigma\in\Sigma: \sigma_5>0\}.$ By construction
these sets are a partition of $\Sigma.$

\paragraph{\bf Notations:}
$||.||$ denotes the Euclidean vector norm;
$\mathbb{D}^n:=\left\{{\bf x}\in\mathbb{R}^n: ||{\bf
x}||<1\right\}$ denotes the n-dimensional unitary disc; and
$\mathbb{H}^n:=\left\{{\bf x}\in\mathbb{R}^n: x_n\geq 0\right\}$
denotes the n-dimensional Euclidean semi-space.

\begin{defn}[(Topological) manifold with boundary]\label{MB}
Let $M$ be a Hausdorff space provided with a numerable basis. Let
$p\in M.$ If there exists a positive number $m$ (possibly
depending on $p$), a neighborhood $V(p)$ of $p$ and an
homeomorphism $h: V(p)\rightarrow \mathbb{H}^m$ such that
$h(V(p))\subset \mathbb{H}^m$ is an open set of $\mathbb{H}^m$,
then, $M$ is a (topological) manifold with boundary.
\end{defn}

\begin{defn}[Boundary]\label{B}
Let $M$ be a manifold with boundary. The boundary of $M$,
$\partial M,$ is defined by $\partial M:=\left\{p\in M:
h(p)\in\mathbb{H}^{m-1}\times \{0\}\right\}.$
\end{defn}

\begin{defn}[Interior]\label{Int}
Let $M$ be a manifold with boundary. The interior of $M$,
$\text{Int} M,$ is defined by $\text{Int}
M:=M\setminus\partial M.$
\end{defn}

\begin{prop}\label{thm1}
$\Sigma_0$ is a manifold with boundary.
\end{prop}

{\bf Proof.} Since $\Sigma_0\subset\mathbb{R}^4$ is a closed set
with respect the usual topology of $\mathbb{R}^4,$ it is a
Hausdorff space equipped with a numerable basis. The rest of the
proof requires the construction of a set of local charts.

Let us define the sets $V_j:=\left\{\sigma\in\mathbb{R}^4:
\sigma_j>0\right\}\cap\Sigma_0,\, j=3,4.$ These sets are open with
respect to the induced topology in $\Sigma_0.$

Let be defined the projection maps
$$h_3:V_3\rightarrow\mathbb{H}^3,\, \sigma\rightarrow
h_3(\sigma)=(\sigma_1,\sigma_2,\sigma_4),$$ and
$$h_4:V_4\rightarrow\mathbb{H}^3,\, \sigma\rightarrow
h_4(\sigma)=(\sigma_1,\sigma_2,\sigma_3).$$ These maps satisfy
$h_j(V_j)=\mathbb{R}\times (\mathbb{H}^2\cap \mathbb{D}^2),
j=3,4,$ which are open sets of $\mathbb{H}^3.$ Their inverses are
given by
$$h_3^{-1}:\mathbb{R}\times (\mathbb{H}^2\cap \mathbb{D}^2)\rightarrow V_3,
\\ (\sigma_1,\sigma_2,\sigma_4)\rightarrow \left(\sigma_1,\sigma_2,\sqrt{1-\sigma_2^2-\sigma_4^2},\sigma_4\right),$$
and $$h_4^{-1}:\mathbb{R}\times (\mathbb{H}^2\cap \mathbb{D}^2)\rightarrow V_4,\, (\sigma_1,\sigma_2,\sigma_3)\rightarrow\left(\sigma_1,\sigma_2,\sigma_3,\sqrt{1-\sigma_2^2-\sigma_3^2}\right).$$ It is clear that they are homoemorphisms.

Observe that $\left\{(V_3,h_3), (V_4,h_4)\right\}$ does not cover
the sets with $\sigma_3=\sigma_4=0.$ The construction is completed by defining the sets $$V_1^\pm:=\left\{\sigma\in\Sigma_0:
\sigma_2=\pm 1, \sigma_3=\sigma_4=0\right\}$$ which are disjoint copies of $\mathbb{R},$ and thus they are 1-dimensional manifolds.
$\blacksquare$

By definition $(V_3,h_3), (V_4,h_4)$ are topological manifolds. It
can be proved that, in fact, each one is a topological manifold
with boundary. The boundaries are $$\partial
V_3=\left\{\sigma\in V_3: \sigma_4=0\right\},$$ and
$$\partial V_4=\left\{\sigma\in V_4: \sigma_3=0\right\}.$$
Both are homeomorphic to $\mathbb{R}\times (-1,1).$

Let be defined $V_1:=V_1^- \cup V_1^+.$ Observe that
\ben\Sigma_0=\left(\Sigma_0\setminus V_1\right)\cup
V_1=\text{Int}\left(\Sigma_0\setminus V_1\right)\cup(\partial
V_3\cup\partial V_4)\cup
V_1\nonumber\\=\text{Int}\left(\Sigma_0\setminus
V_1\right)\cup\left(\partial\Sigma_0\right)_1\cup\left(\partial\Sigma_0\right)_2,\label{relation}\een
where we have defined $\left(\partial\Sigma_0\right)_1=\partial
V_4\cup V_1=\left\{\sigma\in\Sigma: \sigma_3=0\right\}$ and
$\left(\partial\Sigma_0\right)_2=\partial V_3\cup
V_1=\left\{\sigma\in\Sigma: \sigma_4=0\right\}.$

From the above arguments and expression \eqref{relation} we have:

\begin{rem}

\begin{itemize}
\item The interior of $\Sigma_0$ is given by
$\text{Int}\Sigma_0=\text{Int}\left(\Sigma_0\setminus V_1\right)$
which is a 3-dimensional manifold (without boundary). \item The
boundary of $\Sigma_0$ is the union of two 2-dimensional
topological manifolds with boundary given by
$\left(\partial\Sigma_0\right)_1$ and
$\left(\partial\Sigma_0\right)_2.$ \item
$\left(\partial\Sigma_0\right)_1$ and
$\left(\partial\Sigma_0\right)_2$  share the same boundary $V_1$
which is the union of two disjoint copies of $\mathbb{R}.$
\end{itemize}

\end{rem}

\begin{prop}\label{thm2}
$\Sigma_+$ is a topological manifold with boundary.
\end{prop}

{\bf Proof.} It is easy to check that $\Sigma_+\subset\mathbb{R}^5$ is a
Hausdorff space equipped with a numerable basis. The rest of the
proof requires the construction of a set of local charts.

Let us define the sets $W_j:=\left\{\sigma\in\mathbb{R}^5:
\sigma_j>0\right\}\cap\Sigma_+,\, j=3,4.$ These sets are open with
respect to the induced topology in $\Sigma_+.$

Let us define the maps $$g_3:W_3\rightarrow\mathbb{H}^4,\, \sigma\rightarrow g_3(\sigma)=\left(\sigma_1,\sigma_2,\sigma_5,\sigma_4\right),$$ and $$g_4:W_4\rightarrow\mathbb{H}^4,\, \sigma\rightarrow g_4(\sigma)=\left(\sigma_1,\sigma_2,\sigma_5,\sigma_3\right).$$ These maps satisfy $g_j(W_j)=\mathbb{R}\times (\mathbb{H}^3\cap \mathbb{D}^3),\, j=3,4$ which are open set of $\mathbb{H}^4.$ Their inverses are given by
$$g_3^{-1}:\mathbb{R}\times (\mathbb{H}^3\cap \mathbb{D}^3)\rightarrow W_3,\, \left(\sigma_1,\sigma_2,\sigma_5,\sigma_4\right)\rightarrow \left(\sigma_1,\sigma_2,\sqrt{1-\sigma_2^2-\sigma_4^2-\sigma_5^2},\sigma_4,\sigma_5\right),$$ and $$g_4^{-1}:\mathbb{R}\times (\mathbb{H}^3\cap \mathbb{D}^3)\rightarrow W_4,\, \left(\sigma_1,\sigma_2,\sigma_5,\sigma_3\right)\rightarrow \left(\sigma_1,\sigma_2,\sigma_3,\sqrt{1-\sigma_2^2-\sigma_3^2-\sigma_5^2},\sigma_5\right).$$  It is clear that they are homeomorphism.

Observe that $\left\{(W_3,g_3), (W_4,g_4)\right\}$ do not cover
the sets with $\sigma_3=\sigma_4=0.$ The construction is
completed by defining the set $W_1:=\left\{\sigma\in\Sigma_+:
\sigma_1^2+\sigma_5^2=1\right\}.$ Using the projection map
$(\sigma_1,\sigma_2,\sigma_5)\stackrel{g_1}{\rightarrow}
(\sigma_1,\sigma_2)$ it is easily proved that $W_1$ is
homeomorphic to the open set $\mathbb{R}\times (-1,1).$
$\blacksquare$

We have that $W_3$ is a manifold with boundary. Its boundary is
the set $\partial W_3:=\{\sigma\in\Sigma: \sigma_3>0, \sigma_4=0,
\sigma_5>0\}$ and its interior is the set $\{\sigma\in\Sigma:
\sigma_3>0, \sigma_4>0, \sigma_5>0\}.$ Also, $W_4$ is a manifold
with boundary. Its boundary is the set $\partial
W_4:=\{\sigma\in\Sigma: \sigma_3=0, \sigma_4>0, \sigma_5>0\}$ with
same interior as $W_3$ and $W_1$ is a manifold without boundary
which is homeomorphic to $\mathbb{R}\times (-1,1).$

Let us define the sets $$\left(\partial
\Sigma_+\right)_1:=\partial W_4 \cup W_1$$ and
$$\left(\partial \Sigma_+\right)_2:=\partial W_3 \cup W_1.$$
Following the same arguments yielding to \eqref{relation} we get
the formula \ben\Sigma_+=\text{Int}\left(\Sigma_+\setminus
W_1\right)\cup\left(\partial\Sigma_+\right)_1\cup\left(\partial\Sigma_+\right)_2\label{relation2}.\een
Using the above arguments and relation \eqref{relation2} we have
the following

\begin{rem}

\begin{itemize}
\item The interior of $\Sigma_+$ is given by
$\text{Int}\Sigma_+=\text{Int}\left(\Sigma_+\setminus W_1\right)$
which is a 4-dimensional manifold (without boundary). \item The
boundary of $\Sigma_+$ is the union of two 3-dimensional
topological manifolds with boundary given by
$\left(\partial\Sigma_+\right)_1$ and
$\left(\partial\Sigma_+\right)_2.$ \item
$\left(\partial\Sigma_+\right)_1$ and
$\left(\partial\Sigma_+\right)_2$ share the same boundary $W_1$
which is a 2-dimensional manifold without boundary homeomorphic to
$\mathbb{R}\times (-1,1).$
\end{itemize}

\end{rem}

\subsubsection{Monotonic functions}

The construction of monotonic functions in the state space is an important tool in any phase space analysis. The existence of such functions can rule out periodic orbits, homoclinic orbits, and other complex behavior in invariant sets. If so, the dynamics is dominated by critical points (and possibly, heteroclinic orbits joining it). Additionally, some global results can be obtained. In table \ref{monotonic} are displayed several monotonic functions for the flow of \eqref{eq0phi}-\eqref{eq0x4} in $\Sigma$.

\begin{table}[!htb]
\caption{\label{monotonic} Monotonic functions for the flow of
\eqref{eq0phi}-\eqref{eq0x4} in \eqref{Sigma}.}
\begin{tabular}[t]{|l|c|c|c|c|}
\hline
$Z$&$\mathrm{d}Z/\mathrm{d}\tau$&Invariant set&Restrictions$^{\rm a}$ \\[1ex]
\hline
\hline &&& \\[-2ex]
$Z_1=\left(\frac{\sigma_3}{\sigma_4}\right)^2 V(\sigma_1) \chi(\sigma_1)^{2-\frac{3\gamma}{2}}$&$-\gamma Z_1$
&$\sigma_3>0, \sigma_4>0$ & $\gamma\in \left(0,\frac{4}{3}\right)\cup\left(\frac{4}{3},2\right)$\\[1ex]
\hline
\hline &&& \\[-2ex]
$Z_2=\left(\frac{\sigma_5}{\sigma_3}\right)^2 \chi(\sigma_1)^{-2+\frac{3\gamma}{2}}$
&$-\left(\frac{4}{3}-\gamma\right) Z_2$&$\sigma_3>0, \sigma_5>0$ & $\gamma\in \left(0,\frac{4}{3}\right)\cup\left(\frac{4}{3},2\right)$\\[1ex]
\hline
\hline &&& \\[-2ex]
$Z_3=\left(\frac{\sigma_5}{\sigma_4}\right)^2 V(\sigma_1)$&$-\frac{4}{3} Z_3$&$\sigma_4>0, \sigma_5>0$ & none\\[1ex]
\hline
\hline &&& \\[-2ex]
$Z_4=\frac{\sigma_2^2}{1-\sigma_2^2}$&$-\frac{2}{3} Z_4$&$\sigma_2\neq 0, \sigma_3=\sigma_4=0, \sigma_5>0$ & none\\[1ex]
\hline
\end{tabular}
\begin{center}
$^{\rm a}$ We assume the general conditions $\chi, V\in
C^3,\,\chi(\sigma_1)>0,V(\sigma_1)>0$.
\end{center}
\end{table}

\begin{rem}
From the definition of $Z_1$ it follows that it is a monotonic
decreasing function for the flow of \eqref{eq0phi}-\eqref{eq0x4}
restricted to the invariant set $\sigma_3>0, \sigma_4>0.$ Applying
the Monotonicity Principle (theorem \ref{theorem 4.12}) it follows
that the past attractor the flow of \eqref{eq0phi}-\eqref{eq0x4}
restricted to the invariant set $\sigma_3>0, \sigma_4>0$ is
contained in the set where $\sigma_4=0$ and the future attractor
in contained in the set where $\sigma_3=0.$
\end{rem}

\begin{rem}
Using the same argument follows from the definition of $Z_2$ that
the past attractor of the flow of \eqref{eq0phi}-\eqref{eq0x4}
restricted to the invariant set $\sigma_3>0, \sigma_5>0$ is
contained in the invariant set where $\sigma_3=0$ and the future
asymptotic attractor is contained in the invariant set
$\sigma_5=0$ provided $\gamma<\frac{4}{3}.$ If
$\gamma>\frac{4}{3}$ the asymptotic behavior is the reverse of the
previously described.
\end{rem}

\begin{rem}
From the definition of $Z_3$ follows that the past attractor of
the flow of \eqref{eq0phi}-\eqref{eq0x4} restricted to the
invariant set $\sigma_4>0, \sigma_5>0$ is contained in the
invariant set where $\sigma_4=0$ and the future asymptotic
attractor is contained in the invariant set $\sigma_5=0.$
\end{rem}

\begin{rem}
From the definition of $Z_4$ follows that the past attractor of
the flow of \eqref{eq0phi}-\eqref{eq0x4} restricted to the
invariant set $\sigma_2\neq 0, \sigma_3=\sigma_4=0, \sigma_5>0$ is
contained in the invariant set where $\sigma_2=0$ (i.e., where
$\sigma_5=1$) and the future asymptotic attractor is contained in
the invariant set $\sigma_2=\pm 1.$
\end{rem}

\subsubsection{Critical points with $\phi$ bounded}\label{Qualitative}

Let us make a preliminary analysis of the linear stability of the critical points of the flow of \eqref{eq0phi}-\eqref{eq0x4} defined in $\Sigma.$ It is a classic result that the linear stability of the critical points does not change under homeomorphisms.

Since $\Sigma$ is a 4-dimensional manifold (with boundary) we will consider the projection of $\Sigma$ in a real 4-dimensional manifold (with boundary).

Let be defined the projection map

\ben && \pi: \Sigma \rightarrow \Omega \nonumber\\
&&(\sigma_1, \sigma_2,\sigma_3,\sigma_4,\sigma_5)\rightarrow
(\sigma_1, \sigma_2,\sigma_3,\sigma_5)\een where
\be\Omega:=\left\{\sigma\in\mathbb{R}^4:
\sigma_2^2+\sigma_3^2+\sigma_5^2\leq 1, \sigma_j\geq 0, j= 3,
5\right\}.\ee

The flow of \eqref{eq0phi}-\eqref{eq0x4} defined on $\Sigma$ is topologically equivalent (under $\pi$) to the flow of
%\footnote{The relations between the new variables and the field variables is as follows: $\sigma_1=\phi,\, \sigma_2=\frac{\dot
%\phi}{\sqrt{6} H}, \,\sigma_3=\frac{\sqrt{\rho}}{\sqrt{3}H},
%\,\sigma_5=\frac{\sqrt{\rho_r}}{\sqrt{3}H}.$}

\begin{align}
&\sigma_1'=\sqrt{\frac{2}{3}} \sigma_2 \label{eqsigma1}\\
&\sigma_2'=\sigma_2^3+\frac{1}{6}\left(3\gamma \sigma_3^2+4 \sigma_5^2-6\right)\sigma_2-\frac{(1-\sigma_2^2-\sigma_3^2-\sigma_5^2)}{\sqrt{6}}\frac{\mathrm{d}\ln V(\sigma_1)}{\mathrm{d}\sigma_1}+\frac{\left(4-3\gamma\right)\sigma_3^2}{2\sqrt{6}}\frac{\mathrm{d}\ln \chi(\sigma_1)}{\mathrm{d}\sigma_1},\label{eqsigma2}\\
&\sigma_3'=\frac{1}{6}\sigma_3\left(6\sigma_2^2+3\gamma\left(\sigma_3^2-1\right)+4 \sigma_5^2\right)-\frac{\left(4-3\gamma\right)\sigma_2 \sigma_3}{2\sqrt{6}}\frac{\mathrm{d}\ln \chi(\sigma_1)}{\mathrm{d}\sigma_1},\label{eqsigma3}\\
&\sigma_5'=\frac{1}{6}\sigma_5\left(6\sigma_2^2+3\gamma \sigma_3^2+4 \sigma_5^2-4\right),\label{eqsigma5}
\end{align}
defined in $\Omega.$

\begin{table}[!htb]
\begin{center}
\caption{\label{crit0A} Location of the critical points of the
flow of \eqref{eqsigma1}-\eqref{eqsigma5} defined in $\Omega$.}
\begin{tabular}[t]{|l|c|c|c|c|}
\hline
Label&$\sigma_1$$^{\rm a}$ &$\sigma_2$&$\sigma_3$&$\sigma_5$\\[1ex]
\hline
\hline &&& \\[-2ex]
$Q_1$&$\sigma_{1c}: \chi'(\sigma_{1c})=0$&$0$&$1$&$0$\\[1ex]
\hline
\hline &&& \\[-2ex]
$Q_2$&$\sigma_{1c}: V'(\sigma_{1c})=0$&$0$&$0$&$0$\\[1ex]
\hline
\hline &&& \\[-2ex]
$Q_3$&$\sigma_{1c}\in\mathbb{R}$&$0$&$0$&$1$\\[1ex]
\hline
\end{tabular}
\end{center}
\begin{center}
$^{\rm a}$ We are assuming $V(\sigma_{1c})\neq 0,\,\chi(\sigma_{1c})\neq 0$ and $\gamma\in \left(0,\frac{4}{3}\right)\cup\left(\frac{4}{3},2\right).$
\end{center}
\end{table}

The system \eqref{eqsigma1}-\eqref{eqsigma5} defined in $\Omega$ admits three classes of critical points located at  $\Omega,$ denoted by $Q_j,\, j=1,2,3.$  In table \ref{crit0A} are displayed the coordinates of such points. The dynamics near the fixed points (and its stability properties) is dictated by the signs of the real parts of the eigenvalues of the Jacobian matrix evaluated at each critical point as follows:

\begin{enumerate}

\item The eigenvalues of the linearization around $Q_1$ are
$-\Delta_2,\,\gamma,\,\Delta_1\pm\sqrt{\Delta_1^2
+\Delta_2\frac{\chi''(\phi_1)}{\chi(\phi_1)}},$ where
$\Delta_1=(-2+\gamma)/{4}< 0,$ and
$\Delta_2=\left(4-3\gamma\right)/6.$ Then, the local stability of
critical point $P_1$ is as follows:

\begin{enumerate}
\item If $-\frac{\Delta_1^2 \chi (\sigma_{1c} )}{\Delta_2}\leq \chi
   ''(\sigma_{1c} )<0$ and $0< \gamma< \frac{4}{3}$ there exist a 3-dimensional
   stable manifold and a 1-dimensional unstable manifold of $Q_1.$
\item If $0<\gamma<\frac{4}{3}$ and $\chi
   ''(\sigma_{1c} )>0$ or $\frac{4}{3}<\gamma<2$ and $0<\chi
   ''(\sigma_{1c} )<-\frac{\Delta_1^2 \chi (\sigma_{1c} )}{\Delta_2},$
   there exist a 2-dimensional stable manifold and a 2-dimensional unstable manifold of $Q_1.$
\item If $0<\gamma<\frac{4}{3}$ and $\chi
   ''(\sigma_{1c} )=0$ there exist a 2-dimensional stable manifold,
   a 1-dimensional unstable manifold and a 1-dimensional center manifold of $Q_1.$
\item If $\frac{4}{3}<\gamma<2$ and $\chi
   ''(\sigma_{1c} )=0$ there exist a 1-dimensional stable manifold, a 2-dimensional
   unstable manifold and a 1-dimensional center manifold of $Q_1.$
\end{enumerate}

\item The eigenvalues of the linearization around $Q_2$ are:
$-\frac{2}{3}, -\frac{\gamma}{2}, -\frac{1}{2}\pm
\frac{1}{2}\sqrt{1-\frac{4}{3}\frac{V''(\sigma_{1c})}{V(\sigma_{1c})}}.$
Then, the local stability of critical point $P_2$ is as follows:
\begin{enumerate}
\item If $V''(\sigma_{1c})< 0$ there exists a 3-dimensional stable
manifold and a 1-dimensional unstable manifold of $Q_2.$ \item If
$V''(\sigma_{1c})= 0$ the eigenvalues are
$-1,-\frac{2}{3},0,-\frac{\gamma }{2}.$ Then there exists a
1-dimensional center manifold tangent to the $\sigma_1$-axis and a
3-dimensional stable manifold tangent to the 3-dimensional surface
$$\left\{(\sigma_1,\sigma_2,\sigma_3,\sigma_5)\in\mathbb{R}^4:
\sigma_1=-\sqrt{\frac{2}{3}}\sigma_2\right\}.$$ \item If
$0<V''(\sigma_{1c})\leq \frac{3}{4}V(\sigma_{1c})$ the stable manifold of $Q_2$ is 4-dimensional (the critical point is asymptotically stable and it is an stable node). If $V''(\sigma_{1c})> \frac{3}{4}V(\sigma_{1c})$ the stable manifold of $Q_2$ is 4-dimensional (the critical point is asymptotically stable and it is an stable focus).
\end{enumerate}

\item The eigenvalues of the linearization around $Q_3$ are
$\frac{4}{3},-\frac{1}{3},0,\frac{1}{6} (4-3 \gamma ).$ Then, the
local stability of critical point $Q_3$ is as follows:
\begin{enumerate}
\item There exists a 1-dimensional center manifold tangent to the
$\sigma_1$-axis \item If $\gamma<\frac{4}{3},$ $Q_3$ has a
2-dimensional unstable manifold and a 1-dimensional stable
manifold tangent to the line
$$\left\{(\sigma_1,\sigma_2,\sigma_3,\sigma_5)\in\mathbb{R}^4:
\sigma_1=-\sqrt{\frac{2}{3}}\sigma_2,\,\sigma_3=\sigma_5=0\right\}.$$
 \item If $\gamma>\frac{4}{3},$ $Q_3$ has a 1-dimensional unstable manifold and a 2-dimensional stable
manifold tangent to the 2-dimensional
subspace $$\left\{(\sigma_1,\sigma_2,\sigma_3,\sigma_5)\in\mathbb{R}^4: \sigma_1=-\sqrt{\frac{2}{3}}\sigma_2,\,\sigma_5=0\right\}.$$

\end{enumerate}

\end{enumerate}

Let us comment on the physical interpretation of the critical points listed above. The critical points $Q_1$ represent matter-dominated cosmological solutions with infinite curvature ($H\rightarrow +\infty$). Since $\chi'(\sigma_{1c})=0$ (i.e., $\sigma_{1c}$ is an stationary point of the coupling function) and $\chi'(\sigma_{1c})\neq 0$ they are solutions with minimally coupled scalar field and negligible kinetic energy. The potential function does not influence its dynamical character.

The critical points $Q_2$ represent a \emph{de Sitter} cosmological solution. When one of these critical points is approached, the energy density of dark matter and the kinetic energy of the scalar field go to zero. In this case the potential energy of the scalar field becomes dominant. Hence, the universe would be expanding forever in a \emph{de Sitter} phase.

The critical points $Q_3$ represent radiation-dominated cosmological solutions. These solutions are very important during the radiation era.

\subsubsection{Normal form expansion for \eqref{eqsigma1}-\eqref{eqsigma5} around $Q_2.$}

In this section we present normal form expansions for \eqref{eqsigma1}-\eqref{eqsigma5} around $Q_2.$ We exclude from the normal forms analysis the case where  $V'(\sigma_{1c})=0$ and $V''(\sigma_{1c})>0$ (i.e., $V(\phi)$ has a local minimum at $\sigma_{1c}$) since from the linear analysis (see point 2c in subsection \ref{Qualitative}) follows the asymptotic stability of $Q_2.$

Let $V(\phi)$ and $\chi(\phi)$ be smooth ($C^\infty$) functions. The barotropic index of the fluid satisfies $0<\gamma<2, \gamma\neq \frac{4}{3}.$

{\bf Case 1}. Let $\sigma_{1c}$ such that
$V'(\sigma_{1c})=V''(\sigma_{1c})=0$ and $V^{(3)}(\sigma_{1c})\neq
0.$ \footnote{In this case $\sigma_{1c}$ is not an extremum point of
$V(\phi)$ but an inflection point.}

Let us consider the coordinate transformation $(\sigma_1,\sigma_2,\sigma_3,\sigma_5)\rightarrow
(\sigma_1+\sigma_{1c},\sigma_2,\sigma_3,\sigma_5).$ The Taylor expansion up to third order of the of the system arising from \eqref{eqsigma1}-\eqref{eqsigma5} under such coordinate transformation around the origin reads

\begin{align} &\sigma_1'=\sqrt{\frac{2}{3}}\sigma_2 +{\cal O}(3),\nonumber\\
&\sigma_2'=-\sigma_2-\frac{\sigma_1^2}{2\sqrt{6}}\frac{V^{(3)}(\sigma_{1c})}{V(\sigma_{1c})}+\frac{(4-3\gamma)\sigma_3^2}{2\sqrt{6}}
\frac{\chi'(\sigma_{1c})}{\chi(\sigma_{1c})}+{\cal
O}(3),\nonumber\\
&\sigma_3'=-\frac{\gamma}{2}\sigma_3-\frac{(4-3\gamma)\sigma_2\sigma_3}{2\sqrt{6}}
\frac{\chi'(\sigma_{1c})}{\chi(\sigma_{1c})}+{\cal
O}(3),\nonumber\\
&\sigma_5'=-\frac{2}{3}\sigma_5+{\cal O}(3).
\label{thirdorder}\end{align}

\begin{enumerate}
\item Let be $\gamma\neq 1.$

Applying the successive coordinate transformations

\be
\left(\begin{array}{c}\sigma_1\\\sigma_2\\\sigma_3\\\sigma_5\end{array}\right)\rightarrow \left(\begin{array}{c}\sigma_3-\sqrt{\frac{2}{3}}\sigma_1\\\sigma_1\\\sigma_5\\\sigma_2\end{array}\right)
\ee

\be
\left(\begin{array}{c}\sigma_1\\\sigma_2\\\sigma_3\\\sigma_5\end{array}\right)\rightarrow\left(
\begin{array}{c}
 \sigma_1+\frac{(3 \gamma -4) \chi '(\sigma_{1c}) \sigma_5^2}{2
   \sqrt{6} (\gamma -1) \chi (\sigma_{1c})}+\frac{\left(2
   \sigma_1^2-3 \sigma_3^2\right)
   V^{(3)}(\sigma_{1c})}{6 \sqrt{6} V(\sigma_{1c})} \\
 \sigma_2 \\
 \sigma_3+\frac{(3 \gamma -4) \chi '(\sigma_{1c}) \sigma_5^2}{6
   \gamma  \chi (\sigma_{1c})}+\frac{\sigma_1
   \left(\sigma_1-2 \sqrt{6} \sigma_3\right)
   V^{(3)}(\sigma_{1c})}{18 V(\sigma_{1c})} \\
 \sigma_5+\frac{(4-3 \gamma ) \sigma_1 \sigma_5 \chi
   '(\sigma_{1c})}{2 \sqrt{6} \chi (\sigma_{1c})}
\end{array}
\right)
\ee

The system \eqref{thirdorder} becomes

\begin{align} &\sigma_1'=\sigma_1 \left(\frac{\sigma_3
   V^{(3)}(\sigma_{1c})}{3 V(\sigma_{1c})}-1\right)+{\cal O}(3),\nonumber\\
   &\sigma_2'=-\frac{2 \sigma_2}{3}+{\cal O}(3),\nonumber\\
   &\sigma_3'=-\frac{\sigma_3^2 V^{(3)}(\sigma_{1c})}{6 V(\sigma_{1c})}+{\cal O}(3),\nonumber\\
   &\sigma_5'=-\frac{\gamma  \sigma_5}{2}+{\cal O}(3).\label{normalthirdorder1}\end{align}

Neglecting the terms of third order
in \eqref{normalthirdorder1} the exact solution of the approximated system passing through $(\sigma_{10}, \sigma_{20}, \sigma_{30},
\sigma_{50})$ at $\tau=\tau_0$ is given by \begin{align}
&\sigma_1(\tau)=e^{-\left(\tau-\tau_0\right)} \sigma_{10} \left(1+\frac{ (\tau-\tau_0) \sigma_{30}  V^{(3)}(\sigma_{1c})}{6 V(\sigma_{1c})}\right)^2,\nonumber\\
&\sigma_2(\tau)=e^{-\frac{2}{3}
   (\tau-\tau_0)} \sigma_{20},\nonumber\\
&\sigma_3(\tau)=\sigma_{30}\left(1+\frac{ (\tau-\tau_0) \sigma_{30}  V^{(3)}(\sigma_{1c})}{6 V(\sigma_{1c})}\right)^{-1},\nonumber\\
&\sigma_5(\tau)=e^{\frac{1}{2}
   (\tau_0-\tau) \gamma } \sigma_{50}.\label{solution1}
\end{align} From solution \eqref{solution1} it is easy to check that
\be\lim_{\tau\rightarrow
+\infty}(\sigma_1(\tau),\sigma_2(\tau),\sigma_3(\tau),\sigma_4(\tau))=(0,0,0,0)\ee
at the rate $(1+{\cal O(\tau)})^{-1}.$ This means that, under the conditions discussed here, the origin of coordinates is asymptotically stable for the flow of the original system \eqref{thirdorder}.
 \item Let be $\gamma=1.$ Applying the successive coordinate transformations

\be
\left(\begin{array}{c}\sigma_1\\\sigma_2\\\sigma_3\\\sigma_5\end{array}\right)\rightarrow \left(
\begin{array}{c}
 \sigma_5-\sqrt{\frac{2}{3}} \sigma_1 \\
 \sigma_1 \\
 \sigma_3 \\
 \sigma_2
\end{array}
\right)
\ee

\be
\left(\begin{array}{c}\sigma_1\\\sigma_2\\\sigma_3\\\sigma_5\end{array}\right)\rightarrow\left(
\begin{array}{c}
 \sigma_1+\frac{\left(2 \sigma_1^2-3 \sigma_5^2\right) V^{(3)}(\sigma_{1c})}{6 \sqrt{6} V(\sigma_{1c})} \\
 \sigma_2 \\
 \sigma_3+\frac{\sigma_1 \sigma_3 \chi '(\sigma_{1c})}{2 \sqrt{6} \chi (\sigma_{1c})} \\
 \sigma_5+\frac{1}{18} \left(\frac{\sigma_1 \left(\sigma_1-2 \sqrt{6} \sigma_5\right) V^{(3)}(\sigma_{1c})}{V(\sigma_{1c})}-\frac{3 \sigma_3^2 \chi '(\sigma_{1c})}{\chi (\sigma_{1c})}\right)
\end{array}
\right)
\ee

The system \eqref{thirdorder}  becomes  \begin{align}
&\sigma_1'=\sigma_1 \left(\frac{\sigma_5
   V^{(3)}(\sigma_{1c})}{3 V(\sigma_{1c})}-1\right)+\frac{\chi '(\sigma_{1c}) \sigma_3^2}{2 \sqrt{6} \chi (\sigma_{1c})}+{\cal O}(3),\nonumber\\
&\sigma_2'=-\frac{2 \sigma_2}{3}+{\cal O}(3),\nonumber\\
&\sigma_3'=-\frac{\sigma_3}{2}+{\cal O}(3),\nonumber\\
&\sigma_5'=-\frac{\sigma_5^2 V^{(3)}(\sigma_{1c})}{6
V(\sigma_{1c})}+{\cal O}(3). \label{normalthirdorder2}\end{align}
Neglecting the terms of third order in \eqref{normalthirdorder2}
the exact solution of the approximated system passing through
$(\sigma_{10}, \sigma_{20}, \sigma_{30}, \sigma_{50})$ at
$\tau=\tau_0$ is given by
\begin{align} &\sigma_1(\tau)=e^{-\left(\tau-\tau_0\right)} \sigma_{10} \left(1+\frac{ (\tau-\tau_0) \sigma_{50}  V^{(3)}(\sigma_{1c})}{6 V(\sigma_{1c})}\right)^2+\nonumber\\&+\frac{e^{-\left(\tau-\tau_0\right)} (\tau-\tau_0) \chi '(\sigma_{1c}) \sigma_{30}^2}{2 \sqrt{6} \chi (\sigma_{1c})}\left(1+\frac{ (\tau-\tau_0) \sigma_{50}  V^{(3)}(\sigma_{1c})}{6 V(\sigma_{1c})}\right),\nonumber\\
   &\sigma_2(\tau)=e^{-\frac{2}{3} (\tau-\tau_0)} \sigma_{20},\nonumber\\
   &\sigma_3(\tau)=e^{-\frac{1}{2} (\tau-\tau_0)} \sigma_{30},\nonumber\\
   &\sigma_5(\tau)=\left(1+\frac{ (\tau-\tau_0) \sigma_{50}  V^{(3)}(\sigma_{1c})}{6 V(\sigma_{1c})}\right)^{-1}.\label{solution2}
\end{align} From solution \eqref{solution2} it is easy to check that
\be\lim_{\tau\rightarrow
+\infty}(\sigma_1(\tau),\sigma_2(\tau),\sigma_3(\tau),\sigma_5(\tau))=(0,0,0,0)\ee
at the rate $(1+{\cal O(\tau)})^{-1}.$ This means that, under the conditions discussed here, the origin of coordinates is asymptotically stable for the flow of the original system \eqref{thirdorder}.
\end{enumerate}

{\bf Case 2}. Let the function $V(\phi)$ to have a degenerate
local minimum at $\sigma_{1c}$ (for $n=2$).

Let us consider the coordinate transformation $(\sigma_1,\sigma_2,\sigma_3,\sigma_5)\rightarrow
(\sigma_1+\sigma_{1c},\sigma_2,\sigma_3,\sigma_5).$ The Taylor expansion up to fourth order of the of the system arising from \eqref{eqsigma1}-\eqref{eqsigma5} under such coordinate transformation around the origin reads

\begin{align}
&\sigma_1'=\sqrt{\frac{2}{3}}\sigma_2 +{\cal O}(4),\nonumber\\
&\sigma_2'=-\frac{V^{(4)}(\sigma_{1c}) \sigma_1^3}{6 \sqrt{6} V(\sigma_{1c})}+\frac{(3
   \gamma -4) \sigma_3^2 \chi '(\sigma_{1c})^2
   \sigma_1}{2 \sqrt{6} \chi (\sigma_{1c})^2}+\frac{(4-3
   \gamma ) \sigma_3^2 \chi ''(\sigma_{1c})
   \sigma_1}{2 \sqrt{6} \chi (\sigma_{1c})}+\nonumber\\&+\sigma_2^3-\sigma_2+\frac{1}{6}
   \sigma_2 \left(3 \gamma  \sigma_3^2+4 \sigma_5^2\right)+\frac{(4-3 \gamma ) \sigma_3^2 \chi
   '(\sigma_{1c})}{2 \sqrt{6} \chi (\sigma_{1c})}+{\cal O}(4),\nonumber\\
&\sigma_3'=\frac{(4-3 \gamma ) \sigma_1 \sigma_2
   \sigma_3 \chi '(\sigma_{1c})^2}{2 \sqrt{6} \chi
   (\sigma_{1c})^2}+\frac{(3 \gamma -4) \sigma_2
   \sigma_3 \chi '(\sigma_{1c})}{2 \sqrt{6} \chi
   (\sigma_{1c})}-\frac{\gamma  \sigma_3}{2}+\nonumber\\&+\frac{1}{6}
   \sigma_3 \left(6 \sigma_2^2+3 \gamma  \sigma_3^2+4 \sigma_5^2\right)+\frac{(3 \gamma -4) \sigma_1 \sigma_2 \sigma_3 \chi ''(\sigma_{1c})}{2 \sqrt{6} \chi (\sigma_{1c})}+{\cal O}(4),\nonumber\\
&\sigma_5'=\frac{1}{6}
   \sigma_5 \left(6 \sigma_2^2+3 \gamma  \sigma_3^2+4 \sigma_5^2\right)-\frac{2 \sigma_5}{3}+ {\cal O}(4).\label{fourthorder}
\end{align}

\begin{enumerate}
\item Let be $\gamma\neq 1.$
Applying the successive coordinate transformations

\be
\left(\begin{array}{c}\sigma_1\\\sigma_2\\\sigma_3\\\sigma_5\end{array}\right)\rightarrow \left(\begin{array}{c}\sigma_3-\sqrt{\frac{2}{3}}\sigma_1\\\sigma_1\\\sigma_5\\\sigma_2\end{array}\right)
\ee

\be
\left(\begin{array}{c}\sigma_1\\\sigma_2\\\sigma_3\\\sigma_5\end{array}\right)\rightarrow\left(
\begin{array}{c}
 \sigma_1+\frac{(3 \gamma -4) \chi '(\sigma_{1c}) \sigma_5^2}{2
   \sqrt{6} (\gamma -1) \chi (\sigma_{1c})} \\
 \sigma_2 \\
 \sigma_3+\frac{(3 \gamma -4) \chi '(\sigma_{1c}) \sigma_5^2}{6
   \gamma  \chi (\sigma_{1c})}\\
 \sigma_5+\frac{(4-3 \gamma ) \sigma_1 \sigma_5 \chi
   '(\sigma_{1c})}{2 \sqrt{6} \chi (\sigma_{1c})}
\end{array}
\right)
\ee

\begin{align}
&\left(\begin{array}{c}\sigma_1\\\sigma_2\\\sigma_3\\\sigma_5\end{array}\right)\rightarrow\nonumber\\&\left(
\begin{array}{c}
\Xi_1-\frac{1}{2}
   \sigma_1 \left(\sigma_1^2+\sigma_2^2+\sigma_5^2\right)-\frac{\left(2 \sigma_1^3-6 \sqrt{6}
   \sigma_3 \sigma_1^2+3 \sqrt{6} \sigma_3^3\right) V^{(4)}(\sigma_{1c})}{108 V(\sigma_{1c})} \\
 -\frac{1}{2} \sigma_2 \left(\sigma_1^2+\sigma_2^2+\sigma_5^2\right) \\
 \Xi_2+\frac{\sigma_1 \left(-2 \sigma_1^2-\frac{12 \sigma_2^2}{7}-\frac{3 \gamma  \sigma_5^2}{\gamma
   +1}\right)}{3 \sqrt{6}}-\frac{\sigma_1 \left(2 \sqrt{6} \sigma_1^2-27 \sigma_3 \sigma_1+27 \sqrt{6}
   \sigma_3^2\right) V^{(4)}(\sigma_{1c})}{486 V(\sigma_{1c})} \\
 \Xi_3-\frac{1}{2} \sigma_5 \left(\sigma_1^2+\sigma_2^2+\sigma_5^2\right)+\frac{(3 \gamma
   -4) \sigma_1 \left(\sigma_1-\sqrt{6} \sigma_3\right) \sigma_5 \chi ''(\sigma_{1c})}{12 \chi
   (\sigma_{1c})}
\end{array}
\right)
\end{align}
where

$\Xi_1= -\frac{(3 \gamma -4) \left(3 ((\gamma -3) \gamma +2) \sigma_1+\sqrt{6} \gamma  \sigma_3\right) \chi '(\sigma_{1c})^2
   \sigma_5^2}{12 (\gamma -1) \gamma  \chi (\sigma_{1c})^2}-\frac{(3 \gamma -4) \left(2 (\gamma -1) \sigma_1-\sqrt{6}
   \gamma  \sigma_3\right) \chi ''(\sigma_{1c}) ^2}{12 (\gamma -1) \gamma  \chi (\sigma_{1c})},$

   $\Xi_2=\frac{(3 \gamma -4) \left(-\frac{\sqrt{6} (\gamma -2) \sigma_1}{\gamma +1}-\frac{2 \sigma_3}{\gamma }\right) \chi
   '(\sigma_{1c})^2 \sigma_5^2}{12 \chi (\sigma_{1c})^2}+\frac{(3 \gamma -4) \left(\frac{3 \sigma_3}{\gamma
   }-\frac{\sqrt{6} \sigma_1}{\gamma +1}\right) \chi ''(\sigma_{1c}) \sigma_5^2}{18 \chi (\text{$\sigma
   $1c})}$

   and

   $\Xi_3=\frac{(3 \gamma -4) \sigma_5 \left(8 \sigma_5^2+\gamma  \left((\gamma -1) \sigma_1 \left((3 \gamma -8) \text{$\sigma
   $1}+4 \sqrt{6} \sigma_3\right)-6 \sigma_5^2\right)\right) \chi '(\sigma_{1c})^2}{48 (\gamma -1) \gamma  \chi
   (\sigma_{1c})^2},$ the system \eqref{fourthorder} becomes

   \begin{align}
   &\sigma_1'=\frac{1}{6} \sigma_1 \left(\frac{\sigma_3^2 V^{(4)}(\sigma_{1c})}{V(\sigma_{1c})}-6\right)+{\cal O}(4),\nonumber\\
   &\sigma_2'=-\frac{2
  \sigma_2}{3}+{\cal O}(4),\nonumber\\
   &\sigma_3'=-\frac{\sigma_3^3 V^{(4)}(\sigma_{1c})}{18 V(\sigma_{1c})}+{\cal O}(4),\nonumber\\
   &\sigma_5'=-\frac{\gamma  \sigma_5}{2}+{\cal O}(4).\label{fourthorder1}
   \end{align}

\item Let be $\gamma=1.$ Applying the
successive coordinate transformations

\be
\left(\begin{array}{c}\sigma_1\\\sigma_2\\\sigma_3\\\sigma_5\end{array}\right)\rightarrow \left(
\begin{array}{c}
 \sigma_5-\sqrt{\frac{2}{3}} \sigma_1 \\
 \sigma_1 \\
 \sigma_3 \\
 \sigma_2
\end{array}
\right)
\ee

\be
\left(\begin{array}{c}\sigma_1\\\sigma_2\\\sigma_3\\\sigma_5\end{array}\right)\rightarrow\left(
\begin{array}{c}
 \sigma_1\\
 \sigma_2 \\
 \sigma_3+\frac{\sigma_1 \sigma_3 \chi
   '(\sigma_{1c})}{2 \sqrt{6} \chi (\sigma_{1c})}\\
 \sigma_5-\frac{\sigma_3^2 \chi '(\sigma_{1c})}{6 \chi
   (\sigma_{1c})}
\end{array}
\right)
\ee

\begin{align}
&\left(\begin{array}{c}\sigma_1\\\sigma_2\\\sigma_3\\\sigma_5\end{array}\right)\rightarrow\nonumber\\&
\left(
\begin{array}{c}
 \Xi_4-\frac{1}{2} \sigma_1 \left(\sigma_1^2+\sigma_2^2+\sigma_3^2\right)-\frac{\left(2
   \sigma_1^3-6 \sqrt{6} \sigma_5 \sigma_1^2+3 \sqrt{6} \sigma_5^3\right) V^{(4)}(\sigma_{1c})}{108 V(\sigma_{1c})} \\
 -\frac{1}{2} \sigma_2 \left(\sigma_1^2+\sigma_2^2+\sigma_3^2\right) \\
 \Xi_5-\frac{1}{2}
   \sigma_3 \left(\sigma_1^2+\sigma_2^2+\sigma_3^2\right)-\frac{\sigma_1
   \sigma_3 \left(\sigma_1-\sqrt{6} \sigma_5\right) \chi ''(\sigma_{1c})}{12 \chi (\sigma_{1c})}
   \\
 \Xi_6-\frac{\sigma_1 \left(28 \sigma_1^2+24 \sigma_2^2+21 \sigma_3^2\right)}{42
   \sqrt{6}}-\frac{\sigma_1 \left(2 \sqrt{6} \sigma_1^2-27 \sigma_5 \sigma_1+27 \sqrt{6}
   \sigma_5^2\right) V^{(4)}(\sigma_{1c})}{486
   V(\sigma_{1c})}
\end{array}
\right)
\end{align} where $\Xi_4=-\frac{\sigma_1 \chi '(\sigma_{1c})^2 \sigma_3^2}{4 \chi (\sigma_{1c})^2}+\frac{\sigma_1 \chi
   ''(\sigma_{1c}) \sigma_3^2}{6 \chi (\sigma_{1c})},$ $\Xi_5=\frac{\sigma_3 \left(5 \sigma_1^2-4 \sqrt{6}
   \sigma_5 \sigma_1+2 \sigma_3^2\right) \chi
   '(\sigma_{1c})^2}{48 \chi (\sigma_{1c})^2}$ and $\Xi_6=-\frac{\left(\sqrt{6} \sigma_1-4 \sigma_5\right) \chi
   '(\sigma_{1c})^2 \sigma_3^2}{24 \chi (\sigma_{1c})^2}+\frac{\left(\sqrt{6} \sigma_1-6 \sigma_5\right) \chi ''(\sigma_{1c}) \sigma_3^2}{36 \chi
   (\sigma_{1c})},$
the system \eqref{fourthorder} becomes

\begin{align}
&\sigma_1'=\sigma_5 \left(\frac{\chi ''(\sigma_{1c})}{2
   \sqrt{6} \chi (\sigma_{1c})}-\frac{\chi '(\sigma_{1c})^2}{2 \sqrt{6} \chi (\sigma_{1c})^2}\right) \sigma_3^2+\sigma_1 \left(\frac{\sigma_5^2
   V^{(4)}(\sigma_{1c})}{6 V(\sigma_{1c})}-1\right)+{\cal O}(4),\nonumber\\
&\sigma_2'=   -\frac{2 \sigma_2}{3}+{\cal O}(4),\nonumber\\
&\sigma_3'=-\frac{\sigma_3}{2}+{\cal O}(4),\nonumber\\
&\sigma_5'=-\frac{\sigma_5^3 V^{(4)}(\sigma_{1c})}{18
   V(\sigma_{1c})}+{\cal O}(4).\label{fourthorder2}
\end{align}

\end{enumerate}

Having at hand the normal forms expansions \eqref{fourthorder1} and \eqref{fourthorder2} we can extract the following information.
Let $V(\sigma_1)$ to have  a strict degenerate local minimum at $\sigma_{1c}$ (for $n=2$) with $V(\sigma_{1c})>0$. Let $\gamma\neq 1.$ Neglecting the terms of fourth order in \eqref{fourthorder1} the exact solution of the approximated system passing through
$(\sigma_{10}, \sigma_{20}, \sigma_{30}, \sigma_{50})$ at
$\tau=\tau_0$ is given by

\begin{align} & \sigma_1(\tau)= \left(\frac{(\tau-\tau_0)
V^{(4)}(\sigma_{1c}) \sigma_{30}^2}{9
   V(\sigma_{1c})}+1\right)^{\frac{3}{2}}e^{\tau_0-\tau}\sigma_{10},\nonumber\\
&\sigma_2(\tau)=e^{-\frac{2}{3} (\tau-\tau_0)} \sigma_{20},\nonumber\\
&\sigma_3(\tau)=\left(\frac{(\tau-\tau_0) V^{(4)}(\sigma_{1c})
\sigma_{30}^2}{9
   V(\sigma_{1c})}+1\right)^{-\frac{1}{2}}\sigma_{30},\nonumber\\
&\sigma_5(\tau)=e^{-\frac{\gamma}{2} (\tau-\tau_0)} \sigma_{50}
.\label{solution4a}
\end{align}
From solution \eqref{solution4a} it is easy to check that
\be\lim_{\tau\rightarrow
+\infty}(\sigma_1(\tau),\sigma_2(\tau),\sigma_3(\tau),\sigma_5(\tau))=(0,0,0,0)\ee
at the rate $(1+{\cal O(\tau)})^{-1/2}.$

Let $\gamma=1.$ Neglecting the terms of third order in \eqref{fourthorder2} the
exact solution of the approximated system passing through
$(\sigma_{10}, \sigma_{20}, \sigma_{30}, \sigma_{50})$ at
$\tau=\tau_0$ is given by

\begin{align} & \sigma_1(\tau)= \left(\frac{(\tau-\tau_0)
V^{(4)}(\sigma_{1c}) \sigma_{50}^2}{9
   V(\sigma_{1c})}+1\right)^{\frac{1}{2}}e^{\tau_0-\tau}\times\nonumber\\
&\left(\frac{(\tau_0-\tau) \sigma_{50} \chi
   '(\sigma_{1c})^2 \sigma_{30}^2}{2 \sqrt{6} \chi
   (\sigma_{1c})^2}+\frac{(\tau-\tau_0) \sigma_{50} \chi
   ''(\sigma_{1c}) \sigma_{30}^2}{2 \sqrt{6} \chi
   (\sigma_{1c})}+\sigma_{10}+\frac{(\tau-\tau_0)
   \sigma_{10} \sigma_{50}^2 V^{(4)}(\sigma_{1c})}{9
   V(\sigma_{1c})}\right),\nonumber\\
&\sigma_2(\tau)=e^{-\frac{2}{3} (\tau-\tau_0)} \sigma_{20},\nonumber\\
&\sigma_3(\tau)=e^{-\frac{1}{2} (\tau-\tau_0)} \sigma_{30},\nonumber\\
&\sigma_5(\tau)=\left(\frac{(\tau-\tau_0) V^{(4)}(\sigma_{1c})
\sigma_{50}^2}{9
   V(\sigma_{1c})}+1\right)^{-\frac{1}{2}}\sigma_{50}.\label{solution4}
\end{align}
From solution \eqref{solution4} it is easy to check that
\be\lim_{\tau\rightarrow
+\infty}(\sigma_1(\tau),\sigma_2(\tau),\sigma_3(\tau),\sigma_5(\tau))=(0,0,0,0)\ee
$(1+{\cal O(\tau)})^{-1/2}.$

Hence, for all $\gamma\in (0,4/3)\cup(4/3,2),$ the origin of coordinates is asymptotically stable for the flow of the original system \eqref{fourthorder}. Since the remainder of the Taylor expansion \eqref{fourthorder} is ${\cal O}(|{\sigma}|^4)$ as ${\sigma}\rightarrow {\bf 0},$ uniformly in $\tau_0\leq \tau<\infty,$ follows the asymptotic stability of the origin.
This analysis is in agreement with the result of proposition
\ref{thmIII} which states that if $V(\sigma_1)$ have  a strict degenerate local minimum at $\sigma_{1c}$ with $V(\sigma_{1c})>0,$
then $Q_2:=(\sigma_{1c},0,0,0)$ is asymptotically stable.

\subsubsection{A generalization of theorem 3.2 of \cite{Leon:2008de} p. 8.}

Theorem 3.2 of \cite{Leon:2008de} p. 8, states, essentially, that if the potential and the coupling function are sufficiently smooth functions, then for almost all the points $p$ lying in $\Sigma_0,$ then the scalar field diverges through the past orbit of $p.$ This result can generalized as follows.

\paragraph{Generalization}

\begin{thm}\label{thm4}
Assume that $\chi(\phi)$ and $V(\phi)$ are positive functions of
class $C^3,$ such that $\chi$ has at most a finite number of
stationary points and does not tend to zero in any compact set of $\mathbb{R}$. Let $\gamma\in \left(0,\frac{4}{3}\right)\cup\left(\frac{4}{3},2\right)$ and let $p$ be a point in
$\Sigma_+.$ Let $O^{-}(p)$ be the past orbit of $p$ under the flow
of \eqref{eq0phi}-\eqref{eq0x4} restricted to $\Sigma_+$. Then,
$\phi$ is always unbounded on $O^{-}(p)$ for almost all $p$.
\end{thm}

\paragraph{\bf Proof of theorem \ref{thm4}} In order to prove the theorem it is sufficient to consider interior points of $\Sigma_+.$ Also, in order to apply results concerning future attractors and $\omega$-limit sets we perform the time reversal $\tau\rightarrow -\tau.$ Thus we get the system

\begin{align}
&\sigma_1'=-\sqrt{\frac{2}{3}} \sigma_2 \label{eqrevphi}\\
&\sigma_2'=-\sigma_2^3-\frac{1}{6}\left(3\gamma \sigma_3^2+4
\sigma_5^2-6\right)\sigma_2
+\frac{\sigma_4^2}{\sqrt{6}}\frac{\mathrm{d}\ln
V(\sigma_1)}{\mathrm{d}\sigma_1}
-\frac{\left(4-3\gamma\right)\sigma_3^2}{2\sqrt{6}}\frac{\mathrm{d}\ln \chi(\sigma_1)}{\mathrm{d}\sigma_1},\label{eqrevx1}\\
&\sigma_3'=-\frac{1}{6}\sigma_3\left(6\sigma_2^2+3\gamma\left(\sigma_3^2-1\right)+4
\sigma_5^2\right)
+\frac{\left(4-3\gamma\right)\sigma_2 \sigma_3}{2\sqrt{6}}\frac{\mathrm{d}\ln \chi(\sigma_1)}{\mathrm{d}\sigma_1},\label{eqrevx2}\\
&\sigma_4'=-\frac{1}{6}\sigma_4\left(6\sigma_2^2+3\gamma
\sigma_3^2+4 \sigma_5^2\right)
-\frac{\sqrt{6}}{6}\sigma_2 \sigma_4 \frac{\mathrm{d}\ln V(\sigma_1)}{\mathrm{d}\sigma_1},\label{eqrevx3}\\
&\sigma_5'=-\frac{1}{6}\sigma_5\left(6\sigma_2^2+3\gamma
\sigma_3^2+4 \sigma_5^2-4\right).\label{eqrevx4}
\end{align}
where the prime denotes now derivative with respect to $-\tau.$

Let
$p_0:=(\sigma_{10},\sigma_{20},\sigma_{30},\sigma_{40},\sigma_{50})\in
\text{Int} \Sigma_+$ such that there exist a real positive number
$K$ with $|\sigma_1|<K$ for all
$p:=(\sigma_1,\sigma_2,\sigma_3,\sigma_4,\sigma_5)\in O^+(p_0),$
where $O^+(p_0)$ denotes the positive (future) orbit for the flow
of \eqref{eqrevphi}-\eqref{eqrevx4}.  Then, for all $p\in
O^+(p_0)$ we have $$-1\leq \sigma_2\leq 1,\, 0\leq \sigma_3\leq
1,\,0\leq \sigma_4\leq 1,\, 0\leq \sigma_5\leq 1.$$ Hence
$O^+(p_0)$ is contained in a compact set of (the closure of)
$\Sigma_+.$

Since $O^+(p_0)$ is a positive invariant set, then using proposition \ref{omegalimitsetproperties} we ensure the existence of a non empty, closed, connected and invariant $\omega$-limit of $p_0$ denoted by $\omega(p_0).$

First we demonstrate by contradiction that that $\sigma_3$ and $\sigma_4$ cannot be simultaneously zero at $\omega(p_0).$ Suppose that $\omega(p_0)$ is contained in the set where
$\sigma_3=\sigma_4=0.$ Let us define the function $M_1=Z_4^{-1}$
(see table \ref{monotonic} for the definition of $Z_4$) defined in
the invariant set $\sigma_3=\sigma_4=0, 0<\sigma_5<1.$ From the definition of $M_1$ and applying the Monotonicity principle (theorem \ref{theorem 4.12}) follows that the future asymptotic attractor of the flow of \eqref{eqrevphi}-\eqref{eqrevx4} restricted to the invariant set $\sigma_3=\sigma_4=0$ is contained in the invariant set $\sigma_2=\pm 1.$ Thus, from \eqref{eqrevphi} follows that $\phi\rightarrow \mp \infty$ as $\omega(p_0)$ is approached, a contradiction.

Second, let be defined in $Int \Sigma_+$ the function $M_2=Z_1^{-1}$ (see table \ref{monotonic} for the definition of $Z_1$). The derivative of $M_2$ along any orbit of \eqref{eqrevphi}-\eqref{eqrevx4} is given by $M_2'=-\gamma M_2.$
Then $M_2$ is a $C^3$ monotonic decreasing function for the flow taking values in the interval $(0,+\infty).$ Since $\sigma_3$ and
$\sigma_4$ cannot tend to zero simultaneously in $\omega(p)$ for $p\in O^+(p_0),$ then the function $M_2$ tends asymptotically to a well defined limit. By construction $M_2(p)\rightarrow 0$ if and
only if $p\rightarrow q$ with $q$ satisfying $\sigma_4=0$ (we are using here the condition that $\chi(\phi)$ does not tend to zero in any compact of $\mathbb{R}$) and $M_2(p)\rightarrow +\infty$ if and only if $p\rightarrow q$ with $q$ satisfying $\sigma_3=0.$
Thus, applying the Monotonicity principle (theorem \ref{theorem
4.12}) follows that
$$\omega(p_0)\subset \left\{p\in\Sigma_+: |\sigma_1|<K, \sigma_3>0, \sigma_4=0\right\}=S_1.$$
Let $q_0\in\omega(p_0).$ By the invariance of the $\omega$-limit set follows that $\omega(q_0)=\omega(p_0).$

Observe that $g_3(S_1)=\left\{\sigma\in W_3: |\sigma_1|<K,
\sigma_4=0\right\}$ where $g_3$ is defined in the proof of proposition \ref{thm2}.

Let us define the projection map $$g:
(\sigma_1,\sigma_2,\sigma_3,\sigma_5)\rightarrow
(\sigma_1,\sigma_2,\sigma_5)$$ and let $\sigma_0=g\circ g_3(q_0)$
then the flow of \eqref{eqrevphi}-\eqref{eqrevx4} in a neighborhood of $q_0$ contained in $S_1$, is topologically equivalent to the flow of
\begin{align}
&\sigma_1'=-\sqrt{\frac{2}{3}} \sigma_2,\nonumber\\
&\sigma_2'=\frac{1}{2}\left(1-\sigma_2^2\right)\left((2-\gamma)\sigma_2-\frac{(4-3\gamma)}{\sqrt{6}}\frac{\mathrm{d}\ln\chi(\sigma_1)}{\mathrm{d}\sigma_1}\right)-\frac{\left(4-3\gamma\right)\sigma_5^2}{6}\left(\sigma_2-\frac{\sqrt{6}}{2}\frac{\mathrm{d}\ln\chi(\sigma_1)}{\mathrm{d}\sigma_1}\right),\nonumber\\
&\sigma_5'=-\frac{1}{6} \sigma_5 \left(3(2-\gamma)\sigma_2^2-(4-3\gamma)(1-\sigma_5^2)\right), \label{systthm4}
\end{align}
in a neighborhood of $\sigma_0$ contained in
$$S=\left\{(\sigma_1,\sigma_2,\sigma_5): -K<\sigma_1<K,
\sigma_2^2+\sigma_5^2<1, \sigma_5>0\right\}.$$ Since the vector field is $C^2$ we can extent the flow of \eqref{systthm4} to the closure of $S$ (denoted by $\bar{S}$).

Let us investigate all possible compact, non empty, and connected invariant sets of \eqref{proxi1}-\eqref{proxi2} located in the closure of $S$ (these ones can be candidates to the $\omega$-limit $\omega(\sigma_0)$).

Let us consider two cases:

\begin{enumerate}
\item[i)] $0<\gamma<\frac{4}{3}.$ Let be defined in $S$, the function \be M_3(\sigma)=\frac{\left(1-\sigma_2^2-\sigma_5^2\right)^2 \chi
   (\sigma_1)^{4-3 \gamma }}{\sigma_5^4}.\ee  The derivative of $M_3$ through an arbitrary orbit of \eqref{systthm4} is given by $$M_3'=-2\left(\frac{4}{3}-\gamma\right)M_3.$$ Then
$M_3$ is a $C^3$ monotonic decreasing function for the flow taking
values in the interval $(0,+\infty).$ By construction $M_3(p)\rightarrow 0$ if and only if
$p\rightarrow q$ with $q$ satisfying $\sigma_2^2+\sigma_5^2=1$ (since $\chi$ does tends to zero in $[-K,K]$) and $M_3(p)\rightarrow
+\infty$ if and only if $p\rightarrow q$ with $q$ satisfying
$\sigma_5=0.$  Thus, applying the Monotonicity principle (theorem \ref{theorem 4.12})  follows that $$\omega(\sigma_0)\subset\left\{\sigma\in \bar{S}\setminus S: \sigma_2^2+\sigma_5^2=1\right\}.$$ Let $q_0\in \omega(\sigma_0).$ By the invariance of the $\omega$-limit follows that $\omega(\sigma_0)=\omega(q_0).$

Let us define the projection map $$g': (\sigma_1,\sigma_2,\sigma_5)\rightarrow (\sigma_1,\sigma_2)$$ and let $\sigma_0'=g'(q_0)$ then the flow of \eqref{systthm4} in a neighborhood of $q_0$ contained in $S$, is topologically equivalent to the flow of
\begin{align}
&\sigma_1'=-\sqrt{\frac{2}{3}}\sigma_2,\nonumber\\
&\sigma_2'=\frac{1}{3}\sigma_2\left(1-\sigma_2\right)\left(1+\sigma_2\right).\label{casei)}
\end{align} in a neighborhood of $\sigma_0'$ (contained in
$S':=(-K,K)\times(-1,1)$).

Let be defined in $S'$ the function
$$M_4(\sigma)=\frac{1-\sigma_2^2}{\sigma_2^2}$$ which satisfies
$M_4'=-\frac{2}{3}M_4$ along an arbitrary orbit of \eqref{casei)}.
Thus $M_4$ is a $C^3$ monotonic decreasing function in $S'$.
Applying the monotonicity principle (\ref{theorem 4.12}) follows
that $$\omega(\sigma_0')\subset\left\{\sigma\in \bar{S'}\setminus
S': \sigma_2^2=1\right\}.$$ Thus $\omega(\sigma_0')$ is contained
in one of the invariant sets of \eqref{proxi1}-\eqref{proxi2}
given by $\sigma_2=\pm 1$ but this would imply the divergence of
$\phi.$ A contradiction.

\item[ii)] $\frac{4}{3}<\gamma<2.$ Let be defined in $S$, the function \be M_5(\sigma)=\frac{\sigma_5^2 \chi
   (\sigma_1)^{3 \gamma -4}}{\left(1-\sigma_2^2-\sigma_5^2\right)^2}.\ee  The derivative of $M_5$ through an arbitrary orbit of \eqref{systthm4} is given by $$M_5'=-2\left(\gamma-\frac{4}{3}\right)M_5.$$ Then
$M_5$ is a $C^3$ monotonic decreasing function for the flow taking values in the interval $(0,+\infty).$ By construction $M_5(p)\rightarrow 0$ if and only if
$p\rightarrow q$ with $q$ satisfying $\sigma_5=0$ (since $\chi$ does not tends to zero in $[-K,K]$) and $M_5(p)\rightarrow
+\infty$ if and only if $p\rightarrow q$ with $q$ satisfying
$\sigma_2^2+\sigma_5^2=1.$ Applying the Monotonicity principle (theorem \ref{theorem 4.12}) follows that $$\omega(\sigma_0)\subset \left\{\sigma\in \bar{S}\setminus S: \sigma_5=0\right\}.$$ Let $q_0\in \omega(\sigma_0).$ By the invariance of the $\omega$-limit follows that $\omega(\sigma_0)=\omega(q_0).$

Let us define the projection
map \be h: (\sigma_1,\sigma_2,\sigma_5)\rightarrow (\sigma_1,\sigma_2)\label{proj}.\ee Let $\sigma_0'=h(q_0).$ Then, then the flow of \eqref{systthm4} in a neighborhood of $q_0$ contained in $S$, is topologically equivalent to the flow of

\begin{align}
&\sigma_1'=-\sqrt{\frac{2}{3}}\sigma_2,\label{proxi1}\\
&\sigma_2'=\frac{1}{2}\left(1-\sigma_2^2\right)\left((2-\gamma)\sigma_2-\frac{(4-3\gamma)}{\sqrt{6}}\frac{\mathrm{d}\ln\chi(\sigma_1)}{\mathrm{d}\sigma_1}\right),\label{proxi2}
\end{align} in a neighborhood of $\sigma_0'$ (contained in
$S'$).

Let us investigate the possible compact invariant sets of
\eqref{proxi1}-\eqref{proxi2} located in the closure of $S'$ which can be candidates to the $\omega$-limit $\omega(\sigma_0').$

First  $\omega(\sigma_0')$ cannot be contained in the invariant sets of \eqref{proxi1}-\eqref{proxi2} given by $\sigma_2=\pm 1$ because this would imply the divergence of $\phi,$ a contradiction. Second, combining the results of the Poincar\'e-Bendixon Theorem (theorem \ref{PBT}) and Dulac's criterion (theorem \ref{DC}) with $B(\xi)=(1-\sigma_2^2)^{-1}$
follows that the only possible compact invariant sets are the critical points with $\sigma_1$ bounded (or heteroclinic orbits joining
such critical points).

Let us consider $\chi(\sigma_1)$ other than exponential.
\footnote{As we will see next in section \ref{application1}, the following analysis applies also to exponential coupling functions.} In this case the system
\eqref{proxi1}-\eqref{proxi2} admits a (possibly empty) family of
critical points
$$Q:=\left\{(q_1,0)\in[-K,K]\times[-1,1]: \chi'(q_1)=0\right\}.$$
If $Q=\emptyset,$ i.e.,  $\chi'(q_1)\neq 0$ for all $|q_1|<K,$
then the future orbit $O^+(\sigma_0)$ tends to a point with
$\sigma_1=\pm 1.$ From this follows that $\phi$ is unbounded (a
contradiction) and the proof is done.

Let us assume that $Q\neq \emptyset.$ Let $q\in Q.$ The eigenvalues of the Jacobian matrix $\frac{\partial f^i}{\partial
\sigma_j}(q), i,j=1,2$ are $\mu^\pm=\Delta_1\pm
\sqrt{\Delta_1^2+\Delta_2 \frac{\chi''(q)}{\chi(q)}},$ where
$\Delta_1=\frac{2-\gamma}{4}>0, \Delta_2=\frac{4-3\gamma}{6}.$
Hence, at least one of its associated eigenvalues has positive
real part. Let be defined the sets $Q^\pm =\{q\in Q:\pm
\chi''(q)>0\}$ and $Q^0=\{q\in Q: \chi(q)=0\}.$ At least one of
these sets is not empty. Let be define $R=\{p\in [-K, K]\times
[-1,1]: \lim_{\tau\rightarrow\infty}\Phi_\tau(p)=q\}.$ There are
the following cases \footnote{We denote the stable, unstable and
center manifolds of $q$ by ${\cal E}^s(q), {\cal E}^u(q)$ and
${\cal E}^c(q)$ respectively. By $leb(A)$ we denote de Lebesgue
measure of $A\subset\mathbb{R}^2.$}
\begin{itemize}
\item  $q\in Q^+, \,
\frac{4}{3}<\gamma<2,$ then ${\cal E}^u(q)$ is 2-dimensional implying
$R=\emptyset.$ \item $q\in Q^-, \, \frac{4}{3}<\gamma<2,$ then ${\cal E}^u(q)$ is 1-dimensional and
${\cal E}^s(q)$ is 1-dimensional. Then $R\subset N,$ $leb (N)=0.$
\item $q\in Q^0,$ then ${\cal E}^c(q)$ is 1-dimensional and ${\cal E}^u(q)$
is 1-dimensional in such way that $R\subset {\cal E}^c(q),$ $leb ({\cal
E}^c(q))=0.$
\end{itemize}

Therefore, all solutions future asymptotic to $q$ (and then with
$\phi$ bounded towards the future) must lie on an stable manifold or center manifold of dimension $r<2,$ and then contained in a subset of $[-K, K]\times [-1,1]$ with zero Lebesgue measure. Since there are at most a finite number of such $q$ the result of the theorem follows. $\blacksquare$
\end{enumerate}

Theorem \ref{thm4} allow us to conclude that in order to investigate the generic asymptotic behavior of the system \eqref{eq0phi}-\eqref{eq0x4} restricted to $\Sigma_+$ it is sufficient to study the region where $\phi=\pm\infty.$

\subsubsection{Analysis in the limit $\phi\rightarrow\infty.$}

In this section we will investigate the flow as
$\phi\rightarrow \infty$ following the nomenclature and formalism introduced in \cite{Foster:1998sk} (see also \cite{Giambo:2008sa}). Analogous results hold as $\phi\rightarrow-\infty.$

\begin{defn}[Function well-behaved at infinity \cite{Foster:1998sk}]\label{WBI}
Let $V:\mathbb{R}\rightarrow \mathbb{R}$ be a $C^2$ non-negative function. Let there exist
some $\phi_0>0$ for which $V(\phi)> 0$
 for all $\phi>\phi_0$ and some number $N$ such that the function
$W_V:[\phi_0,\infty)\rightarrow \mathbb{R}$,
$$ W_V(\phi)=\frac{\partial_\phi V(\phi)}{V(\phi)} - N $$
 satisfies
\begin{equation}
\lim_{\phi\rightarrow\infty}W_V(\phi)=0.\label{Lim}
\end{equation}
Then we say that $V$ is Well Behaved at Infinity (WBI) of exponential order $N$.
\end{defn}

It is important to point out that $N$ may be 0, or even negative. Indeed the class of WBI functions of order 0 is of particular interest, containing all non-negative polynomials as remarked in \cite{Foster:1998sk}.

In order to classify the smoothness of WBI functions at infinity it is introduced the definition

\begin{defn}\label{bar}
Let be some coordinate transformation $\varphi=f(\phi)$ mapping a neighborhood of infinity to a neighborhood of the origin. If
$g$ is a function of $\phi$,
 $\overline{g}$ is the function of $\varphi$ whose domain is the range of
$f$ plus the origin, which takes the values;

$$
\overline{g}(\varphi)=\left\{\begin{array}{rcr} g(f^{-1}(\varphi))&,&\varphi>0\\
                                     \lim_{\phi\rightarrow\infty} g(\phi)&,&\varphi=0 \end{array}\right.
$$
\end{defn}
\begin{defn}[Class k WBI functions \cite{Foster:1998sk}]\label{CkWBI}
A $C^k$ function $V$ is class k WBI if it is WBI and if there exists $\phi_0>0$ and a coordinate transformation
$\varphi=f(\phi)$ which maps the interval $[\phi_0,\infty)$ onto
$(0, \epsilon]$, where $\epsilon=f(\phi_0)$ and $\lim_{\phi\rightarrow\infty} f=0$, with  the following additional
properties:
\begin{tabbing}
i)\hspace{0.4cm}\=  $f$ is $C^{k+1}$ and strictly decreasing.\\
ii)            \>the functions $\overline{W}_V(\varphi)$ and $\overline{f'}(\varphi)$ are $C^k$ on
the  closed interval $[0,\epsilon]$.\\
iii)           \> ${\displaystyle \frac{\mathrm{d}\overline{W}_V}{\mathrm{d}\varphi}(0)=\frac{\mathrm{d}\overline{f'}}{\mathrm{d}\varphi}(0)=0.}$
\end{tabbing}
\end{defn}

We designate the set of all class k WBI functions ${\cal E}^k_+.$ In table \ref{WBItransf} are displayed simple examples of WBI behavior at large $\phi.$

\begin{table}
\begin{center}
\caption{\label{WBItransf} Simple examples of WBI behavior at large $\phi$. $n$ and $\lambda$ are arbitrary constants. Adapted from \cite{Foster:1998sk}.}
\begin{tabular}{l|l|l|l|l}
\hline
 $V(\phi)$&$W_V(\phi)$ & $\varphi=f(\phi)$&$\overline{W}_V(\varphi)$ &$\overline{f'}(\varphi)$ \\ \hline
$\left|\frac{\lambda}{n}\right|\phi^{n}$&$n\phi^{-1}$
 &$\phi^{-\frac{1}{2}}$
 &$n \varphi^2$&
$-\frac{1}{2}\varphi^3$\\[5pt]
$e^{\lambda\phi}$&0&$\phi^{-1}$&0&$-\varphi^2$\\[3pt]
$2e^{\lambda\sqrt{\phi}}$ &$\lambda\phi^{-\frac{1}{2}}$
&$\phi^{-\frac{1}{4}}$ &$\lambda \varphi^2$&$-\frac{1}{4}\varphi^5$\\[3pt]
$\left(A+(\phi-B)^2\right)e^{-\mu\phi}$ & $\frac{2\left(\phi-B\right)}{A+(B-\phi)^2}$&$\phi^{-\frac{1}{2}}$
 &$-\frac{2\varphi^2\left(B\varphi^2-1\right)}{A \varphi^4+\left(B\varphi^2-1\right)^2}$&
$-\frac{1}{2}\varphi^3$\\[3pt]
$\left(1-e^{-\lambda^2\phi}\right)^2$ & $ -\frac{2 \lambda ^2}{1-e^{\lambda ^2 \phi }}$
&$\phi^{-1}$& $-\frac{2 \lambda ^2}{1-e^{\frac{\lambda ^2}{\varphi }}}$&$-\varphi^2$\\[3pt]
 $\ln{\phi}$&$(\phi\ln\phi)^{-1}$&$(\ln\phi)^{-1}$&$\varphi e^{-\frac{1}{\varphi}}$&$-\varphi e^{-\frac{2}{\varphi}}$\\[3pt]

$\phi^2\ln{\phi}$&$2\phi^{-1} +(\phi\ln\phi)^{-1}$&
$(\ln\phi)^{-1}$&$(2+\varphi)e^{-\frac{1}{\varphi}}$&$-\varphi e^{-\frac{2}{\varphi}}$\\[3pt]
\hline
\end{tabular}

\end{center}
\end{table}

By assuming that $V,\chi\in {\cal E}^3_+,$ with exponential orders $N$ and $M$ respectively, we can define a dynamical system well suited to investigate the dynamics near the initial singularity. We will investigate the critical points therein. Particularly those representing scaling solutions and those associated with the initial singularity.

Let $\Sigma_\epsilon=\left\{(\sigma_1,
\sigma_2,\sigma_3,\sigma_4,\sigma_5)\in \Sigma:
\sigma_1>\epsilon^{-1}\right\}$ where $\epsilon$ is any positive constant which is chosen sufficiently small so as to avoid any points where $V$ or $\chi=0,$ thereby ensuring that $\overline{W}_V(\varphi)$ and $\overline{W}_{\chi}(\varphi)$ are well-defined. \footnote{See \ref{bar} for the definition of functions with bar.}

Let be defined the projection map

\ben && \pi_1: \Sigma_\epsilon \rightarrow \Omega_\epsilon \nonumber\\
&&(\sigma_1, \sigma_2,\sigma_3,\sigma_4,\sigma_5)\rightarrow
(\sigma_1, \sigma_2,\sigma_4,\sigma_5)\een where
\be\Omega_\epsilon:=\left\{\sigma\in\mathbb{R}^4:\sigma_1>\epsilon^{-1},
\sigma_2^2+\sigma_4^2+\sigma_5^2\leq 1, \sigma_j\geq 0, j= 4,
5\right\}.\ee

Let be defined in $\Omega_\epsilon$ the coordinate transformation
$(\sigma_1, \sigma_2,\sigma_4,\sigma_5)
\stackrel{\varphi=f(\sigma_1)}{\longrightarrow} (\varphi,
\sigma_2,\sigma_4,\sigma_5)$ where $f(\sigma_1)$ tends to zero as
$\sigma_1$ tends to $+\infty$ and has been chosen so that the conditions i)-iii) of definition \ref{CkWBI} are satisfied with
$k=2.$

The flow of \eqref{eq0phi}-\eqref{eq0x4} defined on
$\Sigma_\epsilon$ is topologically equivalent (under $f\circ
\pi_1$) to the flow of the 4-dimensional dynamical system
%\footnote{The relations between the new variables and the field variables is as follows: $\sigma_1=\phi,\, \sigma_2=\frac{\dot
%\phi}{\sqrt{6} H}, \,\sigma_4=\frac{\sqrt{V(\phi)}}{\sqrt{3}H},
%\,\sigma_5=\frac{\sqrt{\rho_r}}{\sqrt{3}H}.$}

\begin{align}
&\varphi'=\sqrt{\frac{2}{3}} \overline{f'} \sigma_2,\label{eqvphi}\\
&\sigma_2'= \sigma_2^3+\left(\frac{2 \sigma_5^2}{3}-1\right)
\sigma_2-\frac{\left(\overline{W}_V+N\right)
\sigma_4^2}{\sqrt{6}}+\nonumber\\&+\left(\frac{\left(\overline{W}_{\chi}+M\right)
(4-3 \gamma )}{2 \sqrt{6}}+\frac{\sigma_2 \gamma }{2}\right)
   \left(1-\sigma_2^2-\sigma_4^2-\sigma_5^2\right)\label{eqxrad}\\
&\sigma_4'=\frac{1}{6} \sigma_4 \left(\sqrt{6} \left(\overline{W}_V+N\right) \sigma_2+3(2-\gamma)\sigma_2^2+3\gamma(1-\sigma_4^2)+(4-3\gamma)\sigma_5^2\right), \label{eqyrad}\\
&\sigma_5'=\frac{1}{6} \sigma_5 \left(3(2-\gamma)\sigma_2^2-3\gamma \sigma_4^2-(4-3\gamma)(1-\sigma_5^2)\right),
\label{eqzrad}
\end{align}
defined in the phase space \footnote{For notational simplicity we will denote the image of $\Omega_\epsilon$ under $f$ by the same symbol.} \be\Omega_\epsilon=\{(\varphi, \sigma_2,
\sigma_4,\sigma_5)\in\mathbb{R}^4: 0\leq\varphi\leq
f(\epsilon^{-1}), \sigma_2^2+\sigma_4^2+\sigma_5^2\leq 1,
\sigma_4\geq 0, \sigma_5\geq 0\}.\label{PhaseSpace}\ee It can be easily proved that \eqref{PhaseSpace} defines a manifold with boundary of dimension 4. Its boundary, $\partial\Psi,$  is the union of the sets $\{p\in\Omega_\epsilon:
\varphi=0\},\,\{p\in\Omega_\epsilon:
\varphi=f(\epsilon^{-1})\},\,\{p\in\Omega_\epsilon:
\sigma_4=0\},\,\{p\in\Omega_\epsilon: \sigma_5=0\}$ with the
unitary 3-sphere.

\paragraph{Critical points of the flow of \eqref{eqvphi}-\eqref{eqzrad} in the phase space \eqref{PhaseSpace}.}

The system \eqref{eqvphi}-\eqref{eqzrad} admits the following critical points

\begin{enumerate}
\item The critical point $P_1$ with coordinates $\varphi=0,\sigma_2=-1,\sigma_4=0,\sigma_5=0$ always exists. The eigenvalues of the linearisation around the critical point are $0, \frac{1}{3},1-\frac{N}{\sqrt{6}},\frac{M (4-3 \gamma )}{\sqrt{6}}-\gamma +2.$ This means that the critical point is non-hyperbolic thus the Hartman-Grobmann theorem does not apply. However, by applying the Invariant Manifold theorem, we obtain that:
\begin{enumerate}
\item $P_1$ has a 1-dimensional center manifold tangent to the $\varphi$-axis provided $N\neq\sqrt{6}$ and $M\neq -\frac{\sqrt{6} (\gamma -2)}{3 \gamma -4}$ (otherwise the center manifold would be 2- or 3-dimensional).
\item $P_1$ admits a 3-dimensional unstable manifold and a 1-dimensional center manifold for
\begin{enumerate}\item  $N<\sqrt{6},\, 0<\gamma<\frac{4}{3},\,M>-\frac{\sqrt{6} (\gamma -2)}{3 \gamma -4};$ or
\item $N<\sqrt{6},\, \frac{4}{3}<\gamma<2,\,M<-\frac{\sqrt{6} (\gamma -2)}{3 \gamma -4}.$
\end{enumerate} In this case the center manifold of $P_1$ acts as a local source for an open set of orbits in \eqref{PhaseSpace}.
\item $P_1$ admits a 2-dimensional unstable manifold, a 1-dimensional stable manifold and a 1-dimensional center if
\begin{enumerate}\item  $N>\sqrt{6},\, 0<\gamma<\frac{4}{3},\,M>-\frac{\sqrt{6} (\gamma -2)}{3 \gamma -4};$ or
\item $N>\sqrt{6},\, \frac{4}{3}<\gamma<2,\,M<-\frac{\sqrt{6} (\gamma -2)}{3 \gamma -4};$ or
\item  $N<\sqrt{6},\, 0<\gamma<\frac{4}{3},\,M<-\frac{\sqrt{6} (\gamma -2)}{3 \gamma -4};$ or
\item  $N<\sqrt{6},\, \frac{4}{3}<\gamma<2,\,M>-\frac{\sqrt{6} (\gamma -2)}{3 \gamma -4}.$
\end{enumerate}
\item $P_1$ admits a 1-dimensional unstable manifold, a 2-dimensional stable manifold and a 1-dimensional center manifold for \begin{enumerate}\item  $N>\sqrt{6},\, 0<\gamma<\frac{4}{3},\,M<-\frac{\sqrt{6} (\gamma -2)}{3 \gamma -4};$ or
\item  $N>\sqrt{6},\, \frac{4}{3}<\gamma<2,\,M>-\frac{\sqrt{6} (\gamma -2)}{3 \gamma -4}.$
\end{enumerate}
\end{enumerate}

\item The critical point $P_2$ with coordinates $\varphi=0,\sigma_2=1,\sigma_4=0,\sigma_5=0$ always exists. The eigenvalues of the linearisation around the critical point are $0, \frac{1}{3},1+\frac{N}{\sqrt{6}},-\gamma +\frac{M (3 \gamma -4)}{\sqrt{6}}+2.$
As before, let us determine conditions on the free parameters for the existence of center, unstable and stable manifolds for $P_2$.
\begin{enumerate}
\item If $N\neq-\sqrt{6}$ and $M\neq \frac{\sqrt{6} (\gamma -2)}{3 \gamma -4}$ there exists a 1-dimensional center manifold tangent to the $\varphi$-axis, otherwise the center manifold would be 2- or 3-dimensional.
\item $P_2$ has a 3-dimensional unstable manifold a a 1-dimensional center manifold (tangent the $\varphi$-axis) if \begin{enumerate}\item $N>-\sqrt{6},\,0<\gamma<\frac{4}{3},\,M<\frac{\sqrt{6} (\gamma -2)}{3 \gamma -4};$ or \item $N>-\sqrt{6},\,\frac{4}{3}<\gamma<2,\,M>\frac{\sqrt{6} (\gamma -2)}{3 \gamma -4}.$
\end{enumerate} In this case the center manifold of $P_2$ acts as a local source for an open set of orbits in \eqref{PhaseSpace}.
\item $P_2$ has a 2-dimensional unstable manifold a 1-dimensional stable and a 1-dimensional center manifold if
\begin{enumerate}\item $N<-\sqrt{6},\,0<\gamma<\frac{4}{3},\,M<\frac{\sqrt{6} (\gamma -2)}{3 \gamma -4};$ or
\item $N<-\sqrt{6},\,\frac{4}{3}<\gamma<2,\,M>\frac{\sqrt{6} (\gamma -2)}{3 \gamma -4};$ or
\item $N>-\sqrt{6},\,0<\gamma<\frac{4}{3},\,M>\frac{\sqrt{6} (\gamma -2)}{3 \gamma -4};$ or
\item $N>-\sqrt{6},\,\frac{4}{3}<\gamma<2,\,M<\frac{\sqrt{6} (\gamma -2)}{3 \gamma -4}.$
\end{enumerate}
\item $P_2$ has a 1-dimensional unstable manifold a 2-dimensional stable and a 1-dimensional center manifold if
\begin{enumerate}
\item $N<-\sqrt{6},\,0<\gamma<\frac{4}{3},\,M>\frac{\sqrt{6} (\gamma -2)}{3 \gamma -4};$ or
\item $N<-\sqrt{6},\,\frac{4}{3}<\gamma<2,\,M<\frac{\sqrt{6} (\gamma -2)}{3 \gamma -4}.$
\end{enumerate}
\end{enumerate}
\item The critical point $P_3$ with coordinates $\varphi=0,\sigma_2=\frac{M (3 \gamma -4)}{\sqrt{6} (\gamma -2)},\sigma_4=0,\sigma_5=0$ exists for
\begin{enumerate}\item $0<\gamma <\frac{4}{3},\,-\frac{\sqrt{6} (\gamma -2)}{3 \gamma -4}\leq M\leq \frac{\sqrt{6} (\gamma -2)}{3 \gamma
   -4};$ or \item $\frac{4}{3}<\gamma <2,\,\frac{\sqrt{6} (\gamma -2)}{3 \gamma -4}\leq M\leq -\frac{\sqrt{6} (\gamma
   -2)}{3 \gamma -4}.$\end{enumerate}
The eigenvalues of the linearisation are

$0,\,\lambda_1=-\frac{(3 \gamma -4) \left((3 \gamma -4) M^2-2 \gamma +4\right)}{12 (\gamma -2)},\,\lambda_2=\frac{-M^2 (4-3 \gamma )^2+6 (\gamma
   -2) \gamma +2 M N (3 \gamma -4)}{12 (\gamma -2)},\,\lambda_3=\frac{6 (\gamma -2)^2-M^2 (4-3 \gamma )^2}{12 (\gamma -2)}.$
As before, let us determine conditions on the free parameters for the existence of center, unstable and stable manifolds for $P_3$.
\begin{enumerate}
\item For $\gamma, N$ and $M$ such that $\lambda_{1}\neq 0, \lambda_{2}\neq 0, \lambda_{3}\neq 0$ the center manifold is 1-dimensional and tangent to the $\varphi$-axis. Otherwise the center manifold coud be 2-, or 3-dimensional (it is never 4-dimensional).
\item $P_3$ admits a 1-dimensional center manifold and a 3-dimensional stable manifold for
\begin{enumerate}\item $0<\gamma<\frac{4}{3},\,-\frac{\sqrt{2} \sqrt{\gamma -2}}{\sqrt{3 \gamma -4}}<M<0,\,\ N>\frac{M^2 (4-3 \gamma
   )^2-6 (\gamma -2) \gamma }{2 M (3 \gamma -4)};$ or
\item $0<\gamma<\frac{4}{3},\,0<M<\frac{\sqrt{2} \sqrt{\gamma -2}}{\sqrt{3 \gamma -4}},\,
   N<\frac{M^2 (4-3 \gamma )^2-6 (\gamma -2) \gamma }{2 M (3 \gamma -4)}.$
\end{enumerate}
\item In the cases
\begin{enumerate}
\item $0<\gamma <\frac{4}{3},\,-\frac{\sqrt{6} (\gamma -2)}{3 \gamma -4}<M<-\frac{\sqrt{2} \sqrt{\gamma -2}}{\sqrt{3 \gamma -4}},N<\frac{M^2 (4-3 \gamma
   )^2-6 (\gamma -2) \gamma }{2 M (3 \gamma -4)};$ or
\item $0<\gamma <\frac{4}{3},\,\frac{\sqrt{2} \sqrt{\gamma -2}}{\sqrt{3 \gamma -4}}<M<\frac{\sqrt{6} (\gamma -2)}{3 \gamma -4},N>\frac{M^2 (4-3 \gamma
   )^2-6 (\gamma -2) \gamma }{2 M (3 \gamma -4)};$ or
\item $\frac{4}{3}<\gamma <2,\frac{\sqrt{6} (\gamma -2)}{3 \gamma -4}<M<0,N>\frac{M^2 (4-3 \gamma )^2-6 (\gamma -2) \gamma }{2 M
   (3 \gamma -4)};$ or
\item $\frac{4}{3}<\gamma <2, M=0, N\in\mathbb{R};$ or
\item $\frac{4}{3}<\gamma <2,0<M<-\frac{\sqrt{6} (\gamma -2)}{3 \gamma -4},N<\frac{M^2 (4-3 \gamma )^2-6 (\gamma -2) \gamma }{2 M
   (3 \gamma -4)},$  the unstable manifold is 2-dimensional (hence the stable manifold and the center manifold are both 1-dimensional).
\item Otherwise, $P_3$ has a 1-dimensional unstable manifold. Thus, it is never a local source since its unstable manifold is of dimension less than $3.$

\end{enumerate}

\end{enumerate}
\item The critical point $R_1$ with coordinates $\varphi=0,\sigma_2=0,\sigma_4=0,\sigma_5=1$ always exists. The eigenvalues of the linearisation are

$0,\frac{2}{3},-\frac{1}{3},\frac{4}{3}-\gamma.$ The center manifol is 1-dimensional and tangent to the $\varphi$-axis. The unstable (stable) manifold is 1-dimensional (2-dimensional) if $\gamma>\frac{4}{3}$ otherwise it is 2-dimensional (1-dimensional).
\item The critical point $R_2$ with coordinates $\sigma_2=\frac{\sqrt{\frac{2}{3}}}{M},\sigma_4=0,\sigma_5=\frac{\sqrt{\frac{4-2 \gamma }{M^2}+3 \gamma -4}}{\sqrt{3 \gamma -4}}$ exists for $0<\gamma<\frac{4}{3},\,M^2\geq \frac{2\left(\gamma -2\right)}{3 \gamma -4}.$ The eigenvalues of the linearisation are

$0,-\frac{M+\sqrt{3 M^2 (4 \gamma -5)-8 (\gamma -2)}}{6 M},\frac{\sqrt{3 M^2 (4 \gamma -5)-8 (\gamma -2)}-M}{6
   M},\frac{1}{3} \left(\frac{N}{M}+2\right).$
Let us determine conditions on the free parameters for the existence of center, unstable and stable manifolds for $R_2$.
\begin{enumerate}
\item $R_2$ has a 3-dimensional stable manifold and a 1-dimensional center manifold if
\begin{enumerate}
\item $0<\gamma <\frac{5}{4},-2 \sqrt{\frac{2}{3}} \sqrt{\frac{\gamma -2}{4 \gamma -5}}\leq M<-\sqrt{2} \sqrt{\frac{\gamma -2}{3
   \gamma -4}},N>-2 M;$ or
\item $0<\gamma <\frac{5}{4},\sqrt{2} \sqrt{\frac{\gamma -2}{3 \gamma -4}}<M\leq 2 \sqrt{\frac{2}{3}} \sqrt{\frac{\gamma -2}{4
   \gamma -5}},N<-2 M;$ or
\item $\frac{5}{4}\leq \gamma <\frac{4}{3},M<-\sqrt{2} \sqrt{\frac{\gamma -2}{3 \gamma -4}},N>-2 M;$  or
\item $\frac{5}{4}\leq \gamma <\frac{4}{3},M>\sqrt{2} \sqrt{\frac{\gamma -2}{3 \gamma -4}},N<-2 M;$ or
\item $0<\gamma <\frac{5}{4},M<-2 \sqrt{\frac{2}{3}} \sqrt{\frac{\gamma -2}{4 \gamma -5}}, N>-2 M;$ or
\item $0<\gamma <\frac{5}{4},M>2 \sqrt{\frac{2}{3}} \sqrt{\frac{\gamma -2}{4 \gamma -5}}, N<-2 M.$
\end{enumerate}
\item By reversing the sign of the last inequality, i.e., the inequality solved for $N$, in the previous six cases we obtain conditions for $R_2$ having a 2-dimensional stable manifold, a 1-dimensional unstable manifold and a 1-dimensional center manifold.
\end{enumerate}
\item The critical point $P_4$ with coordinates $\varphi=0, \sigma_2=-\frac{N}{\sqrt{6}},\sigma_4=\sqrt{1-\frac{N^2}{6}}, \sigma_5=0$ exists whenever $N^2<6.$ The eigenvalues of the linearisation are

$0,\frac{1}{6} \left(N^2-6\right),\frac{1}{6} \left(N^2-4\right),\frac{1}{3} N (2 M+N)-\frac{1}{2} (M N+2) \gamma.$ The conditions for the existence of stable, unstable and center manifolds is as follows.
\begin{enumerate}
\item The center manifold is 1-dimensional and the stable manifold is 3-dimensional provided
\begin{enumerate}
\item $N=0,\,M\in\mathbb{R},\gamma\neq \frac{4}{3};$ or
\item $0<\gamma <\frac{4}{3},-2<N<0,M>\frac{2 \left(N^2-3 \gamma \right)}{N (3 \gamma -4)};$ or
\item $0<\gamma <\frac{4}{3},0<N<2,M<\frac{2 \left(N^2-3 \gamma \right)}{N (3 \gamma -4)};$ or
\item $\frac{4}{3}<\gamma <2,-2<N<0,M<\frac{2 \left(N^2-3 \gamma \right)}{N (3 \gamma -4)};$ or
\item $\frac{4}{3}<\gamma <2,0<N<2,M>\frac{2 \left(N^2-3 \gamma \right)}{N (3 \gamma -4)}.$
\end{enumerate}
\item The stable manifold is 2-dimensional, the unstable manifold is 1-dimensional and the center manifold is 1-dimensional provided
\begin{enumerate}
\item $0<\gamma <\frac{4}{3},-\sqrt{6}<N<-2,M>\frac{2 \left(N^2-3 \gamma \right)}{N (3 \gamma -4)};$ or
\item $0<\gamma <\frac{4}{3},-2<N<0,M<\frac{2 \left(N^2-3 \gamma \right)}{N (3 \gamma -4)};$ or
\item $0<\gamma <\frac{4}{3},0<N<2,M>\frac{2 \left(N^2-3 \gamma \right)}{N (3 \gamma -4)};$ or
\item $0<\gamma <\frac{4}{3},2<N<\sqrt{6},M<\frac{2 \left(N^2-3 \gamma \right)}{N (3 \gamma -4)};$ or
\item $\frac{4}{3}<\gamma <2,-\sqrt{6}<N<-2,M<\frac{2 \left(N^2-3 \gamma \right)}{N (3 \gamma -4)};$ or
\item $\frac{4}{3}<\gamma <2,-2<N<0,M>\frac{2 \left(N^2-3 \gamma \right)}{N (3 \gamma -4)};$ or
\item $\frac{4}{3}<\gamma <2,0<N<2,M<\frac{2 \left(N^2-3 \gamma \right)}{N (3 \gamma -4)};$ or
\item $\frac{4}{3}<\gamma <2,2<N<\sqrt{6},M>\frac{2 \left(N^2-3 \gamma \right)}{N (3 \gamma -4)}.$
\end{enumerate}
\item The stable manifold is 1-dimensional, the unstable manifold is 2-dimensional and the center manifold is 1-dimensional provided
\begin{enumerate}
\item $0<\gamma <\frac{4}{3},-\sqrt{6}<N<-2,M<\frac{2 \left(N^2-3 \gamma \right)}{N (3 \gamma -4)};$ or
\item $0<\gamma <\frac{4}{3},2<N<\sqrt{6},M>\frac{2 \left(N^2-3 \gamma \right)}{N (3 \gamma -4)};$ or
\item $\frac{4}{3}<\gamma <2,-\sqrt{6}<N<-2,M>\frac{2 \left(N^2-3 \gamma \right)}{N (3 \gamma -4)};$ or
\item $\frac{4}{3}<\gamma <2,2<N<\sqrt{6},M<\frac{2 \left(N^2-3 \gamma \right)}{N (3 \gamma -4)}.$
\end{enumerate}
\end{enumerate}
\item The critical point $R_3$ with coordinates $\varphi=0,\sigma_2=-\frac{2 \sqrt{\frac{2}{3}}}{N},\sigma_4=\frac{2}{\sqrt{3} |N|},\sigma_5=\frac{\sqrt{N^2-4}}{|N|}$ exists for $N^2\geq 4.$
The eigenvalues of the linearisation are

$0,\frac{1}{6} \left(-\frac{\sqrt{64 N^2-15 N^4}}{N^2}-1\right),\frac{1}{6} \left(\frac{\sqrt{64 N^2-15
   N^4}}{N^2}-1\right),-\frac{(2 M+N) (3 \gamma -4)}{3 N}.$
The conditions for the existence of stable, unstable and center manifolds are as follows.
\begin{enumerate}
\item The stable manifold is 3-dimensional and the center manifold is 1-dimensional provided
\begin{enumerate}
\item $0<\gamma <\frac{4}{3},N<-\frac{8}{\sqrt{15}}, M>-\frac{N}{2};$ or
\item $0<\gamma <\frac{4}{3},-\frac{8}{\sqrt{15}}\leq N<-2,M>-\frac{N}{2};$ or
\item $0<\gamma <\frac{4}{3},2<N\leq \frac{8}{\sqrt{15}},M<-\frac{N}{2};$ or
\item $0<\gamma <\frac{4}{3},N>\frac{8}{\sqrt{15}},M<-\frac{N}{2};$ or
\item $\frac{4}{3}<\gamma <2,N<-\frac{8}{\sqrt{15}},M<-\frac{N}{2};$ or
\item $\frac{4}{3}<\gamma <2,-\frac{8}{\sqrt{15}}\leq N<-2,M<-\frac{N}{2};$ or
\item $\frac{4}{3}<\gamma <2,2<N\leq \frac{8}{\sqrt{15}},M>-\frac{N}{2};$ or
\item $\frac{4}{3}<\gamma <2,N>\frac{8}{\sqrt{15}},M>-\frac{N}{2}.$
\end{enumerate}
\item By reversing the sign of the last inequality, i.e., the inequality solved for $M$, in the previous eight cases we obtain conditions for $R_3$ having a 2-dimensional stable manifold, a 1-dimensional unstable manifold and a 1-dimensional center manifold.
\end{enumerate}
\item The critical point $P_5$ with coordinates

$\varphi=0,\sigma_2=\frac{\sqrt{6} \gamma }{M (3 \gamma -4)-2 N},$ $\sigma_4=\frac{\sqrt{M^2 (4-3 \gamma )^2+M N (8-6 \gamma )-6 (\gamma -2) \gamma }}{2 N+M (4-3 \gamma )},\sigma_5=0$ exists for $2 (2 M+N)>3 M \gamma,\, M (3 \gamma -4) (M (3 \gamma -4)-2 N)\geq 6 (\gamma -2) \gamma,$
and
$\frac{3 (M N+2) \gamma -2 N (2
   M+N)}{(2 N+M (4-3 \gamma ))^2}\leq 0.$ The eigenvalues of the linearisation are

$0,\frac{12 M+6 N-3 (3 M+N) \gamma +\sqrt{3} \sqrt{f(\gamma ,M,N)}}{6 (M (3 \gamma -4)-2 N)},\frac{3 N (\gamma -2)+3 M (3
   \gamma -4)+\sqrt{3} \sqrt{f(\gamma ,M,N)}}{6 (2 N+M (4-3 \gamma ))},\frac{(2 M+N) (3 \gamma -4)}{6 N+3 M (4-3 \gamma
   )},$
   where $f(\gamma ,M,N)=2 M^3 N (3 \gamma -4)^3+2 M N \left(4 N^2-6 \gamma ^2+3 \gamma -6\right) (3 \gamma -4)-M^2 \left(8 N^2-12 \gamma
   -3\right) (4-3 \gamma )^2+3 (\gamma -2) \left(N^2 (9 \gamma -2)-24 \gamma ^2\right).$ The stability conditions of $P_5$ are very complicated to display them here. Thus we must rely on numerical experimentation. We can obtain, however, some analytic results. For instance, there exists at least a 1-dimensional center manifold. The unstable manifold is always of dimension lower than 3. Thus the critical point is never a local source. If all the eigenvalues, apart form the zero one, have negative reals parts, then the center manifold of $P_5$ acts as a local sink. This means that the orbits in the stable manifold approach the center manifold of $P_5$ when the time goes forward.
\item The critical point $P_6$ with coordinates

$\varphi=0,\sigma_2=\frac{\sqrt{6} \gamma }{M (3 \gamma -4)-2 N},$ $\sigma_4=-\frac{\sqrt{M^2 (4-3 \gamma )^2+M N (8-6 \gamma )-6 (\gamma -2) \gamma }}{2 N+M (4-3 \gamma )},\sigma_5=0$ exists for $M (3 \gamma -4) (M (3 \gamma -4)-2 N)\geq 6 (\gamma -2) \gamma ,\,2 (2 M+N)<3 M \gamma,$
and
$\frac{3 (M N+2) \gamma -2 N (2
   M+N)}{(2 N+M (4-3 \gamma ))^2}\leq 0.$ The eigenvalues of the linearisation are the same displayed in the previous point. However the stability conditions are rather different (since the existence conditions are different from those of $P_5$). As before, the stability conditions are very complicated to display them here, but similar conclusions concerning the center and unstable manifold, as for $P_5,$ are obtained. For get further information about its stability we must to resort to numerical experimentation.
\end{enumerate}

\paragraph{Physical description of the solutions and connection with observables.}

Let us now present the formalism of obtaining the physical description of a critical point, and also connect with the basic observables relevant for a physical discussion. These will allow us to describe the cosmological behavior of each critical point, in the next section.

Firstly, around a critical point we obtain first-order expansions for $H,a,\phi,$ and $\rho$ and $\rho_r$ in terms of $t$, considering equations: \eqref{Raych}; the definition of the scale factor $a$ in terms of the Hubble factor $H$; the definition of $\sigma_2;$  the matter conservation equations \eqref{consm} and
\eqref{consr}, respectively, given by
\begin{align}
&& 2 \dot H(t)=H(t)^2 \left(3 (\gamma -2) {\sigma_2^\star}^2+3 \gamma  \left({\sigma_4^\star}^2+{\sigma_5^\star}^2-1\right)-4
   {\sigma_5^\star}^2\right),\nonumber\\&& \dot a(t)=a(t) H(t),\nonumber\\&& \dot\phi(t)=\sqrt{6} {\sigma_2^\star} H(t),\nonumber\\&& \dot\rho(t)=-\frac{3}{2} H(t)^3
   \left(\sqrt{6} M (3 \gamma -4) {\sigma_2^\star}-6 \gamma \right) \left({\sigma_2^\star}^2+{\sigma_4^\star}^2+{\sigma_5^\star}^2-1\right),\nonumber\\&& \dot\rho_r(t)=-12 {\sigma_5^\star}^2 H(t)^3,\label{APPROX}
\end{align}
where the star-upperscript denotes the evaluation at a specific  critical point.
The equation \be \ddot \phi(t)= \frac{3}{2} H(t)^2 \left(M (3 \gamma -4) \left({\sigma_2^\star}^2+{\sigma_4^\star}^2+{\sigma_5^\star}^2-1\right)-2 \left(N {\sigma_4^\star}^2+\sqrt{6} {\sigma_2^\star}\right)\right),\label{consistency}\ee derived from the equation of motion for the scalar field \eqref{KG} should be used as a consistency test for the above procedure.  Solving the differential equations \eqref{APPROX} and substituting the resulting expressions in the equation \eqref{consistency} results in
\begin{align}-6 M (3 \gamma -4) \left({\sigma_2^\star}^2+{\sigma_4^\star}^2+{\sigma_5^\star}^2-1\right)+12 N {\sigma_4^\star}^2+\nonumber\\+2 \sqrt{6}
   {\sigma_2^\star} \left(3 \gamma  \left({\sigma_2^\star}^2+{\sigma_4^\star}^2+{\sigma_5^\star}^2-1\right)-6 {\sigma_2^\star}^2-4 {\sigma_5^\star}^2+6\right)=0.\end{align}

This integrability condition should be (at least asymptotically) fulfilled.

\begin{table*}
\caption{ \label{tab2b} Observable cosmological quantities, and
physical behavior of the solutions, at the critical points of the
cosmological system. We use the notations $M_1(\gamma)=\frac{\sqrt{2 \gamma  (3 \gamma -8)+8}}{4-3 \gamma },$ $M_2(\gamma)=\frac{\sqrt{6} \sqrt{(\gamma -3) \gamma +2}}{4-3 \gamma }$
 }    \centering
\begin{tabular}{cccc}
  \hline   \hline
  \ \ Cr.P.  \ \  &\  $q$ \ & $w_{\text{eff}}$   & Solution/description\\
  \hline \hline
$P_1$ & 2 & 1 &  Decelerating. \\
   \hline\\
$P_2$ & 2 & 1 & Decelerating. \\
     \hline\\
$P_3$ & $\frac{-M^2 (4-3 \gamma )^2+2 \gamma  (3 \gamma -8)+8}{4 (\gamma -2)}$ & $-\frac{M^2 (4-3 \gamma
   )^2}{6 (\gamma -2)}+\gamma -1$ &  Accelerating for   \\
   \vspace{0.2cm}
   &  & &   $0<\gamma <\frac{2}{3},\, -M_1(\gamma)<M<M_1(\gamma)$ \\ \hline \\
$P_4$ & $\frac{1}{2} \left(N^2-2\right)$& $\frac{1}{3}
   \left(N^2-3\right)$ &  Accelerating for   \\
   \vspace{0.2cm}
   &  & &   $-\sqrt{2}<n<\sqrt{2}$ \\
   \vspace{0.2cm}
   &  & &   powerlaw-inflationary \\ \hline \\
$P_5$ & $\frac{3 (M+N) \gamma -2 (2 M+N)}{2 N+M (4-3 \gamma )}$ & $\frac{M
   (4-3 \gamma )-2 N (\gamma -1)}{M (3 \gamma -4)-2 N}$ &  Accelerating for   \\
   \vspace{0.2cm}
   &  & & $\frac{3 (M+N) \gamma -2 (2 M+N)}{2 N+M (4-3 \gamma )}<0$   \\
   \vspace{0.2cm}
   &  & &  matter-kinetic-potential scaling   \\ \hline \\
$P_6$ & $\frac{3 (M+N) \gamma -2 (2 M+N)}{2 N+M (4-3 \gamma )}$ & $\frac{M
   (4-3 \gamma )-2 N (\gamma -1)}{M (3 \gamma -4)-2 N}$ &  Accelerating for   \\
   \vspace{0.2cm}
   &  & & $\frac{3 (M+N) \gamma -2 (2 M+N)}{2 N+M (4-3 \gamma )}<0$   \\
   \vspace{0.2cm}
   &  & &  matter-kinetic-potential scaling   \\ \hline \\
   $R_1$ & 1 & $\frac{1}{3}$ & Decelerating. Radiation-dominated. \\
     \hline\\
$R_2$ & 1 & $\frac{1}{3}$ & Decelerating. \\
   \vspace{0.2cm}
   &  & &  radiation-kinetic-potential \\
     \hline\\
$R_3$ & 1 & $\frac{1}{3}$ & Decelerating. \\
   \vspace{0.2cm}
   &  & &   radiation-kinetic-potential scaling. \\
     \hline\\
\end{tabular}

\end{table*}

Instead of apply this procedure to a generic fixed point here, we submit the reader to section \ref{applications} for some worked examples where this procedure has been applied. However we will discuss on some cosmological observables.

We can calculate the deceleration parameter $q$ defined as usual as \cite{WE}
\begin{equation}
\label{qq}q=-\frac{a \ddot a}{a^2}.
\end{equation}
Additionally, we can calculate the effective (total) equation-of-state parameter of the universe $w_{\text{eff}}$, defined conventionally as
\begin{equation}
\label{weff}
w_{\text{eff}}\equiv\frac{p_{\text{tot}}}{\rho_{\text{tot}}},
\end{equation}
where $p_{\text{tot}}$ and $\rho_{\text{tot}}$ are respectively the
total isotropic pressure and the total energy density. Therefore, in terms of the auxiliary variables we have
\begin{eqnarray}
\label{qq2}
q&=& -\frac{3}{2} (\gamma -2) \sigma_2^2-\frac{3 \gamma  \sigma_4^2}{2}+\frac{1}{2} (4-3
   \gamma ) \sigma_5^2+\frac{1}{2} (3 \gamma -2)\\
w_{\text{eff}}&=&(2-\gamma ) \sigma_2^2-\gamma  \sigma_4^2+\frac{1}{3} (4-3 \gamma ) \sigma_5^2+\gamma -1. \label{weff2}
\end{eqnarray}

First of all, for each critical point described in the last section we calculate the effective (total) equation-of-state parameter of the universe $w_{\text{eff}}$ using \eqref{weff2}, and the deceleration parameter $q$ using \eqref{qq2}. The results are presented in Table \ref{tab2b}. Furthermore, as usual, for an expanding universe $q<0$ corresponds to accelerating expansion and $q>0$ to decelerating expansion.

\subsubsection{The flow as $\phi\rightarrow-\infty$}\label{minusinfinity}

With the purpose of complementing the global analysis of the system it is necessary investigate its behavior as $\phi\rightarrow-\infty.$ It is an easy task since the system \eqref{eq0phi}-\eqref{eq0x4} is invariant under the transformation of coordinates

$$(\phi, \sigma_2)\rightarrow -(\phi, \sigma_2),\; V\rightarrow U,\; \chi\rightarrow \Xi,$$ where $U(\phi)=V(-\phi)$ and
$\Xi(\phi)=\chi(-\phi).$ Hence, for a particular potential $V,$
and a particular coupling function $\chi$, the behavior of the
solutions of the equations \eqref{eq0phi}-\eqref{eq0x4} around $\phi=-\infty$ is equivalent
(except for the sign of $\phi$) to the behavior of the system near
$\phi=\infty$ with potential and coupling functions $U$ and $\Xi,$
respectively.

If $U$ and $\Xi$ are of class $\mathcal{E}^2_+,$ the preceding  analysis in $\bar{\Sigma}_\epsilon$ can be applied
(with and adequate choice of $\epsilon$).

The set of
class $C^k$ functions well behaved in both $+\infty$ and
$-\infty$ is denoted by $\mathcal{E}^k.$ Latin uppercase letters with subscripts
$+\infty$ and $-\infty,$ are used respectively to indicate the exponential
order of $\mathcal{E}^k$ functions in $+\infty$ and in $-\infty.$

\section{Examples}\label{applications}

In this section we apply the mathematics discussed in previous sections to several worked examples from both analytical and numerical viewpoint.

\subsection{Numerical evidence of the result of theorem \ref{thm4}}\label{application1}

\begin{figure}[ht]
\begin{center}
\mbox{\epsfig{figure=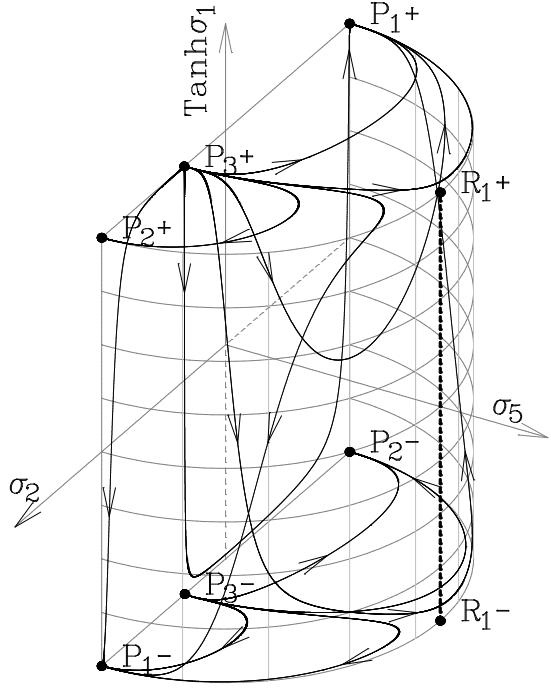,width=9.9cm,angle=0}}
\caption{ \label{fig1}{\it Qualitative dynamics of the flow of \eqref{systthm4} for the choice $M=\sqrt{2/3}$ and  $\gamma=1.$ Observe that $P_1^+$ or $P_1^-$ are the local attractors (sinks). $P_3^+$ is a local source, $P_3^-$ is a saddle for the full dynamics, but it is a local source in the invariant set $\tanh \sigma_1=-1.$ The thick dashed line is an invariant set which is unstable. In fact all its points including $R_1^+$ and $R_1^-$, act as saddle points. They correspond to cosmological radiation-dominated solutions.}}
\end{center}
\end{figure}

For the particular case $\chi(\sigma_1)=e^{M \sigma_1},$ from equation \eqref{proxi2}, follows that $\sigma_2=\sigma_{2c}:=\frac{M(4-3\gamma)}{\sqrt{6}(2-\gamma)}$ is an invariant set. Given $\frac{4}{3}<\gamma <2,$ the existence conditions lead to $M_1(\gamma)\leq M\leq -\frac{\sqrt{6} (\gamma
   -2)}{3 \gamma -4},$ where $M_1(\gamma)=\frac{\sqrt{2 \gamma  (3 \gamma -8)+8}}{4-3 \gamma }.$ For such values $\frac{\partial f_2}{\partial \sigma_2}|_{\sigma_{2c}}=\frac{M^2 (4-3 \gamma )^2-6 (\gamma -2)^2}{12 (\gamma -2)}\geq 0.$ Thus the asymptotic phase configuration $\sigma_1\rightarrow -\text{sgn}{\sigma_{2c}}\infty, \sigma_2\rightarrow \sigma_{2c}$ is never approached (for an open set of orbits) as $\tau\rightarrow \infty.$ For the original system (i.e., taking the time reversal transformation) this means that this asymptotic phase configuration is never approached towards the past.
In figure \ref{fig1} we show the qualitative dynamics of the flow of \eqref{systthm4} for the choice $M=\sqrt{2/3},$ and $\gamma=1.$ In order to compactify the phase space we have introduce the coordinate transformation $\sigma_1\rightarrow \tanh \sigma_1.$ The mentioned asymptotic configuration is represented in figure \ref{fig1} by $P_3^\pm.$

\begin{figure}[ht]
\begin{center}
\mbox{\epsfig{figure=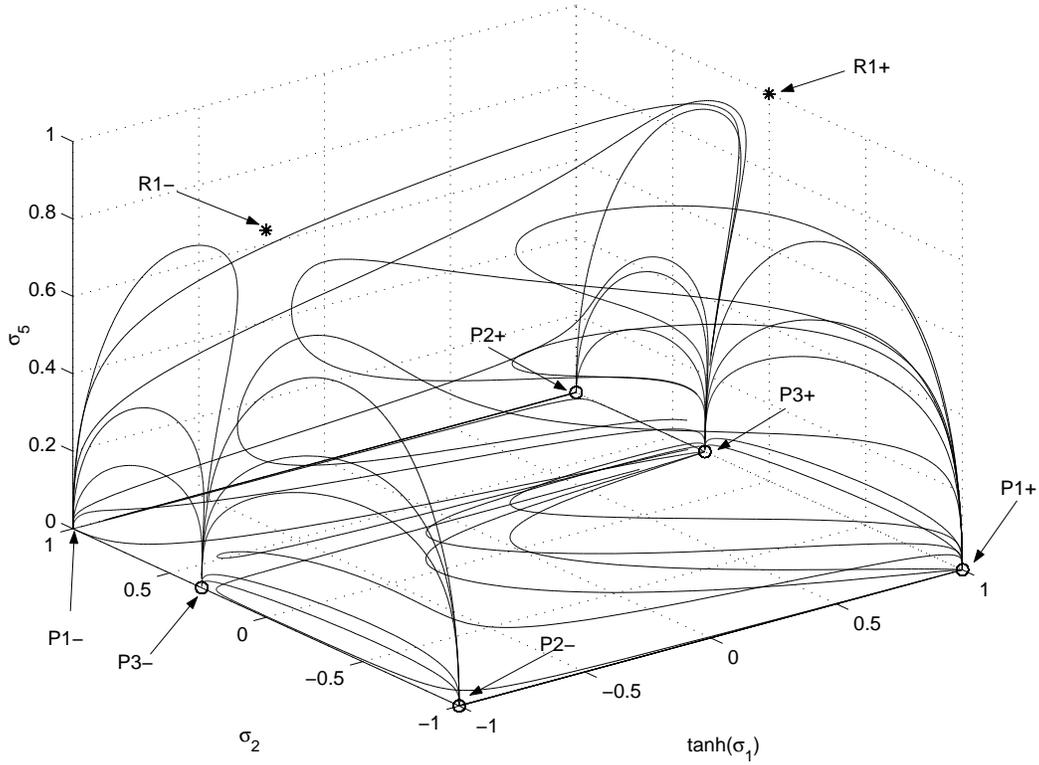,width=14cm,angle=0}}
\caption{ \label{fig1c}{\it The graphic illustrates the result of theorem \ref{thm4}. We set $M=\sqrt{2/3}$ and  $\gamma=1.$ The point with $\tanh \sigma_1=\pm 1$ are the local sinks (thus for the original system \eqref{eq0phi}-\eqref{eq0x4} the scalar field almost always diverges towards the past).}}
\end{center}
\end{figure}

\subsection{Quadratic gravity: $F(R)=R+\alpha R^2$.}

\begin{table}[!hb]
\begin{center}
\caption{\label{crit0} Location of the critical points of the flow
of \eqref{eqvphi}-\eqref{eqzrad} defined in the invariant set
$\left\{p\in\Omega_\epsilon: \varphi=0\right\}$ for $M=\sqrt{2/3}$ and
$N=0.$}
\begin{tabular}[t]{|l|c|c|c|c|}
\hline
Label&$(\sigma_2,\sigma_4,\sigma_5)$&Existence&Stability$^{\rm a}$\\[1ex]
\hline
\hline &&& \\[-2ex]
$P_1$&$(-1,0,0)$&always & unstable for $\left\{\begin{array}{c} 0<\gamma <\frac{4}{3},\, \text{or}\\ \frac{4}{3}<\gamma <\frac{5}{3}\end{array}\right.$ \\[1ex]
     &          &       & saddle, otherwise\\[1ex]
\hline
\hline &&& \\[-2ex]
$P_2$&$(1,0,0)$&always & unstable\\[1ex]
\hline
\hline &&& \\[-2ex]
$P_3$&$\left(1-\frac{2}{3 (2-\gamma)},0,0\right)$& $\left\{\begin{array}{c} 0<\gamma <\frac{4}{3},\, \text{or}\\ \frac{4}{3}<\gamma <\frac{5}{3}\end{array}\right.$ & saddle\\[1ex]
\hline
\hline &&& \\[-2ex]
$R_1$&$(0,0,1)$&always & saddle\\[1ex]
\hline
\hline &&& \\[-2ex]
$P_4$&$\left(0,1,0\right)$& always & stable \\[1ex]
\hline
\end{tabular}
\end{center}
$^{\rm a}$ The stability is analysed for the flow restricted to the invariant set $\varphi=0$.
\end{table}

Quadratic gravity, $F(R)=R+\alpha R^2$, is equivalent to a
non-minimally coupled scalar field with the potential \be V\left(
\phi\right) =\frac{1}{8\alpha}\left( 1-e^{-\sqrt{2/3}\phi}\right)
^{2}\label{pot1}\ee and coupling function
\be\chi(\phi)=e^{\sqrt{\frac{2}{3}}\phi}.\label{coup1}\ee

Observe first that \be W_\chi(\phi)=\chi'(\phi)/\chi(\phi)-\sqrt{2/3}=0 \Rightarrow
\lim_{\phi\rightarrow +\infty} W_\chi(\phi)=0\ee and  \be W_V(\phi)= V'(\phi)/V(\phi)=-\left({\sqrt{{8}/{3}}}\right)/{\left(1-e^{\sqrt{\frac{2}{3}}\phi}\right)} \Rightarrow \lim_{\phi\rightarrow +\infty} W_V(\phi)=0.\ee  In other words, the coupling function \eqref{coup1} and the potential \eqref{pot1} are WBI of exponential orders $M=\sqrt{2/3}$ and $N=0,$ respectively.

It is easy to prove that the coupling function \eqref{coup1} and the potential \eqref{pot1} are at least  ${\cal E}^2_+,$  under the
admissible coordinate transformation \be \varphi=\phi^{-1}=f(\phi)\label{transformQG}.\ee

Using the coordinate transformation \eqref{transformQG} we find

\be\overline{W}_{\chi}(\varphi)=0.\label{WchiQG}\ee

\be\overline{W}_V(\varphi)=\left\{\begin{array}{rcr} -\frac{2 \sqrt{\frac{2}{3}}}{1-e^{{\sqrt{\frac{2}{3}}}/{\varphi }}}&,&\varphi>0\\
                                     0 &,&\varphi=0 \end{array}\right.\label{WVQG}\ee
and

\be\overline{f'}(\varphi)=\left\{\begin{array}{rcr} -\varphi^2&,&\varphi>0\\
                                     0&,&\varphi=0 \end{array}\right.\label{fQG}\ee

In this example, the evolution equations for $\varphi,$ $\sigma_2,$ $\sigma_4,$ and $\sigma_5$ are given by the equations
\eqref{eqvphi}-\eqref{eqzrad} with $M=\sqrt{2/3}$ and $N=0,$ and $\overline{W}_{\chi},$ $\overline{W}_V,$ and $\overline{f'},$
given respectively by (\ref{WchiQG}), (\ref{WVQG}) and (\ref{fQG}). The state space is defined by
$$\Omega_\epsilon=\{(\varphi, \sigma_2,
\sigma_4,\sigma_5)\in\mathbb{R}^4: 0\leq\varphi\leq
 \epsilon, \sigma_2^2+\sigma_4^2+\sigma_5^2\leq 1,
\sigma_4\geq 0, \sigma_5\geq 0\}.$$
Let us analyse the local stability of the critical points of the corresponding system. In the above analysis we are not taking into account perturbations in the $\varphi$-axis. It is obvious, from the previous analysis, that the center manifold of these critical points contains the $\varphi$-axis as a proper eigendirection.
In table \ref{crit0} are summarised the location, existence conditions and stability \footnote{The stability is analysed for the flow restricted to the invariant set $\varphi=0$.} of the critical points.

\begin{figure}[ht]
\begin{center}
\mbox{\epsfig{figure=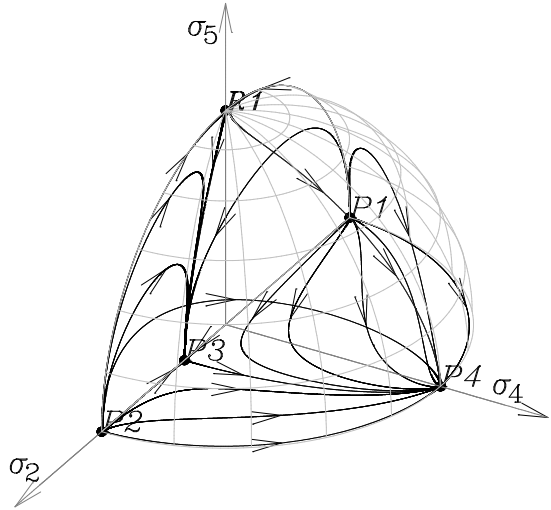,width=9.9cm,angle=0}}
\caption{ \label{fig2}{\it Proyection of some orbits of \eqref{eqvphi}-\eqref{eqzrad} in the invariant set $\varphi=0$ for the coupling function \eqref{coup1} and the potential \eqref{pot1} for $\gamma=1.$ Observe that $P_1$ and $P_2$ are local sources, $R_1,$ and $P_3$ are saddles ($P_3$ is the local attractor in the invariant set $y=0$) and $P_4$ (the de Sitter solution) is the local sink in the invariant set $\varphi=0.$}}

\end{center}
\end{figure}

Let us discuss the stability properties of the critical points displayed in table \ref{crit0}.

The critical point $P_1$ always exists. Its unstable manifold is $3D$ provided $0<\gamma<\frac{4}{3}$ or $\frac{4}{3}<\gamma<\frac{5}{3}.$ Otherwise its unstable manifold is lower dimensional.

The critical point $P_2$ always exists. It has a 3D unstable manifold. Although non-hyperbolic our numerical experiments suggest that it is a local source.

The critical point $P_3$ exists for $0<\gamma<\frac{4}{3}$ or $\frac{4}{3}<\gamma<\frac{5}{3}$ and it is neither a sink nor a local source.

The critical point $R_1$ always exists and it is neither a sink nor a local source.

The critical point $P_4$ (corresponding to the \emph{de Sitter} solution) always exists. Its stable manifold is 3D. Since $P_4$ is nonhyperbolic the linear stability analysis is not conclusive. Thus we need to resort to numerical experimentation or alternatively we can use more sophisticated techniques such as normal forms expansion or center manifold theorem. Due its relevance, the full stability analysis of $P_4$ is deserved to section
\ref{stabilityP4}

Let us discuss some physical properties of the cosmological solutions associated to the critical points displayed in table \ref{crit0}.
\begin{itemize}
\item  $P_{1,2}$  represent kinetic-dominated cosmological solutions. They behave as stiff-like matter. The associated cosmological solution satisfies $H=\frac{1}{3 t-c_1},a=\sqrt[3]{3 t-c_1} c_2,\phi =c_3\pm\sqrt{\frac{2}{3}} \ln \left(3
   t-c_1\right),$ where $c_j,\,j=1,2,3$ are integration constants. These solutions are associated with the local past attractors of the systems for an open set of values of the parameter $\gamma.$
\item  $P_3$ represents matter-kinetic scaling cosmological solutions such that
$H=\frac{3 (\gamma -2)}{t (3 \gamma -8)-3 (\gamma -2) c_1},$ $a= \left(t (3 \gamma -8)-3 (\gamma -2)
   c_1\right){}^{1+\frac{2}{3 \gamma -8}} c_2,\rho =\frac{60-36 \gamma }{\left(t (3 \gamma -8)-3 (\gamma -2)
   c_1\right){}^2}+c_3,$ and $\phi = c_4+\frac{\sqrt{6} (3 \gamma -4) \ln \left(t (3 \gamma -8)-3 (\gamma -2) c_1\right)}{3
   \gamma -8}$ where $c_j,\,j=1,2,3,4$ are integration constants.
\item $R_1$ represents a radiation-dominated cosmological solutions satisfying $H=\frac{1}{2 t-c_1},a=\sqrt{2 t-c_1} c_2,\rho_r=\frac{3}{\left(2 t-c_1\right){}^2}+c_3.$
\item $P_4$ represents a de Sitter solution with $H=\sqrt{\frac{V_0}{3}}, a=c_1 \exp\left[\sqrt{\frac{V_0}{3}} t\right], V(\phi)=V_0.$
\end{itemize}

In the figure \ref{fig2} are are displayed typical orbits of \eqref{eqvphi}-\eqref{eqzrad} in the invariant set $\varphi=0.$ The critical points $P_1$ and $P_2$ are local sources, $R_1,$ and $P_3$ are saddles ($P_3$ is the local attractor in the invariant set $y=0$) and $P_4$ (the de Sitter solution) is the local attractor in the invariant set $\varphi=0$. However, concerning the full dynamics, it is locally asymptotically ustable as we prove in next section by explicit calculation of the center manifold at $P_4.$

\subsubsection{Stability analysis of the \emph{de Sitter} solution in quadratic gravity}\label{stabilityP4}

In order to analyze the stability of \emph{de Sitter} solution we can use center manifold theorem.
Let us proceed as follows. First, in order to remove the trascendental function in $\overline{W}_V,$ let us introduce the new variable $$u=\frac{1}{1-\exp\left[{{\sqrt{\frac{2}{3}}}/{\varphi}}\right]},$$ taking values in the range $$\frac{1}{1-\exp\left[{{\sqrt{\frac{2}{3}}}/{\epsilon}}\right]}\leq u\leq 0.$$
In this way we obtain the new system of ordinary differential equations

\begin{align}
&u'=\frac{2 u^2 \sigma_2}{3}-\frac{2 u \sigma_2}{3},\nonumber\\
&\sigma_2'=   \frac{2 u \sigma_4^2}{3}+\left(1-\frac{\gamma }{2}\right)
   \sigma_2^3+\frac{1}{6} (3 \gamma -4) \sigma_2^2+\sigma_2 \left(-\frac{\gamma  \sigma_4^2}{2}+\frac{1}{6} (4-3 \gamma ) \sigma_5^2+\frac{\gamma
   -2}{2}\right)+\nonumber\\&+\frac{1}{6} (3 \gamma -4) \sigma_4^2+\frac{1}{6} (3
   \gamma -4) \sigma_5^2+\frac{1}{6} (4-3 \gamma ),\nonumber\\
&\sigma_4'= -\frac{2 u
   \sigma_2 \sigma_4}{3}+\left(1-\frac{\gamma }{2}\right)
   \sigma_2^2 \sigma_4-\frac{\gamma  \sigma_4^3}{2}+\sigma_4 \left(\frac{1}{6} (4-3 \gamma ) \sigma_5^2+\frac{\gamma }{2}\right),\nonumber\\
&\sigma_5'= \left(1-\frac{\gamma }{2}\right)
  \sigma_2^2 \sigma_5-\frac{1}{2} \gamma  \sigma_4^2 \sigma_5+\frac{1}{6} (4-3 \gamma ) \sigma_5^3+\frac{1}{6} (3 \gamma -4) \sigma_5\label{quadraticsyst}
\end{align} describing the dynamics of quadratic gravity as $\phi\rightarrow +\infty.$

\begin{figure}[ht]
\begin{center}
\mbox{\epsfig{figure=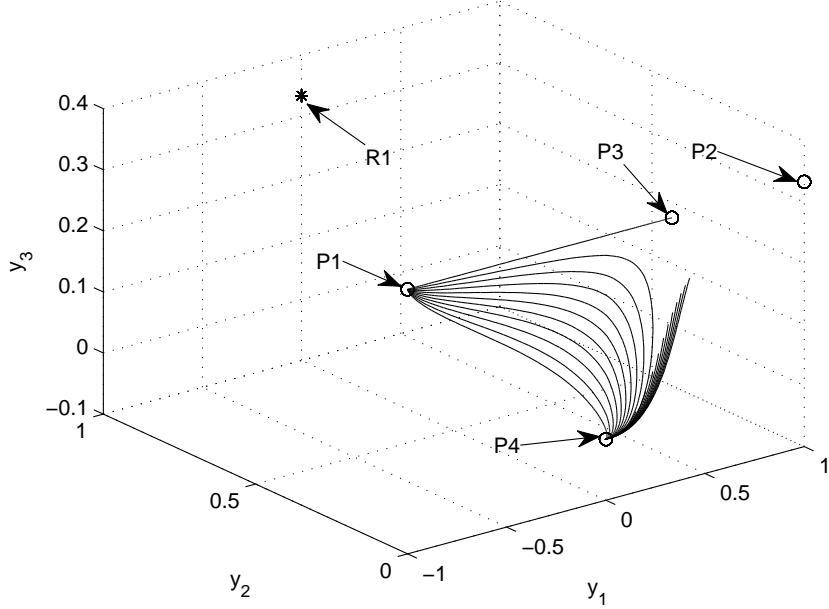,width=12cm,angle=0}}
\caption{ \label{fig2a}{\it Proyection of some orbits of \eqref{center2} in the space $y_1,y_2,y_3$ for the coupling function \eqref{coup1} and the potential \eqref{pot1} for $\gamma=1.$ The graphic shows the behavior in the stable manifold of $P_4.$ The bulk of orbits in front of and at the right hand side of the figure represents a projection of the center(s) manifold(s).}}

\end{center}
\end{figure}

\begin{figure}[ht]
\begin{center}
\mbox{\epsfig{figure=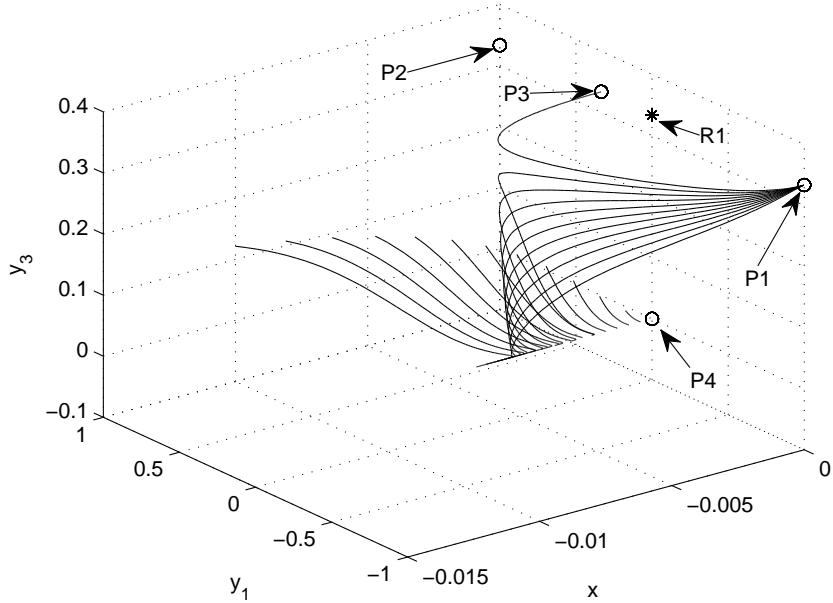,width=12cm,angle=0}}
\caption{ \label{fig2b}{\it  Proyection of some orbits of \eqref{center2} in the space $x,y_1,y_3$ for the coupling function \eqref{coup1} and the potential \eqref{pot1} for $\gamma=1.$ The graphic shows the unstable character of $P_4$ (trajectories starting at $x<0$ move away from the origin).}}

\end{center}
\end{figure}

\begin{figure}[ht]
\begin{center}
\mbox{\epsfig{figure=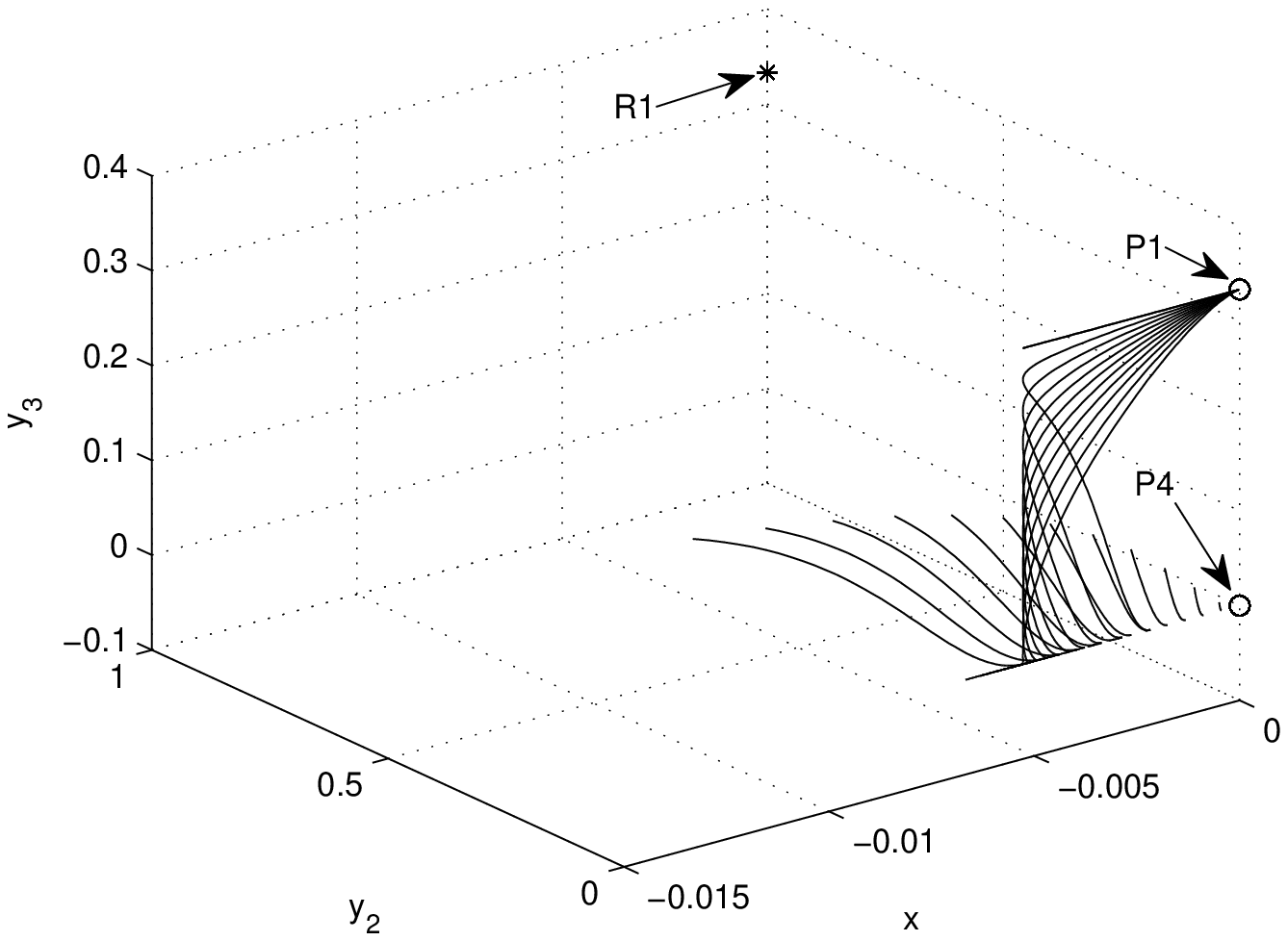,width=12cm,angle=0}}
\caption{ \label{fig2c}{\it  Proyection of some orbits of \eqref{center2} in the space $x,y_2,y_3$ for the coupling function \eqref{coup1} and the potential \eqref{pot1} for $\gamma=1.$ The graphic shows the unstable character of $P_4$ (the orbits depart from the origin for $x<0$).}}
\end{center}
\end{figure}

\begin{prop}\label{centerP4} The equilibrium point $q:=(u,\sigma_2,\sigma_4,\sigma_5)=(0,0,1,0)$ of the system \eqref{quadraticsyst} is locally asymptotically unstable.
\end{prop}
In order to determine the local center manifold of \eqref{quadraticsyst} at $q$ we have to transform the system into a form suitable for the application of the center manifold theorem (see section \ref{sectionCM} for a summary of the techniques involved in the proof).

{\bf Proof.}

{\bf Case $\gamma\neq 1$}

Let be $\gamma\neq 1.$
The Jacobian of \eqref{quadraticsyst} at $q=(0,0,1,0)$ has eigenvalues $0,-1,-\frac{2}{3},$ and $-\gamma$ with
corresponding eigenvectors $(0,0,0,1)^T,\left(0,0,\frac{3}{2},0\right)^T, \left(1,0,1,\frac{1}{3 (\gamma -1)}-1\right)^T,$ and $(0,1,0,0)^T.$ We shift the fixed point to the origin by setting $\hat{\sigma_4}=\sigma_4-1.$ In order to transform the linear part of the vector field into Jordan canonical form, we define new variables $(x,y_1,y_2,y_3)\equiv\mathbf{x}$, by the equations
\begin{align}& u=\frac{3 x}{2},\nonumber\\&\sigma_2=x+y_1+y_3 \left(\frac{1}{3 (\gamma
   -1)}-1\right),\nonumber\\&\hat{\sigma}_4=y_3,\nonumber\\&\sigma_5=y_2,\nonumber\end{align} so that

\be
\left(\begin{array}{c}x'\\y_1'\\y_2'\\y_3'
\end{array}\right)=\left(\begin{array}{cccc}0& 0 &0 &0\\0& -1 &0 &0\\0& 0 &-\frac{2}{3}&0\\0& 0 &0 &-\gamma
\end{array}\right)\left(\begin{array}{c}x\\y_1\\y_2\\y_3
\end{array}\right)+\left(\begin{array}{c}f(x,y_1,y_2,y_3)\\g_1(x,y_1,y_2,y_3)\\g_2(x,y_1,y_2,y_3)\\g_3(x,y_1,y_2,y_3)\end{array}\right)\label{center2}
\ee
where

$f(x,y_1,y_2,y_3)=x^3+x^2 y_1+\frac{x^2 y_3}{3 (\gamma -1)}-x^2 y_3-\frac{2 x^2}{3}-\frac{2 x y_1}{3}-\frac{2 x y_3}{9 (\gamma
   -1)}+\frac{2 x y_3}{3},$
\\\\
$g_1(x,y_1,y_2,y_3)=-\frac{x^3 \gamma }{2}-\frac{3}{2} x^2 y_1 \gamma +2 x^2 y_1+x^2 y_3 \gamma +\frac{x^2 y_3}{3 (\gamma -1)}-\frac{7 x^2
   y_3}{3}+\frac{x^2 \gamma }{6 (\gamma -1)}-\frac{3}{2} x y_1^2 \gamma +3 x y_1^2+2 x y_1 y_3 \gamma +\frac{x
   y_1 y_3}{\gamma -1}-\frac{17 x y_1 y_3}{3}+\frac{2 x y_1}{3}-\frac{1}{2} x y_2^2 \gamma +\frac{2 x
   y_2^2}{3}-\frac{x y_3^2 \gamma ^3}{(\gamma -1)^2}+\frac{16 x y_3^2 \gamma ^2}{3 (\gamma -1)^2}-\frac{157 x y_3^2
   \gamma }{18 (\gamma -1)^2}+\frac{41 x y_3^2}{9 (\gamma -1)^2}-x y_3 \gamma +\frac{2 x y_3}{9 (\gamma -1)}+\frac{4 x
   y_3}{3}-\frac{y_1^3 \gamma }{2}+y_1^3+y_1^2 y_3 \gamma +\frac{y_1^2 y_3}{3 (\gamma -1)}-\frac{7
   y_1^2 y_3}{3}-\frac{3 y_1^2 \gamma }{6-6 \gamma }+\frac{4 y_1^2}{6-6 \gamma }-\frac{1}{2} y_1 y_2^2 \gamma
   +\frac{2 y_1 y_2^2}{3}-\frac{y_1 y_3^2 \gamma ^3}{(\gamma -1)^2}+\frac{10 y_1 y_3^2 \gamma ^2}{3 (\gamma
   -1)^2}-\frac{73 y_1 y_3^2 \gamma }{18 (\gamma -1)^2}+\frac{16 y_1 y_3^2}{9 (\gamma -1)^2}-\frac{y_1 y_3
   \gamma ^3}{(\gamma -1)^2}+\frac{y_1 y_3 \gamma ^2}{(\gamma -1)^2}+\frac{5 y_1 y_3 \gamma }{3 (\gamma -1)^2}-\frac{16
   y_1 y_3}{9 (\gamma -1)^2}+\frac{y_2^2}{18-18 \gamma }+\frac{y_2^2}{6}-\frac{y_3^2 \gamma ^2}{3 (\gamma
   -1)^3}+\frac{5 y_3^2 \gamma }{6 (\gamma -1)^3}-\frac{14 y_3^2}{27 (\gamma -1)^3},$
\\\\$g_2(x,y_1,y_2,y_3)=-\frac{1}{2} x^2 y_2 \gamma +x^2 y_2-x y_1 y_2 \gamma +2 x y_1 y_2+x y_2 y_3 \gamma -\frac{x
   y_2 y_3 \gamma }{3 (\gamma -1)}+\frac{2 x y_2 y_3}{3 (\gamma -1)}-2 x y_2 y_3-\frac{1}{2} y_1^2
   y_2 \gamma +y_1^2 y_2+y_1 y_2 y_3 \gamma -\frac{y_1 y_2 y_3 \gamma }{3 (\gamma
   -1)}+\frac{2 y_1 y_2 y_3}{3 (\gamma -1)}-2 y_1 y_2 y_3-\frac{y_2^3 \gamma }{2}+\frac{2
   y_2^3}{3}-y_2 y_3^2 \gamma +\frac{y_2 y_3^2 \gamma }{3 (\gamma -1)}-\frac{y_2 y_3^2 \gamma }{18
   (\gamma -1)^2}-\frac{2 y_2 y_3^2}{3 (\gamma -1)}+\frac{y_2 y_3^2}{9 (\gamma -1)^2}+y_2 y_3^2-y_2
   y_3 \gamma$ and
\\\\
$g_3(x,y_1,y_2,y_3)=-\frac{1}{2} x^2 y_3 \gamma -\frac{x^2 \gamma }{2}-x y_1 y_3 \gamma +x y_1 y_3-x y_1 \gamma +x y_1+x
   y_3^2 \gamma -\frac{x y_3^2 \gamma }{3 (\gamma -1)}+\frac{x y_3^2}{3 (\gamma -1)}-x y_3^2+x y_3 \gamma -\frac{x
   y_3 \gamma }{3 (\gamma -1)}+\frac{x y_3}{3 (\gamma -1)}-x y_3-\frac{1}{2} y_1^2 y_3 \gamma +y_1^2
   y_3-\frac{y_1^2 \gamma }{2}+y_1^2+y_1 y_3^2 \gamma -\frac{y_1 y_3^2 \gamma }{3 (\gamma -1)}+\frac{2
   y_1 y_3^2}{3 (\gamma -1)}-2 y_1 y_3^2+y_1 y_3 \gamma -\frac{y_1 y_3 \gamma }{3 (\gamma
   -1)}+\frac{2 y_1 y_3}{3 (\gamma -1)}-2 y_1 y_3-\frac{1}{2} y_2^2 y_3 \gamma +\frac{2 y_2^2
   y_3}{3}-\frac{y_2^2 \gamma }{2}+\frac{2 y_2^2}{3}+\frac{y_3^3 \gamma }{3 (\gamma -1)}-\frac{y_3^3 \gamma }{18
   (\gamma -1)^2}-y_3^3 \gamma -\frac{2 y_3^3}{3 (\gamma -1)}+\frac{y_3^3}{9 (\gamma -1)^2}+y_3^3+\frac{y_3^2
   \gamma }{3 (\gamma -1)}-\frac{y_3^2 \gamma }{18 (\gamma -1)^2}-2 y_3^2 \gamma -\frac{2 y_3^2}{3 (\gamma
   -1)}+\frac{y_3^2}{9 (\gamma -1)^2}+y_3^2.$
\\\\
The system \eqref{center2} is written in diagonal form
\begin{align}
x'  &  =Cx+f\left(  x,\mathbf{y}\right) \nonumber\\
\mathbf{y}'  &  =P\mathbf{y}+\mathbf{g}\left(  x,\mathbf{y}\right)  ,
\label{center3}
\end{align}
where $\left(  x,\mathbf{y}\right)  \in\mathbb{R}\times\mathbb{R}^{3},$ $C$ is
the zero $1\times1$ matrix, $P$ is a $3\times3$ matrix with negative eigenvalues and $f,\mathbf{g}$ vanish at $\mathbf{0}$ and have vanishing derivatives at $\mathbf{0.}$ The center manifold theorem \ref{existenceCM} asserts that there exists a 1-dimensional invariant local center manifold $W^{c}\left(
\mathbf{0}\right) $ of \eqref{center3} tangent to the center subspace (the
$\mathbf{y}=\mathbf{0}$ space) at $\mathbf{0}.$ Moreover, $W^{c}\left(
\mathbf{0}\right)  $ can be represented as
\[
W^{c}\left(  \mathbf{0}\right)  =\left\{  \left(  x,\mathbf{y}\right)
\in\mathbb{R}\times\mathbb{R}^{3}:\mathbf{y}=\mathbf{h}\left(  x\right)
,\;\left\vert x\right\vert <\delta\right\}  ;\;\;\;\mathbf{h}\left(  0\right)
=\mathbf{0},\;D\mathbf{h}\left(  0\right)  =\mathbf{0},
\]
for $\delta$ sufficiently small (see definition \ref{CMdef}). The restriction of
(\ref{center3}) to the center manifold is (see definition \ref{vectorfieldCM})
\begin{equation}
x'=f\left( x,\mathbf{h}\left(  x\right)  \right)  . \label{rest}
\end{equation}
According to Theorem \ref{stabilityCM}, if the origin $x=0$ of \eqref{rest}
is stable (asymptotically stable) (unstable) then the origin of \eqref{center3} is also stable (asymptotically stable) (unstable). Therefore, we have to find the local center manifold, i.e.,
the problem reduces to the computation of $\mathbf{h}\left(  x\right).$

Substituting $\mathbf{y}=\mathbf{h}\left(  x\right)  $ in the second
component of \eqref{center3} and using the chain rule, $\mathbf{y
}'=D\mathbf{h}\left(  x\right)  x'$, one can show that the function
$\mathbf{h}\left(  x\right)  $ that defines the local center manifold
satisfies%
\begin{equation}
D\mathbf{h}\left(  x\right)  \left[  f\left(  x,\mathbf{h}\left(  x\right)
\right)  \right]  -P\mathbf{h}\left(  x\right)  -\mathbf{g}\left(
x,\mathbf{h}\left(  x\right)  \right)  =0. \label{h}
\end{equation}
According to Theorem \ref{approximationCM}, equation \eqref{h} can be solved approximately by using an approximation of $\mathbf{h}\left(  x\right)  $
by a Taylor series at $x=0.$ Since $\mathbf{h}\left(  0\right)  =\mathbf{0\ }
$and $D\mathbf{h}\left(  0\right)  =\mathbf{0},$ it is obvious that
$\mathbf{h}\left(  x\right)  $ commences with quadratic terms. We substitute
\[
\mathbf{h}\left(  x\right)  =:\left[
\begin{array}
[c]{c}%
h_{1}\left(  x\right) \\
h_{2}\left(  x\right) \\
h_{3}\left(  x\right)
\end{array}
\right]  =\left[
\begin{array}
[c]{c}%
a_{1}x^{2}+a_{2}x^{3}+O\left(  x^{4}\right) \\
b_{1}x^{2}+b_{2}x^{3}+O\left(  x^{4}\right) \\
c_{1}x^{2}+c_{2}x^{3}+O\left(  x^{4}\right)
\end{array}
\right]
\]
into (\ref{h}) and set the coefficients of like powers of $x$ equal to zero to find the unknowns $a_{1},b_{1},c_{1},...$.

Since $y_2$ absent from the first of \eqref{center3}, we
give only the result for $h_{1}\left(  x\right)$ and $h_{3}\left(  x\right).$ We find $a_1=\frac{\gamma }{6 (\gamma -1)},a_2=-\frac{3 \gamma -5}{9 (\gamma -1)},c_1=-\frac{1}{2},c_2=-\frac{2}{3}.$ \footnote{We find $b_1=b_2=0.$} Therefore, \eqref{rest} yields
\be x'=-\frac{2 x^2}{3}+\frac{5 x^3}{9}+\frac{4 x^4}{9}+O\left(x^5\right).\label{rest1}\ee

It is obvious that the origin $x=0$ of \eqref{rest1} is locally asymptotically unstable (saddle point). Hence, the origin $\mathbf{x}=\mathbf{0}$ of the full four-dimensional system is unstable.

{\bf Case $\gamma=1$}

Let be $\gamma=1.$
The Jacobian of \eqref{quadraticsyst} at $q=(0,0,1,0)$ has eigenvalues $-1,-1,-\frac{2}{3},$ and $0$ with
corresponding eigenvectors $\left(0,0,0,0,\frac{3}{2}\right)^T, \left(1,0,0,1\right)^T, (0,-3,0,0)^T$ and $(0,0,1,0)^T.$ As before we shift the fixed point to the origin by setting $\hat{\sigma_4}=\sigma_4-1$ and define new variables $(x,y_1,y_2,y_3)\equiv\mathbf{x}$, by the equations \begin{align}& u=\frac{3 x}{2},\nonumber\\&\sigma_2=x+y_1,\nonumber\\&\hat{\sigma}_4=-3y_3,\nonumber\\&\sigma_5=y_2\nonumber\end{align} so that

\be
\left(\begin{array}{c}x'\\y_1'\\y_2'\\y_3'
\end{array}\right)=\left(\begin{array}{cccc}0& 0 &0 &0\\0& -1 &0 &1\\0& 0 &-\frac{2}{3}&0\\0& 0 &0 &-1
\end{array}\right)\left(\begin{array}{c}x\\y_1\\y_2\\y_3
\end{array}\right)+\left(\begin{array}{c}f(x,y_1,y_2,y_3)\\g_1(x,y_1,y_2,y_3)\\g_2(x,y_1,y_2,y_3)\\g_3(x,y_1,y_2,y_3)\end{array}\right)\label{center2dust},
\ee
where
$f(x,y_1,y_2,y_3)=x^3+x^2 y_1-\frac{2 x^2}{3}-\frac{2 x y_1}{3},$
\\\\
$g_1(x,y_1,y_2,y_3)=-\frac{x^3}{2}+\frac{x^2 y_1}{2}+\frac{x^2}{2}+\frac{3 x y_1^2}{2}+\frac{x y_1}{3}+\frac{x y_2^2}{6}+\frac{9 x
   y_3^2}{2}-3 x y_3+\frac{y_1^3}{2}-\frac{y_1^2}{6}+\frac{y_1 y_2^2}{6}-\frac{9 y_1 y_3^2}{2}+3
   y_1 y_3-\frac{y_2^2}{6}-\frac{3 y_3^2}{2},$
\\\\$g_2(x,y_1,y_2,y_3)=\frac{x^2 y_2}{2}+x y_1 y_2+\frac{y_1^2 y_2}{2}+\frac{y_2^3}{6}-\frac{9 y_2 y_3^2}{2}+3 y_2
   y_3$ and
\\\\
$g_3(x,y_1,y_2,y_3)=-\frac{x^2 y_3}{2}+\frac{x^2}{6}+\frac{y_1^2 y_3}{2}-\frac{y_1^2}{6}+\frac{y_2^2
   y_3}{6}-\frac{y_2^2}{18}-\frac{9 y_3^3}{2}+\frac{9 y_3^2}{2}.$
\\\\
Observe that the system \eqref{center2dust} is now in the canonical form \eqref{center3}. Then, we proceed to the caculation of the center manifold. The procedure is fairly systematic and since we present it completely in the previous analysis we consider do not repeat it here. Instead, we present the relevant calculations. We obtain $a_1=\frac{2}{3},a_2=\frac{1}{3},b_1=0,b_2=0,c_1=\frac{1}{6},c_2=\frac{2}{9}$ for the Taylor expansion coefficients of \[
\mathbf{h}\left(  x\right)  =:\left[
\begin{array}
[c]{c}%
h_{1}\left(  x\right) \\
h_{2}\left(  x\right) \\
h_{3}\left(  x\right)
\end{array}
\right]  =\left[
\begin{array}
[c]{c}%
a_{1}x^{2}+a_{2}x^{3}+O\left(  x^{4}\right) \\
b_{1}x^{2}+b_{2}x^{3}+O\left(  x^{4}\right) \\
c_{1}x^{2}+c_{2}x^{3}+O\left(  x^{4}\right)
\end{array}
\right].
\] By substituting this values of the unknows  $a_{1},b_{1},c_{1},...$ we obtain that, for $\gamma=1,$ the dynamics of the center manifold in given also by equation \eqref{rest1}. The conclusion is straighforward: the origin $x=0$ of \eqref{rest1} is locally asymptotically unstable (saddle point). Hence, the origin $\mathbf{x}=\mathbf{0}$ of the full four-dimensional system is unstable.

This completes the proof. $\blacksquare$

The result of proposition \ref{centerP4} complements the result of the proposition discussed in \cite{Miritzis:2005hg} p. 5, where it was proved the local asymptotic instability of the de Sitter universe for positively curved FRW models with a perfect fluid matter source and a scalar field which arises in the conformal frame of the $R+\alpha R^2$ theory.

\subsection{$R^n$-gravity.}

Let us consider the model with $F(R)=R^n$ where we have reescaled
the usual multiplicative constant. It can be proved that, for
$n>1,$ $R^n$-gravity is conformally equivalent to a non-minimally
coupled scalar field with a positive potential
\begin{equation}V(\phi)=r(n)
e^{\lambda(n)\phi}\label{Rn}\end{equation} where $r(n)=\frac{1}{2}
(n-1) n^{-\frac{n}{n-1}}$ and
$\lambda(n)=-\frac{\sqrt{\frac{2}{3}} (n-2)}{n-1},$ with coupling
function given by \eqref{coup1}.

In this example, the evolution equations for $\varphi,$ $\sigma_2,$ $\sigma_4,$ and $\sigma_5$ are given by the equations
\eqref{eqvphi}-\eqref{eqzrad} with $M=\sqrt{2/3},\,N=-\frac{\sqrt{\frac{2}{3}} (n-2)}{n-1},$  $\overline{W}_{\chi}(\varphi)=\overline{W}_{V}(\varphi)=0,$ and $\overline{f'},$
given by (\ref{fQG}). The state space is defined by
$$\Omega_\epsilon=\{(\varphi, \sigma_2,
\sigma_4,\sigma_5)\in\mathbb{R}^4: 0\leq\varphi\leq
 \epsilon, \sigma_2^2+\sigma_4^2+\sigma_5^2\leq 1,
\sigma_4\geq 0, \sigma_5\geq 0\}.$$

In table \ref{crit01} and \ref{crit01a} are summarized the location, existence conditions and stability of the critical points.

\begin{table}[!ht]
\begin{center}
\caption{\label{crit01} Location of the critical points of the flow
of \eqref{eqvphi}-\eqref{eqzrad} defined in the invariant set
$\left\{p\in\Omega_\epsilon: \varphi=0\right\}$ for $M=\sqrt{2/3}$ and
$N=-\frac{\sqrt{\frac{2}{3}} (n-2)}{n-1}.$ We use the notations $n_+=\frac{1}{5}(4+\sqrt{6}),N_+=\frac{2}{27} \left(11+2 \sqrt{10}\right),\Gamma(n)=-\frac{2 n(n-2)}{3-9 n+6 n^2},$ and $\Gamma_+(\gamma)=\frac{9 \gamma +\sqrt{9 \gamma ^2+48 \gamma +16}+4}{12 \gamma +4}.$ }
\begin{tabular}[t]{|l|c|c|c|}
\hline
Label&$(\sigma_2,\sigma_4,\sigma_5)$&Existence\\[1ex]
\hline
\hline && \\[-2ex]
$P_1$&$(-1,0,0)$&always  \\[1ex]
\hline
\hline && \\[-2ex]
$P_2$&$(1,0,0)$&always \\[1ex]
\hline
\hline && \\[-2ex]
$P_3$&$\left(\frac{4-3\gamma}{3(2-\gamma)},0,0\right)$& $\left\{\begin{array}{c} 0<\gamma <\frac{4}{3},\, \text{or}\\ \frac{4}{3}<\gamma <\frac{5}{3}\end{array}\right.$ \\[1ex]
\hline
\hline && \\[-2ex]
$R_1$&$(0,0,1)$&always \\[1ex]
\hline
\hline && \\[-2ex]
$P_4$&$\left(\frac{n-2}{3(n-1)},\sqrt{1-\frac{(n-2)^2}{9(n-1)^2}},0\right)$& $n>\frac{5}{4}$ \\[1ex]
    \hline
\hline && \\[-2ex]
$R_3$ & $\left(\frac{2 (n-1)}{n-2},-\frac{\sqrt{2} (n-1)}{n-2}, -\frac{\sqrt{(8-5 n) n-2}}{n-2}\right)$ & $1<n<n_+$ \\[1ex]
\hline
\hline && \\[-2ex]
$P_5$ & $\left(\frac{3 (n-1) \gamma }{3 \gamma  n-2 n-3 \gamma },-\frac{\sqrt{2} \sqrt{n-1} \sqrt{4 n-3 \gamma }}{3 \gamma  n-2 n-3
   \gamma },0\right)$ & $\left\{\begin{array}{c} n=\frac{5}{4}, \, \gamma=\frac{5}{3}\; \text{or}\\
    0<\gamma<\frac{4}{3},\, 1<n\leq \Gamma_+(\gamma),\; \text{or}\\
    \frac{4}{3}<\gamma<\frac{5}{3},\, \frac{3 \gamma }{4}\leq n\leq \Gamma_+(\gamma).
\end{array}\right.$  \\[1ex]
\hline
\hline && \\[-2ex]
$P_6$ & $\left(\frac{3 (n-1) \gamma }{3 \gamma  n-2 n-3 \gamma },\frac{\sqrt{2} \sqrt{n-1} \sqrt{4 n-3 \gamma }}{3 \gamma  n-2 n-3
   \gamma },0\right)$ & $\frac{4}{3}<\gamma \leq \frac{5}{3}, n=\frac{3 \gamma }{4}$ $^{\rm a}$ \\[1ex]
\hline
\end{tabular}
\end{center}
$^{\rm a}$ In this case $P_6$ and $P_3$ coincides. Thus, the critical points are nonhyperbolic. There exists a 1-dimensional stable manifold and a 1-dimensional unstable manifold provided $\frac{4}{3}<\gamma <\frac{5}{3}.$
\end{table}

\begin{table}[!ht]
\begin{center}
\caption{\label{crit01a} Stability of the critical points of the flow
of \eqref{eqvphi}-\eqref{eqzrad} defined in the invariant set
$\left\{p\in\Omega_\epsilon: \varphi=0\right\}$ for $M=\sqrt{2/3}$ and
$N=-\frac{\sqrt{\frac{2}{3}} (n-2)}{n-1}.$ We use the notations $n_+=\frac{1}{5}(4+\sqrt{6}),N_+=\frac{2}{27} \left(11+2 \sqrt{10}\right),\Gamma(n)=-\frac{2 n(n-2)}{3-9 n+6 n^2},$ and $\Gamma_+(\gamma)=\frac{9 \gamma +\sqrt{9 \gamma ^2+48 \gamma +16}+4}{12 \gamma +4}.$ }
\begin{tabular}[t]{|l|c|c|}
\hline
Label&Stability$^{\rm a}$\\[1ex]
\hline
\hline & \\[-2ex]
$P_1$ & unstable for $0<\gamma<\frac{4}{3}, n>\frac{5}{4};$ \\[1ex]
              & or $\frac{4}{3}<\gamma<\frac{5}{3}, n>\frac{5}{4};$\\[1ex]
                & saddle, otherwise\\[1ex]
\hline
\hline &\\[-2ex]
$P_2$& unstable for $\gamma\neq \frac{4}{3}, n>1$\\[1ex]
                     & saddle, otherwise\\[1ex]
\hline
\hline & \\[-2ex]
$P_3$& saddle\\[1ex]
\hline
\hline & \\[-2ex]
$R_1$& saddle\\[1ex]
\hline
\hline & \\[-2ex]
$P_4$ & stable for\\[1ex]
   &               $\left\{\begin{array}{c}
\gamma\neq\frac{4}{3}, n>2 \; \text{or}\\
n_+<n\leq 2, \Gamma(n)<\gamma<\frac{4}{3} \; \text{or}\\

n_+<n\leq 2, \frac{4}{3}<\gamma<2.
\end{array}\right.$ \\[1ex]
                     & saddle, otherwise\\[1ex]
    \hline
\hline & \\[-2ex]
$R_3$   & stable for\\[1ex]
   &               $\left\{\begin{array}{c}
\frac{4}{3}<\gamma<2, N_+\leq n<n_+ \; \text{or}\\
\frac{4}{3}<\gamma<2, 1< n<N_+ .
\end{array}\right.$ \\[1ex]
    &                  saddle, otherwise\\[1ex]
\hline
\hline & \\[-2ex]
$P_5$ & stable for\\[1ex]
   &                  $\left\{\begin{array}{c} \frac{41}{25}\leq n<2,\, 0<\gamma <\Gamma(n)\;\text{or}\\
   1<n\leq N_+,\, 0<\gamma <\frac{4}{3}\;\text{or}\\N_+<n<\frac{41}{25},\, 0<\gamma <\frac{4 \sqrt{96 n^5-272 n^4+230 n^3-50 n^2}}{3 \left(4
   n^3-24 n^2+29 n-9\right)}-\frac{2 \left(10 n^3-19 n^2+13 n\right)}{3 \left(4 n^3-24 n^2+29 n-9\right)}.\end{array}\right.$ \\[1ex]
    &                  saddle, otherwise\\[1ex]
\hline
\end{tabular}
\end{center}
$^{\rm a}$ The stability is analysed for the flow restricted to the invariant set $\varphi=0$.
\end{table}

\begin{figure}[ht]
\begin{center}
\mbox{\epsfig{figure=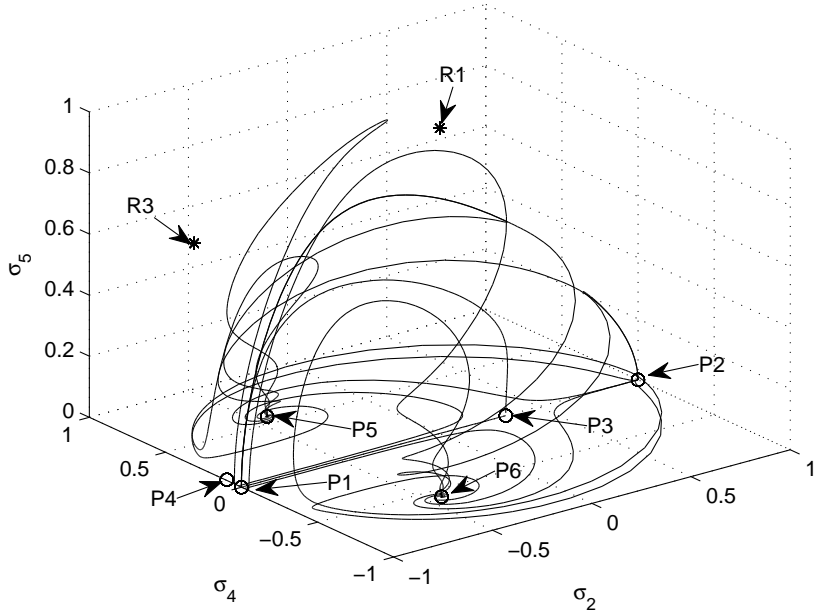,width=12cm,angle=0}}
\caption{ \label{fig3a}{\it Projection in $\varphi=0$ of some orbits of the flow of \eqref{eqvphi}-\eqref{eqzrad} for $M=\sqrt{2/3},\,N=-\frac{\sqrt{\frac{2}{3}} (n-2)}{n-1}.$ We set $n=1.251.$ Observe that $R_1,$ $R_3$ are in the region of physical interest. These are saddle points. The critical points $P_4$ and $P_3$ exist and are saddle points. $P_1$ and $P_2$ are local sources and $P_5$ is a local sink. We display some orbits in the halfspace $\sigma_4<0$ (corresponding to contracting universes) for aesthetical purposes. Observe that $P_6$ mirrors the behavior of $P_5.$}}
\end{center}
\end{figure}

\begin{figure}[ht]
\begin{center}
\mbox{\epsfig{figure=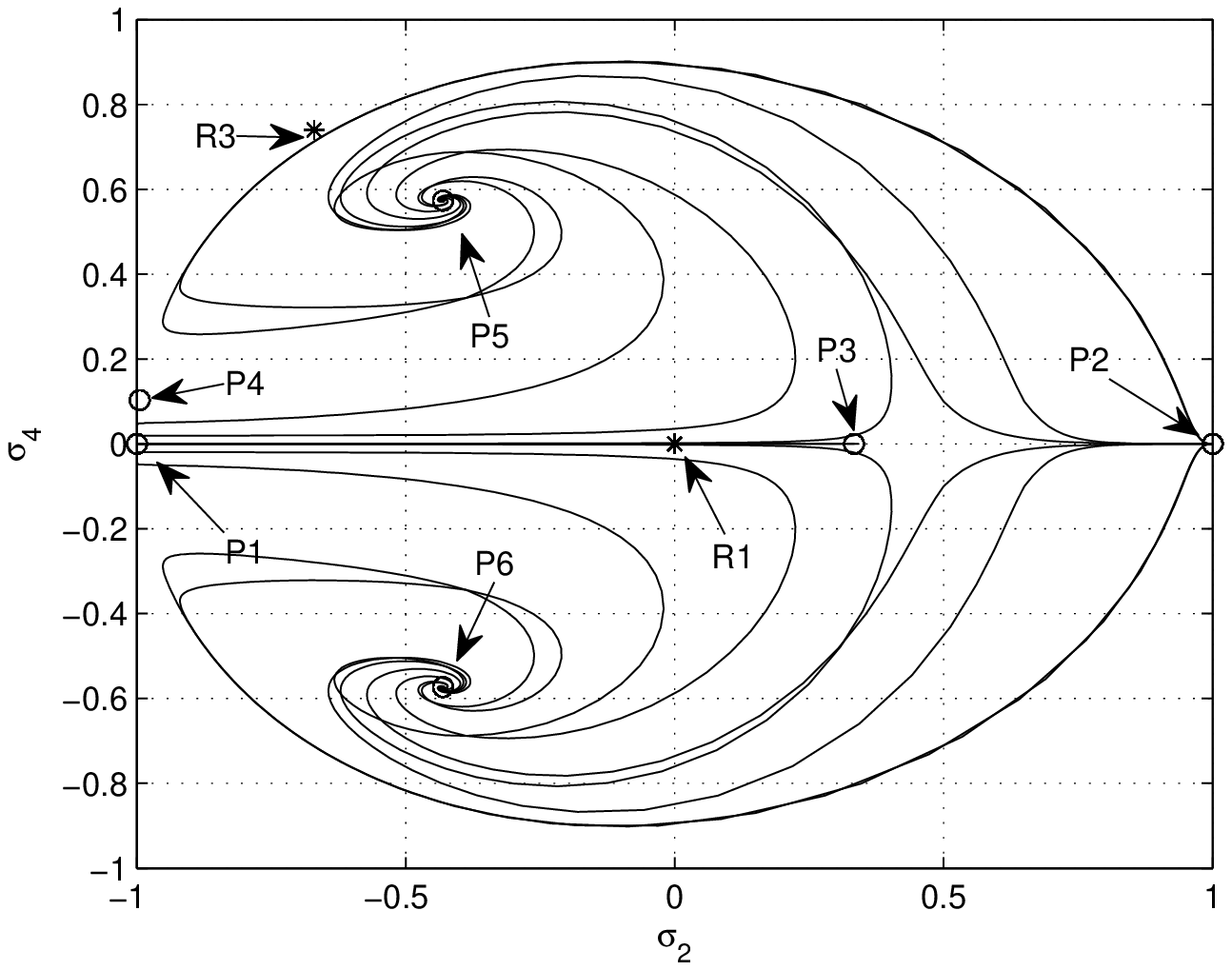,width=9.9cm,angle=0}}
\caption{ \label{fig3b}{\it Projection of the orbits displayed in figure \ref{fig3a} to $\sigma_5=0.$}}
\end{center}
\end{figure}

Let us discuss the stability properties of the critical points displayed in table \ref{crit01a}.

The critical point $P_1$ always exists. Its unstable manifold is 3D provided $0<\gamma<\frac{4}{3},\, n>\frac{5}{4}$ or $\frac{4}{3}<\gamma<2, n>\frac{5}{4}.$ Otherwise its unstable manifold is lower-dimensional.

The critical point $P_3$ always exists and it is a local source.

The critical point $P_3$ exists for $0<\gamma<\frac{4}{3}$ or $\frac{4}{3}<\gamma<2.$ It is neither a source nor a sink.

The critical point $R_1$ always exists and it is neither a source nor a sink.

The critical point $P_4$ exists for $n>\frac{5}{4}.$ Its stable manifold is 3D if $\gamma\neq\frac{4}{3}, n>2$ or
$n_+<n\leq 2, \Gamma(n)<\gamma<\frac{4}{3}$ or $n_+<n\leq 2, \frac{4}{3}<\gamma<2$. Where we have defined the grouping constants $n_+=\frac{1}{5}(4+\sqrt{6}),\Gamma(n)=-\frac{2 n(n-2)}{3-9 n+6 n^2}.$ Otherwise its stable manifold is lower-dimensional.

The critical point $R_3$ exists for $1<n<n_+.$ Thus $P_4$ and $R_3$ are in the same phase portrait for $\frac{5}{4}<n<n_+.$ $R_3$ admits a 3D stable manifold for $\frac{4}{3}<\gamma<2, N_+\leq n<n_+$ or
$\frac{4}{3}<\gamma<2, 1< n<N_+,$ where we have defined $N_+=\frac{2}{27} \left(11+2 \sqrt{10}\right).$ Otherwise its stable manifold is lower-dimensional.

The critical point $P_5$ exists for $n=\frac{5}{4}, \, \gamma=\frac{5}{3}$ or
    $0<\gamma<\frac{4}{3},\, 1<n\leq \Gamma_+(\gamma),$ or
    $\frac{4}{3}<\gamma<\frac{5}{3},\, \frac{3 \gamma }{4}\leq n\leq \Gamma_+(\gamma),$ where we have defined $\Gamma_+(\gamma)=\frac{9 \gamma +\sqrt{9 \gamma ^2+48 \gamma +16}+4}{12 \gamma +4}.$  $P_5$ admits a 3D stable manifold provided  $\frac{41}{25}\leq n<2,\, 0<\gamma <\Gamma(n),$ or
   $1<n\leq N_+,\, 0<\gamma <\frac{4}{3},$ or $N_+<n<\frac{41}{25},\, 0<\gamma <\frac{4 \sqrt{96 n^5-272 n^4+230 n^3-50 n^2}}{3 \left(4
   n^3-24 n^2+29 n-9\right)}-\frac{2 \left(10 n^3-19 n^2+13 n\right)}{3 \left(4 n^3-24 n^2+29 n-9\right)}.$ Otherwise its stable manifold is lower-dimensional.

The critical point $P_6$ exists for $\frac{4}{3}<\gamma\leq\frac{5}{3},\, n=\frac{3\gamma}{4}.$ In this case $P_6$ and $P_3$ coincides; thus, the critical points are non-hyperbolic with a 2D center manifold, a 1D unstable manifold and a 1D stable manifold.

Let us discuss some physical properties of the cosmological solutions associated to the critical points displayed in table \ref{crit01}.
\begin{itemize}
\item  $P_{1,2}$  represent kinetic-dominated cosmological solutions. They behave as stiff-like matter. The associated cosmological solution satisfies $H=\frac{1}{3 t-c_1},a=\sqrt[3]{3 t-c_1} c_2,\phi =c_3\pm\sqrt{\frac{2}{3}} \ln \left(3
   t-c_1\right),$ where $c_j,\,j=1,2,3$ are integration constants. These solutions are associated with the local past attractors of the systems for an open set of values of the parameter $\mu.$
\item  $P_3$ represents matter-dominated cosmological solutions that satisfy $H=\frac{2}{3 t \gamma -2 c_1},a=\left(3 t \gamma -2 c_1\right){}^{\frac{2}{3 \gamma }} c_2,\rho
   =\frac{12}{\left(3 t \gamma -2 c_1\right){}^2}+c_3.$
\item $R_1$ represents a radiation-dominated cosmological
solutions satisfying $H=\frac{1}{2 t-c_1},a=\sqrt{2 t-c_1}
c_2,\rho_r=\frac{3}{\left(2 t-c_1\right){}^2}+c_3.$ \item $P_4$
represents power-law scalar-field dominated inflationary
cosmological solutions. It is easy to obtain the asymptotic exact
solution:
\\
$H=\frac{3 (n-1)^2}{(n-2)^2 t-3 (n-1)^2
c_1},a=\left((n-2)^2 t-3 (n-1)^2 c_1\right){}^{\frac{3
(n-1)^2}{(n-2)^2}} c_3,\phi= c_2+\frac{\sqrt{6} (n-1) \ln
\left((n-2)^2 t-3 (n-1)^2 c_1\right)}{n-2}.$ \item $P_{5}$
represent matter-kinetic-potential scaling solutions satisfying

$H=\frac{3 (n-1) \gamma -2 n}{3 (n-2) t \gamma +(n (2-3 \gamma )+3 \gamma ) c_1},a= \left(3 (n-2) t
   \gamma +(n (2-3 \gamma )+3 \gamma ) c_1\right){}^{\frac{-3 \gamma  n+2 n+3 \gamma }{6 \gamma -3 n \gamma }} c_2,$ and $\rho =
   c_3-\frac{6 (3 \gamma +n (-9 \gamma +n (6 \gamma +2)-4))}{\left(3 (n-2) t \gamma +(n (2-3 \gamma )+3 \gamma )
   c_1\right){}^2},\phi = c_4+\frac{\sqrt{6} (n-1) \log \left(3 (n-2) t \gamma +(n (2-3 \gamma )+3 \gamma )
   c_1\right)}{n-2}.$
\item $R_3$ represent radiation-kinetic-potential scaling solutions satisfying $H= \frac{1}{2 t-c_1},a= \sqrt{2 t-c_1} c_2,\rho_r= c_3-\frac{3 (n (5 n-8)+2)}{(n-2)^2
   \left(c_1-2 t\right){}^2},\phi =c_4+\frac{\sqrt{6} (n-1) \log \left(2 t-c_1\right)}{n-2}.$
\end{itemize}

To complete the section we present in figures \ref{fig3a}, and \ref{fig3b}  a numerical elaboration of the model under consideration.

\subsection{Coupling functions and Potentials of exponential orders $M=0$ and $N=-\mu\neq 0,$ respectively.}

\begin{table}[!htb]
\caption{\label{crit} Location of the critical points of the flow
of \eqref{eqvphi}-\eqref{eqzrad} defined in the invariant set
$\left\{p\in\Omega_\epsilon: \varphi=0\right\}$ for $M=0$ and
$N=-\mu.$}
\begin{tabular}[t]{|l|c|c|c|c|}
\hline
Label&$(\sigma_2,\sigma_4,\sigma_5)$&Existence&Stability$^{\rm a}$\\[1ex]
\hline
\hline &&& \\[-2ex]
$P_1$&$(-1,0,0)$&always & unstable if $\mu>-\sqrt{6}$\\[1ex]
\hline
\hline &&& \\[-2ex]
$P_2$&$(1,0,0)$&always & unstable if $\mu<\sqrt{6}$\\[1ex]
\hline
\hline &&& \\[-2ex]
$P_3$&$(0,0,0)$&always & saddle\\[1ex]
\hline
\hline &&& \\[-2ex]
$R_1$&$(0,0,1)$&always & saddle\\[1ex]
\hline
\hline &&& \\[-2ex]
$P_4$&$\left(\frac{\mu}{\sqrt{6}},\sqrt{1-\frac{\mu^2}{6}},0\right)$&$\mu^2<6$ & stable for $\left\{\begin{array}{c} 0<\gamma<\frac{4}{3},\,\,\mu^2<3\gamma,\, \text{or}\\
\frac{4}{3}<\gamma<2,\,\, \mu^2<2
\end{array},\right.$\\
&&& saddle otherwise\\[1ex]
\hline
\hline &&& \\[-2ex]
$P_{5,6}$&$\left(\sqrt{\frac{3}{2}}\frac{\gamma}{\mu}, \pm \frac{1}{\mu}\sqrt{\frac{3}{2}(2-\gamma)\gamma},0 \right)$&$\mu^2>3\gamma$ & stable for $\left\{\begin{array}{c} 0<\gamma<\frac{2}{9},\,\mu^2>3\gamma,\, \text{or}\\
\frac{2}{9}<\gamma<\frac{4}{3},\,\,
3\gamma<\mu^2<\frac{24\gamma^2}{9\gamma-2},\, \text{or}\\
\frac{2}{9}<\gamma <\frac{4}{3},\,\mu ^2>\frac{24 \gamma ^2}{9 \gamma -2}
\end{array},\right.$\\
&&& saddle otherwise\\[1ex]
\hline
\hline &&& \\[-2ex]
$R_3$ & $\left(\frac{2\sqrt{\frac{2}{3}}}{\mu},\frac{2}{\sqrt{3}|\mu|},\frac{\sqrt{\mu^2-4}}{|\mu|}\right)$& $|\mu|>2$ & stable if $\frac{4}{3}<\gamma<2,$ saddle otherwise \\[1ex]
\hline
\end{tabular}
$^{\rm a}$ The stability is analysed for the flow restricted to the invariant set $\varphi=0$.
\end{table}

As an example let us consider $\chi,V\in {\cal E}^2_+$ of exponential orders $M=0$ and $N=-\mu,$ respectively. This class of potentials contains the cases investigated in \cite{Copeland:1997et,vandenHoogen:1999qq} (there are not considered coupling to matter, i.e., $\chi(\phi)\equiv 1$, in the second case, for flat FRW cosmologies), the case investigated in \cite{Copeland:2009be} (for positive potentials and standard FRW dynamics), the example examined in \cite{Leon:2008de}, etc. In table \ref{crit} are summarised the location, existence conditions and stability of the critical points. \footnote{The stability is analysed for the flow restricted to the invariant set $\varphi=0$, i.e., we are not taking into account perturbations in the $\varphi$-axis.}

Let us discuss some physical properties of the cosmological solutions associated to the critical points displayed in table \ref{crit}.
\begin{itemize}
\item  $P_{1,2}$  represent kinetic-dominated cosmological solutions. They behave as stiff-like matter. The associated cosmological solution satisfies $H=\frac{1}{3 t-c_1},a=\sqrt[3]{3 t-c_1} c_2,\phi =c_3\pm\sqrt{\frac{2}{3}} \ln \left(3
   t-c_1\right),$ where $c_j,\,j=1,2,3$ are integration constants. These solutions are associated with the local past attractors of the systems for an open set of values of the parameter $\mu.$
\item  $P_3$ represents matter-dominated cosmological solutions that satisfy $H=\frac{2}{3 t \gamma -2 c_1},a=\left(3 t \gamma -2 c_1\right){}^{\frac{2}{3 \gamma }} c_2,\rho
   =\frac{12}{\left(3 t \gamma -2 c_1\right){}^2}+c_3.$
\item $R_1$ represents a radiation-dominated cosmological solutions satisfying $H=\frac{1}{2 t-c_1},a=\sqrt{2 t-c_1} c_2,\rho_r=\frac{3}{\left(2 t-c_1\right){}^2}+c_3.$
\item $P_4$ represents power-law scalar-field dominated inflationary cosmological solutions. As $t\rightarrow 0^+$ the potential behaves as $V\sim V_0 \exp[-\mu \phi]$. Thus it is easy to obtain the asymptotic exact solution: $H=\frac{2}{t \mu ^2-2 c_1},a=\left(t \mu ^2-2 c_1\right){}^{\frac{2}{\mu ^2}} c_2,\phi \sim\frac{1}{\mu}\ln\left[\frac{V_0(t\mu^2-2c_1)^2}{2(6-\mu^2)}\right].$
\item $P_{5,6}$ represent matter-kinetic-potential scaling solutions. As before, in the limit $t\rightarrow 0^+$ we obtain the asymptotic expansions: $H=\frac{2}{3 t \gamma -2 c_1},a=\left(3 t \gamma -2 c_1\right){}^{\frac{2}{3 \gamma }} c_2,\phi \sim \frac{1}{\mu}\ln\left[\frac{V_0\mu^2(3t\gamma-2 c_1)^2}{18(2-\gamma)\gamma}\right].$
\item $R_3$ represent radiation-kinetic-potential scaling solutions. As before are deduced the following asymptotic expansions: $H=\frac{1}{2 t-c_1},a=\sqrt{2 t-c_1} c_2,\phi \sim \frac{1}{\mu}\ln\left[\frac{v_0\mu^2(2t-c_1)^2}{4}\right].$
\end{itemize}

To finish this section let us re-examine the example discussed in \cite{Leon:2008de} section B.1 in presence of radiation.

\subsubsection{The dynamics near $\phi=+\infty$ for the model with Power-law coupling and the Albrecht-Skordis potential}

Let us consider the coupling function \be \chi(\phi)=\left(\frac{3\alpha}{8}\right)^{\frac{1}{\alpha}}\chi_0(\phi-\phi_0)^
{\frac{2}{\alpha}},\; \alpha>0,\text{const.},\,\phi_0>0.\label{couplingexample}\ee

Observe that $$\frac{d\ln \chi(\phi)}{d\chi}=\frac{2}{\alpha(\phi-\phi_0)}\neq 0$$ for all finite value of $\phi.$ Since $\ln \chi(\phi)$ has not stationary points thus the early time dynamics is associated to the limit where the scalar field diverges.

This choice produces a coupling BD parameter given by

$$2\omega(\chi)+3=\frac{4}{3}\alpha\left(\frac{\chi}{\chi_0}\right)^\alpha.$$

This types of power law couplings were investigated in \cite{DTorres} from the astrophysical viewpoint. For STTs without potential, the cosmological solutions for the matter domination era (in a Robertson-Walker metric) are $a(t)\propto (\ln t)^{(\alpha-1)/3\alpha} t^{\frac{2}{3}},\; \phi(t)\propto (\ln t)^{\frac{1}{\alpha}}.$ The values of the parameter $\alpha$ in concordance with the predictions of ${}^4 H$
are $\alpha=1, \, 0.33,\, 3$ (see table 4.2 in \cite{DTorres}).

Let us consider also the Albrecht-Skordis potential given by \cite{Albrecht:1999rm}
\begin{equation}
V(\phi )=e^{-\mu \phi }{\left( A+(\phi -B)^2\right).}  \label{Albrecht-Skordis}
\end{equation}

This particularly attractive model of quintessence is driven by
a potential which introduces a small minimum to the exponential potential.
Unlike previous quintessence models, late-time acceleration is achieved without fine tuning of the initial conditions. The authors argue that such potentials arise naturally in the low-energy limit of $M$-theory. The constant parameters, $A$  and $B$, in the potential take values of order $1$ in Planck units, so there is also no fine tuning of the potential  (we suppose also that $\mu\neq 0$). They show that, regardless of the initial conditions, $\rho _\phi $ scales, with  $\rho \propto \rho _\phi \propto t^{-2}$ during the radiation and matter eras, but leads to permanent vacuum domination  and accelerated expansion after a time which can be close to the present.

The extremes of of the potential (\ref{Albrecht-Skordis}) are located at $\phi^{\pm}=\frac{1+B\mu-\sqrt{1-A\mu^2}}{\mu}.$
They are real if $1\geq
\mu ^2A.$ The local minimum (respectively, local maximum)  is located at $\phi^-$ (respectively $\phi^+$) since
$$\pm V''(\phi^{\pm})=- 2 V_0\sqrt{1-A\mu^2}e^{-\left(1+B\mu\pm\sqrt{1-A\mu^2}\right)}<0.$$

As we investigated in section \ref{Qualitative}, the late time
dynamics of the flow of \eqref{eqsigma1}-\eqref{eqsigma5} is associated with the
extremes of the potential (the critical point $P_2=(0,0,0)$). When we restrict ourselves to this
invariant set, we find that the critical point associated to
$\phi^+$ is always a saddle point of the corresponding phase
portrait. The critical point associated to $\phi^-$ could be
either a stable node or a stable spiral if
$$\frac{8(3+2\mu^2)}{\left(3+4\mu^2\right)^2}<A\leq
\frac{1}{\mu^2}$$ or
$$A<\frac{8(3+2\mu^2)}{\left(3+4\mu^2\right)^2}.$$ The early time
dynamics of the flow of \eqref{eqsigma1}-\eqref{eqsigma5} corresponds to the limit
$\phi=+\infty.$

Observe first that \be W_\chi(\phi)=\partial_\phi \chi(\phi)/\chi(\phi)=\frac{2}{\alpha  (\phi -\phi_0)} \Rightarrow
\lim_{\phi\rightarrow +\infty} W_\chi(\phi)=0\ee and  \be W_V(\phi)=\partial_\phi V(\phi)/V(\phi)+\mu=\frac{2(\phi-B)}{A+(B-\phi)^2} \Rightarrow \lim_{\phi\rightarrow +\infty} W_V(\phi)=0.\ee  In other words, the coupling function (\ref{couplingexample}) and the potential (\ref{Albrecht-Skordis}) are WBI of exponential orders $M=0$ and $N=-\mu,$ respectively.

It is easy to prove that Power-law coupling and the Albrecht-Skordis potential are at least  ${\cal E}^2_+,$  under the
admissible coordinate transformation \footnote{We fix here an error in formulas B6-B9 in \cite{Leon:2008de}. With the choice $\vphi=\phi^{-1}$ the resulting barred functions given by B7-B9 there, are not of the desired differentiable class.}

\be \vphi=\phi^{-\frac{1}{2}}=f(\phi)\label{transformAS}.\ee

Using the above coordinate transformation we find

\be\overline{W}_{\chi}(\vphi)=\left\{\begin{array}{rcr} \frac{2 \varphi^2 }{\alpha  (1-\varphi^2  \phi_0)}&,&\vphi>0\\
                                     0&,&\vphi=0 \end{array}\right.\label{WchiAS}\ee

\be\overline{W}_V(\vphi)=\left\{\begin{array}{rcr} -\frac{2\vphi^2(B\vphi^2-1)}{A\vphi^4+(B\vphi^2-1)^2}&,&\vphi>0\\
                                     0 &,&\vphi=0 \end{array}\right.\label{WVAS}\ee
and

\be\overline{f'}(\vphi)=\left\{\begin{array}{rcr} -\frac{1}{2}\vphi^3&,&\vphi>0\\
                                     0&,&\vphi=0 \end{array}\right.\label{fAS}\ee

In this example, the evolution equations for $\varphi,$ $\sigma_2,$ $\sigma_4,$ and $\sigma_5$ are given by the equations
\eqref{eqvphi}-\eqref{eqzrad} with $M=0,\,N=-\mu,$ and $\overline{W}_{\chi}(\varphi),\,\overline{W}_{V}(\varphi)=0,$ and $\overline{f'},$
given by \eqref{WchiAS}, \eqref{WVAS} and \eqref{fAS} respectively. The state space is defined by
$$\Omega_\epsilon=\{(\varphi, \sigma_2,
\sigma_4,\sigma_5)\in\mathbb{R}^4: 0\leq\varphi\leq
 \sqrt{\epsilon}, \sigma_2^2+\sigma_4^2+\sigma_5^2\leq 1,
\sigma_4\geq 0, \sigma_5\geq 0\}.$$

\begin{figure}[ht]
\begin{center}
\mbox{\epsfig{figure=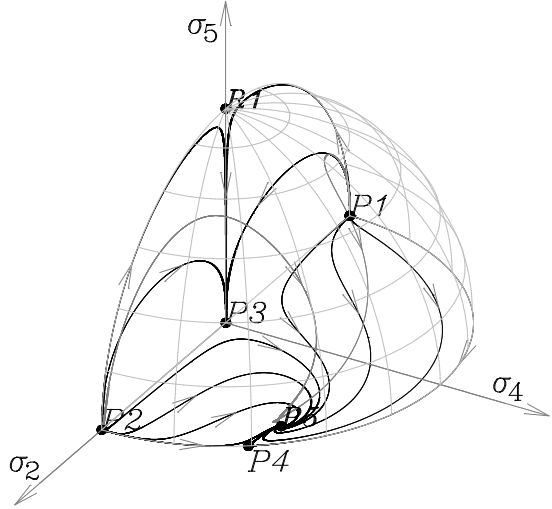,width=9.9cm,angle=0}}
\caption{ \label{fig4}{\it Some orbits in the invariant set $\sigma_2^2+\sigma_4^2+\sigma_5^2\leq 1$ for the choice of $\varphi=0$ for the model with coupling function (\ref{couplingexample})  potential (\ref{Albrecht-Skordis}). We select the values of the parameters: $\epsilon=1.00,$ $\mu= 2.00, A = 0.50, \alpha = 0.33, B = 0.5,$ and $\phi_0=0.$}}
\end{center}
\end{figure}

In figure \ref{fig4} we show some orbits in the invariant set $\sigma_2^2+\sigma_4^2+\sigma_5^2\leq 1$ for $\varphi=0$ for the model with coupling function (\ref{couplingexample})  potential (\ref{Albrecht-Skordis}). We select the values of the parameters: $\epsilon=1.00,$ $\mu= 2.00, A = 0.50, \alpha = 0.33, B = 0.5,$ and $\phi_0=0.$ In this case $P_5$ is the local sink in this invariant set.

\section{Conclusion}

In this chapter we have extended several results about flat FRW models in the conformal (Einstein) frame in scalar-tensor gravity theories including $f(R)$ theories through conformal transformation). Particularly we have considered a cosmological model based on the action \begin{align}&S_{EF}=\int_{M_4} d{ }^4 x \sqrt{|g|}\left\{\frac{1}{2} R-\frac{1}{2}(\nabla\phi)^2-V(\phi)+\chi(\phi)^{-2}
\mathcal{L}_{matter}(\mu,\nabla\mu,\chi(\phi)^{-1}g_{\alpha\beta})\right\},
\end{align}
where $R$ is the curvature scalar, $\phi$ is the a scalar field, and $V$ is the potential of the scalar field and $\chi$ is the coupling function.
We have consider both ordinary matter described by a perfect fluid with equation of state $p=(\gamma-1)\rho$ (coupled to the scalar field) and radiation $\rho_r$ in order to describe the dynamics in a cosmological epoch where matter and radiation coexisted.

We have considered scalar fields with arbitrary (positive) potentials and arbitrary coupling functions from the beginning. Then, we have straightforwardly introduced mild assumptions under such functions (differentiable class, number of critical points, asymptotes, etc.) in order to clarify the structure of the phase space of the dynamical system. We have obtained several analytical results. Also, we have presented several numerical evidences that confirm some of these results.

Our main results are the following.

\begin{enumerate}
\item We have formulated and proved proposition \ref{thmIII} generalizing analogous result in \cite{Giambo':2009cc}. It states that if the potential $V(\phi)$ is such that the (possibly empty set) where it is negative is bounded and the (possibly empty) set of critical points of $V(\phi)$ is finite, then, the critical point $${\bf
p}_*:=\left(\phi_*,y_*=0,\rho_*=0,\rho_r=0,
H=\sqrt{\frac{V(\phi_*)}{3}}\right),$$ where $\phi_*$ is a strict local minimum for $V(\phi),$  is an asymptotically stable
equilibrium point for the flow. From the physical viewpoint this result is relevant since it provides conditions for the asymptotic stability of the \emph{de Sitter} solution.
\item After the introduction of modified normalized variables, we have proved that the phase space has the structure of a manifold with boundaries (see propositions \ref{thm1}, \ref{thm2}). We have devised several monotonic functions for the flow of the dynamical system which allow for the identification of some invariant sets.
\item We have provided normal forms for the vector field around the inflection points and the strict degenerate local minimum of the potential. The normal forms have been used to derive rates at which the critical point under investigation is approached. The results obtained are in agreement with the result in proposition \ref{thmIII}.
\item We have proved theorem \ref{thm4} which is a generalization of the related result in \cite{Leon:2008de,Foster:1998sk,Miritzis:2003ym}. This result state that if $\chi(\phi)$ and $V(\phi)$ are positive functions of
class $C^3,$ such that $\chi$ has at most a finite number of stationary points and does not tend to zero in any compact set of $\mathbb{R}$, then, given $p,$ an interior point of the phase space manifold, the scalar field, $\phi,$ is unbounded through the past orbit $O^{-}(p).$ The relevance from the physical viewpoint of this result is twofold. First, the inclusion of radiation in the cosmic budget does not influence radically the early-time behavior of the scalar field. This result is somewhat expected since for small scale factor $a,$ the energy densities of radiation and the scalar field goes respectively as $\rho_r\sim a^{-4},$ and $\rho_\phi\approx \frac{\dot\phi^2}{2}\sim a^{-6}$ (the last approximation is supported by theorems 4.1 and 4.2 in \cite{Leon:2008de}). Second, the result of theorem \ref{thm4} makes clear that in order to investigate the generic past asymptotic dynamics of the flow we must scan the region of the phase space where $|\phi|\rightarrow\infty$.
\item For the analysis of the system as $\phi\rightarrow\infty$ we have defined a suitable change of variables to bring a neighborhood of $\phi=\infty$ in a bounded set. This method was first introduced in \cite{Foster:1998sk} (see also \cite{Giambo:2008sa}). Having constructed a dynamical system well-suited to investigate the dynamics in a neighborhood of $\phi=\infty$, i.e., near the initial singularity, we have proceeded to the identification of the equilibrium points and to the analysis of their stability by examining the signs of the eigenvalues. We have charaterized the associated cosmological solutions. We have obtained in this regime: radiation-dominated cosmological solutions; power-law scalar-field dominated inflationary cosmological solutions; matter-kinetic-potential scaling solutions and radiation-kinetic-potential scaling solutions.
\item Using the mathematical apparatus developed in the first part of the chapter, we have investigated the important examples of higher order gravity theories $F(R) = R + \alpha R^2$ (quadratic gravity) and $F(R) =R^n.$  In the case of quadratic gravity we have proved in proposition \ref{centerP4}, by an explicit computation of the center manifold, that the equilibrium point corresponding to \emph{de Sitter} solution is locally asymptotically unstable (saddle point). This result complements the result of the proposition discussed in \cite{Miritzis:2005hg} p. 5, where it was proved the local asymptotic instability of the \emph{de Sitter} universe for positively curved FRW models with a perfect fluid matter source and a scalar field which arises in the conformal frame of the $R+\alpha R^2$ theory.
\item Finally, we have investigated a general class of potentials containing the cases investigated in \cite{Copeland:1997et,vandenHoogen:1999qq}. In order to provide a numerical evidence for our analytical results for this class of models, we have re-examined the toy model with power-law coupling and Albrecht-Skordis potential $V(\phi )=e^{-\mu \phi }{\left( A+(\phi -B)^2\right)}$ investigated in \cite{Leon:2008de} in presence of radiation.

\end{enumerate}

\section*{Acknowledgements}

The authors wish to thanks to Dr. John Miritzis which kindly read the original manuscript and provide us with a nice and very valuable review.
This investigation is partially supported by the MES of Cuba. G L was also supported by Programa Nacional de Ciencias B\'asicas (PNCB).

Reviewed by Dr. John Miritzis\\
Department of Marine Sciences\\
University of the Aegean\\
University Hill, Mytilene 81100, Greece\\
Tel. (O)  +30 22510 36812\\
Tel. (H)  +30 22510 27512\\

%\label{lastpage-01}

\end{document}